\documentclass[11pt]{article}

\usepackage[margin=1in]{geometry}
\usepackage{setspace}
\usepackage[pagewise]{lineno}
\usepackage{authblk}
\usepackage{booktabs}
\usepackage{hyperref}
\hypersetup{colorlinks=true,citecolor=blue,filecolor=magenta}
\usepackage[utf8]{inputenc}
\usepackage{subcaption}
\usepackage{caption}
\usepackage{graphicx}
\usepackage{float}
\usepackage{amsmath}
\usepackage[version=4]{mhchem}
\usepackage{siunitx}
\usepackage{longtable,tabularx}
\usepackage[toc,page]{appendix}
\usepackage{apacite}
\usepackage{url}
\usepackage{xargs}
\usepackage[colorinlistoftodos,prependcaption,textsize=tiny]{todonotes}
\newcommandx{\unsure}[2][1=]{\todo[linecolor=red,backgroundcolor=red!25,bordercolor=red,#1]{#2}}
\newcommandx{\change}[2][1=]{\todo[linecolor=blue,backgroundcolor=blue!25,bordercolor=blue,#1]{#2}}
\newcommandx{\info}[2][1=]{\todo[linecolor=teal,backgroundcolor=teal!25,bordercolor=teal,#1]{#2}}
\newcommandx{\improvement}[2][1=]{\todo[linecolor=magenta,backgroundcolor=magenta!25,bordercolor=magenta,#1]{#2}}
\newcommandx{\thiswillnotshow}[2][1=]{\todo[disable,#1]{#2}}

\newcounter{mcounter}
\newcommand\dispmcounter{\refstepcounter{mcounter}\themcounter}
\newcommand{\incfig}{\centering\includegraphics}
\newcommand\abs[1]{\left|#1\right|}

\title{Criteria for dynamic stall onset and vortex shedding in low Reynolds number flows}

\author{Sarasija Sudharsan\footnote{Graduate student, Department of Aerospace Engineering, Iowa State University (ISU).} and Anupam Sharma\footnote{Associate Professor, Department of Aerospace Engineering, ISU.}}

\affil{Iowa State University, Ames, IA, 50011}
\date{November 6, 2023}

\begin{document}

\maketitle
\onehalfspacing

\begin{abstract}

Dynamic stall at low Reynolds numbers, $Re \sim O(10^4)$, exhibits complex flow physics with co-existing laminar, transitional, and turbulent flow regions.
Current state-of-the-art stall onset criteria use parameters that rely on flow properties integrated around the leading edge.
These include the leading edge suction parameter or $LESP$ \shortcite{Ramesh2014} and boundary enstrophy flux or $BEF$ \shortcite{Sudharsan2022}, which have been found to be effective for predicting stall onset at moderate to high $Re$.
However, low $Re$ flows feature strong vortex shedding events occurring across the entire airfoil surface, including regions away from the leading edge, altering the flow field and influencing the onset of stall.
In the present work, the ability of these stall criteria to effectively capture and localize these vortex shedding events in space and time is investigated.
High-resolution large-eddy simulations for an SD7003 airfoil undergoing a constant-rate, pitch-up motion at two $Re$ ($10,000$ and $60,000$) and two pitch rates reveal a rich variety of unsteady flow phenomena, including instabilities, transition, vortex formation, merging, and shedding, which are described in detail. 
While stall onset is reflected in both $LESP$ and $BEF$, local vortex-shedding events are identified only by the $BEF$.
Furthermore, these events can be localized in space and time by considering the contributions to the $BEF$ from different airfoil sections, which holds significant promise for effective flow control.

\end{abstract}



\section{Introduction}
\label{sec:introduction}
Dynamic stall is a topic of great interest in unsteady flows since it can lead to aerodynamic forces and moments severe enough to cause catastrophic structural failure~\shortcite{McCroskey1981,Corke2015}. 
Stall control efforts are most effective before the formation of the dynamic stall vortex (DSV)~\shortcite{Chandrasekhara2007}, a characteristic feature of `deep' dynamic stall.
Therefore, characterizing stall onset is of crucial importance for control efforts to be deployed in a timely manner.
Various criteria for dynamic stall onset based on unsteady aerodynamic coefficients have been explored to formulate first-order, semi-empirical, dynamic stall models~\shortcite{Leishman1989, Sheng2005}.
Several stall criteria have been proposed to narrow down the identification of stall onset to a finer degree in time. 
These include the leading edge suction parameter or the $LESP$~\shortcite{Ramesh2014}, which is pressure-based, and the boundary enstrophy flux or the $BEF$~\shortcite{Sudharsan2022}, which is vorticity-based. 
Reduced-order models of unsteady stall incorporate stall criteria based on the $LESP$ to determine the onset of leading edge vortex shedding. 
In the present work, we analyze the max($\abs{BEF}$) criterion to characterize vortex shedding and unsteady stall at low Reynolds numbers and compare it against the max($LESP$) criterion. 
$BEF$ has previously been explored for chord-based Reynolds numbers, $Re \sim \mathcal{O}\left(10^5 - 10^6\right)$, and has been found to reach its maximum magnitude in advance of DSV formation.
More generally, the max($\abs{BEF}$) criterion signifies the instance of maximum wall shear and indicates imminent vortex formation.

We first summarize the flow physics of low-$Re$ dynamic stall and make the case that the properties of the $BEF$ make it particularly suitable for characterizing unsteady stall at low $Re$.

\begin{figure}
	\centering
	\incfig[width=0.6\textwidth]{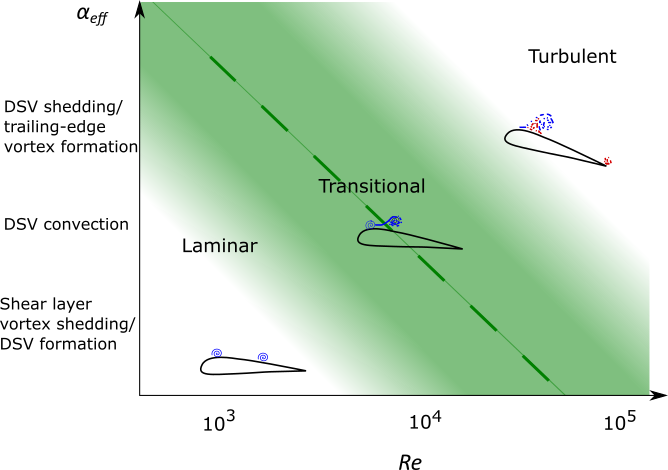}
	\caption{Schematic showing flow regimes prevailing over varying $\alpha_{eff}$ and $Re$}
	\label{fig:laminar_transitional_DSV}
\end{figure}

Low-$Re$ aerodynamics are dominated by a rich variety of coherent vortical structures such as shear layer vortices, a dynamic stall system comprised of multiple vortices, and induced secondary vortex flow structures.
These vortices generally start out as laminar with strong spanwise coherence and undergo transition as the airfoil angle of attack increases.
In their large-eddy simulations (LES) of a plunging airfoil at $Re$ 60,000, \shortciteA{Visbal2011} found that the DSV system and the shear layer vortices independently undergo flow transition. 
In the same study, they observed that the flow remained laminar through the entire plunging cycle at $Re$ 1,000.
Figure~\ref{fig:laminar_transitional_DSV} shows a schematic of the type of flow observed at moderate pitch/plunge rates as a function of $Re$ and the effective angle of attack, $\alpha_{\rm{eff}}$, which is the instantaneous angle of attack accounting for the airfoil motion.
Note that the unsteadiness introduced by the pitching/plunging maneuver could shift the transition boundary.

As $\alpha_{\rm eff}$ increases, spanwise-coherent shear layer vortices are formed, followed by the establishment of a DSV system, which propagates downstream and is eventually shed.
At lower $Re$ and $\alpha_{\rm eff}$, the flow over the airfoil is laminar, though the laminar boundary layer is susceptible to separation when subject to an adverse pressure gradient (APG), resulting in the formation of a laminar separation bubble (LSB).
If the APG remains low, the free shear layer above the bubble does not transition and instead reattaches laminarly.
This leads to a long LSB extending over most of the airfoil with fully laminar flow~\shortcite{Shyy2007}.
However, as $\alpha_{\rm eff}$ or $Re$ is increased, the APG downstream of the suction peak increases, encouraging spanwise instabilities that lead to earlier transition and reattachment, and shorter LSBs.
Therefore, during a pitch-up motion at moderate pitch rates, the DSV system could remain laminar through part of the motion depending on the $Re$, as illustrated by Fig.~\ref{fig:laminar_transitional_DSV}.
Unsteady flow at such low $Re$ can consist of a mix of laminar, transitional, and turbulent flow regimes in space and time.
In contrast, transitional effects are very small at high $Re$ since the shear layer transitions close to the leading edge, with turbulent flow prevailing over most of the airfoil.
It has also been observed that a single coherent DSV is more typical of high $Re$, while a DSV system consisting of one or more laminar/transitional vortices is likely at low $Re$~\shortcite{Galbraith2010}.
At high $Re$, there is a pronounced effect on the moment coefficient, $C_m$, which undergoes severe divergence~\shortcite{Visbal2011}. 
In contrast, multiple, relatively less severe moment stall occurrences are typical at low $Re$.
Dynamic stall vortices are shed faster, and stall occurs earlier.
The early stall is attributed to the stronger viscous response of the boundary layer at low $Re$, leading to the earlier formation of secondary structures having strong circulation that cut off the primary vortices from the feeding shear layer~\shortcite{Widmann2017}.

The structure of the LSB is highly sensitive to $Re$, as determined from experiments on an NACA $66_3$-$018$ airfoil by \shortciteA{OMeara1987}, where the LSB more than doubled in length when the $Re$ was decreased from 140,000 to 50,000.
\shortciteA{ol2009shallow} reported a similar increase in LSB size with reducing $Re$ in their experiments in a free-surface water tunnel where shallow and deep stall cases were investigated for a pitch-plunge motion of a rigid airfoil in the range $10,000 \le Re \le 60,000$.
The vortices shed at low $Re$ also have larger length scales. 
\shortciteA{Visbal2009} observed that the vorticity contained in laminar DSVs was an order of magnitude greater compared to fully transitioned DSVs, which was attributed to increased viscous dissipation and cancellation of primary vorticity by secondary vorticity ejected from the wall in the turbulent case.
The formation of laminar vortices is also accompanied by larger streamwise gradients in momentum thickness and edge velocities~\shortcite{Shyy2007}.
The above observations suggest that the $BEF$, which is sensitive to changes in pressure gradient and vorticity, which are especially pronounced for laminar vortices, would be effective at signaling instances of vortex formation.


The current state-of-the-art criterion used to characterize unsteady stall is the $LESP$ \shortcite{Ramesh2014, Narsipur2020}, which is the camber-wise suction force at the leading edge. 
$LESP$ has been used to trigger leading edge vortex shedding in reduced-order models based on a pre-determined critical value.
$\max(LESP)$ is a standalone criterion that has been used as a proxy for critical $LESP$ in some cases~\shortcite{Deparday2018,Deparday2019} to identify the onset of dynamic stall.
Both criteria, $\max(LESP)$ and $\max(\abs{BEF})$, have previously been observed to reach their peak magnitudes in advance of DSV formation regardless of the stall type (leading-edge, mixed, or trailing-edge) at $Re \sim \mathcal{O}\left(10^5 - 10^6\right)$, in cases where DSV played a significant role in the stall process~\shortcite{Sudharsan2023}.
Therefore, the max($LESP$) criterion is evaluated in the present work for comparison.

While some studies (e.g., \shortciteA{Visbal2009}) have provided descriptions of the unsteady flow field at low $Re$, we focus on the development of transitional instabilities in the flow and how they affect leading edge flow, where stall criteria are typically evaluated.
We assess the applicability of the $\max(\abs{BEF})$ and max($LESP$) criteria to $Re \sim \mathcal{O}(10^4)$, both for signaling imminent DSV formation and localized vortex shedding.
Expanding their applicability would enable the use of a fundamental, standalone parameter in reduced-order dynamic stall models for the prediction of vortex shedding and DSV formation over a wider range of $Re$.
It also holds significant promise for stall control efforts.



\section{Methods and Datasets}
\label{sec:methods}
\begin{figure}
	\centering
    \hspace*{\fill}
    \subcaptionbox{full view}{\incfig[width=0.32\linewidth]{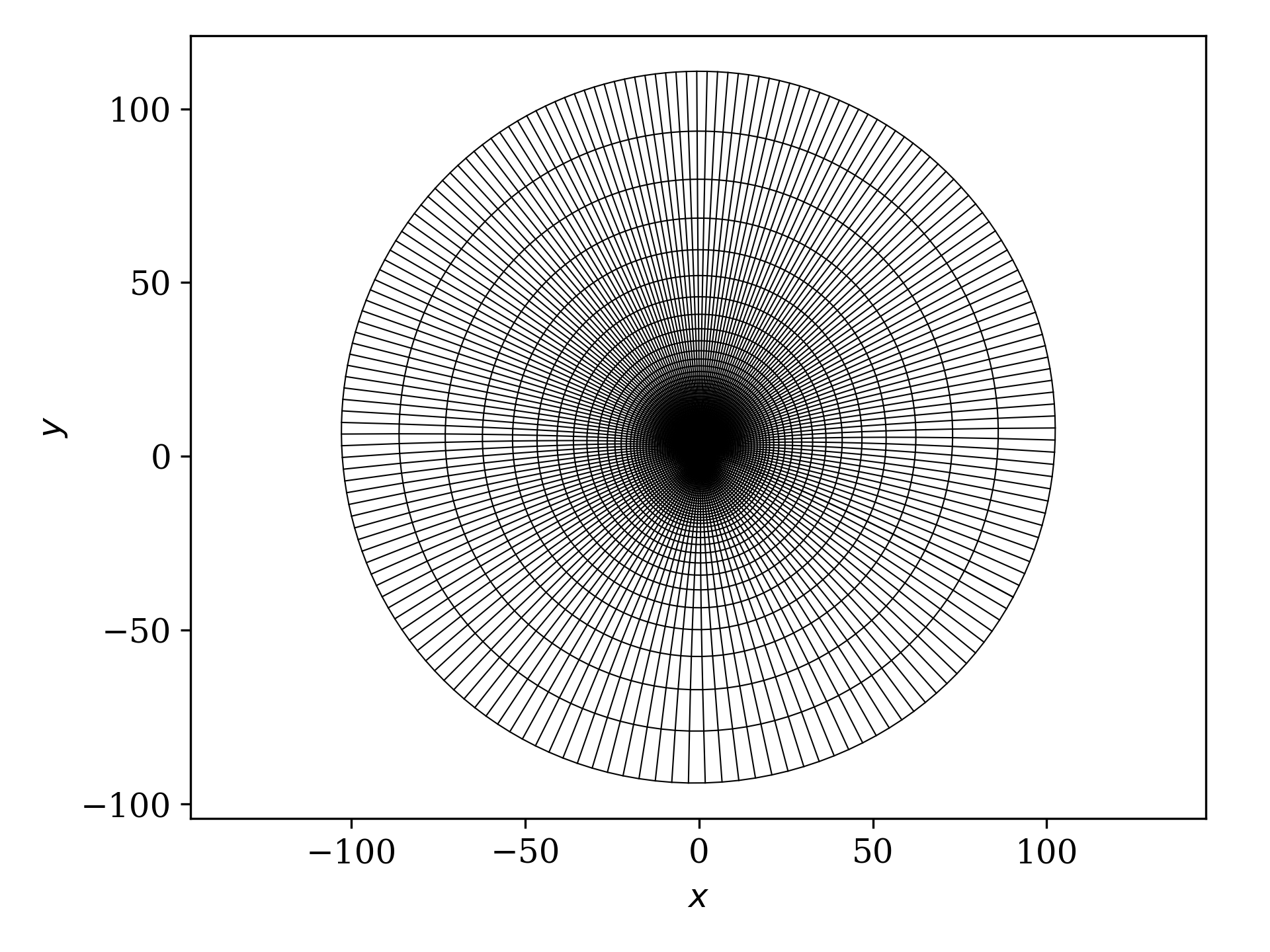}}
    \hfill
    \subcaptionbox{zoomed-in view}{\incfig[width=0.325\linewidth]{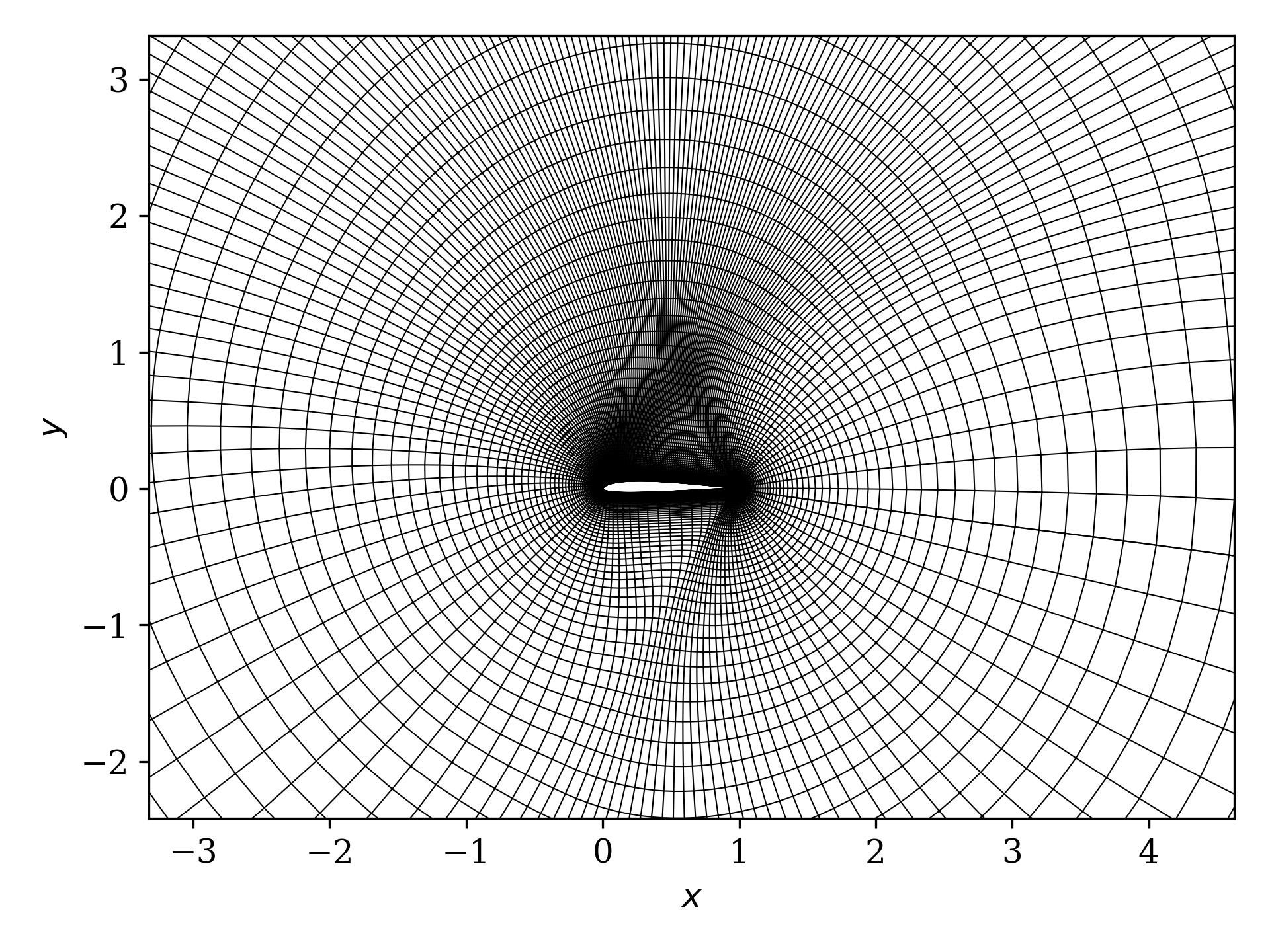}}
    \hfill
	\subcaptionbox{trailing-edge region}{\incfig[width=0.325\linewidth]{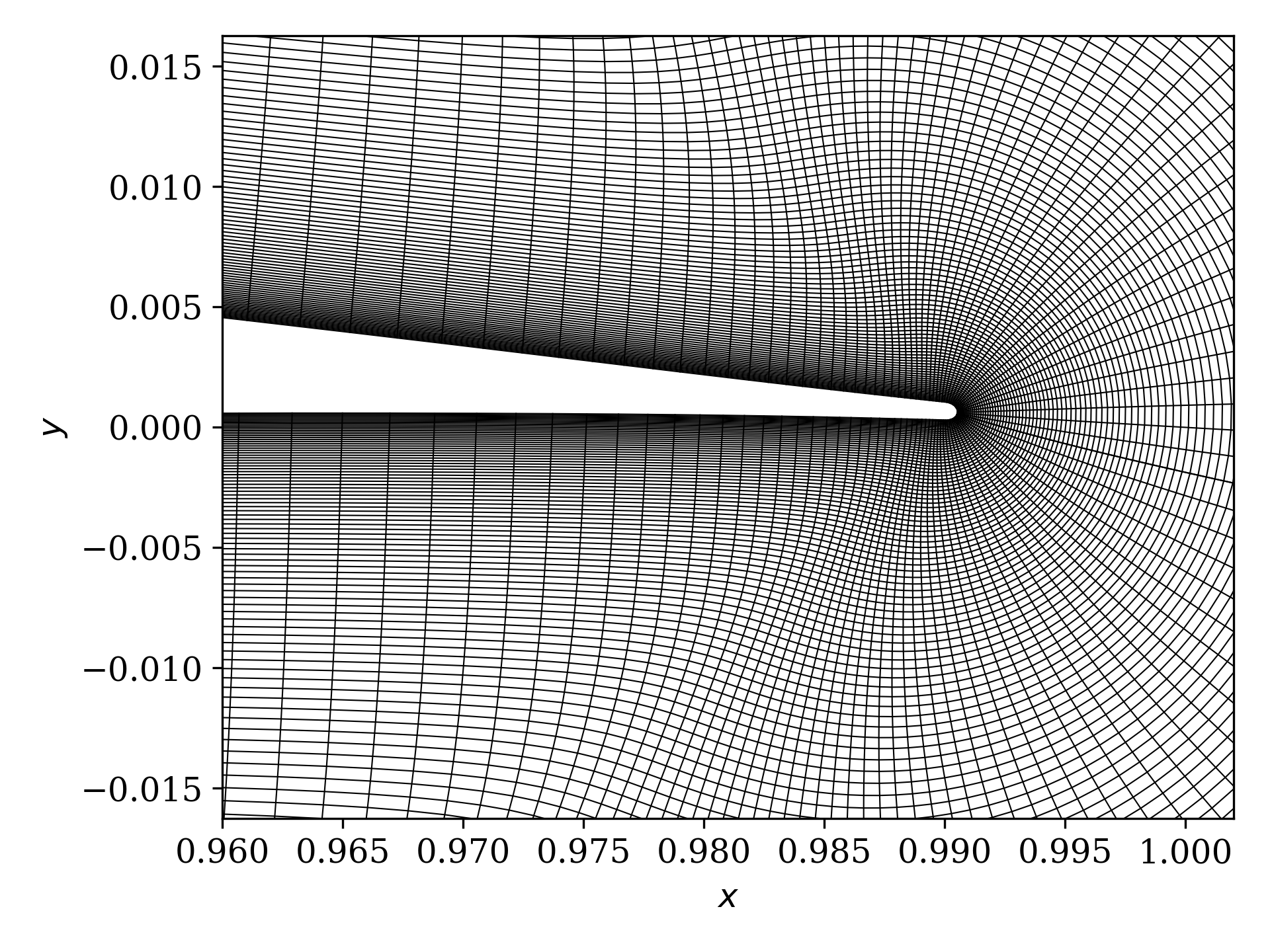}}
    \hspace*{\fill}
	\caption{Grid used in the present study:(a) full view, (b) zoomed-in view, and (c) trailing-edge region. 
    Every third point in the radial and circumferential directions is shown for clarity in panels a and b.
	\label{fig:grid}}
\end{figure}

Our analysis is based on wall-resolved LES carried out using the compressible flow solver FDL3DI \shortcite{gaitonde1998high}. 
The flow over an SD7003 airfoil undergoing a constant-rate, pitch-up motion is simulated at low-to-moderate pitch rates ($\Omega_0^+ = {\Omega_0 c}/{U_{\infty}}$, with freestream velocity $U_{\infty} = 1$) of $0.05$ and $0.25$ at $Re$ of 10,000 and 60,000.
Spanwise-periodic boundary conditions applied to the ends of a span of length $0.2\,c$ are used to simulate an infinite wing geometry.
Extensive studies have been carried out on the effect of grid resolution and spanwise extent for dynamic stall simulations using the FDL3DI solver in the same $Re$ range by \shortciteA{Visbal2011,Visbal2009,Garmann2009}.
The O-grid mesh used in the present study consists of $554 \times 380 \times 101$ points in the circumferential, radial, and spanwise directions, respectively.
The selected discretization is on par with the finest grids simulated in the cited studies.
The simulated airfoil has a unit chord, and the farfield boundary is located about 100 chord lengths away, where freestream conditions are specified.
The airfoil surface is modeled as an adiabatic, no-slip wall.
The spanwise extent must be sufficiently large to prevent the artificial suppression of large-scale spanwise instabilities and avoid numerical artifacts due to the imposed span periodicity. However, the computational cost increases with span size.
A uniform span-wise spacing of $0.002\,c$ between grid points is used over the span-wise extent of $0.2\,c$, which is sufficiently large for the $Re$ and pitch rates under consideration, based on the studies cited above.
Even for the largest $Re$ considered in this study, the $y^+$ values in the static simulations remain well below $1$ over the entire airfoil surface. 
Appendix~\ref{app:staticsim} presents some of the static simulation results.
Figure~\ref{fig:grid} shows images of the grid used in the present study.
A nondimensional time step size ($\Delta t^* = \Delta t \, U_{\infty}/c $) of $1 \times 10^{-4}$ is used for time integration, as is typical in LES simulations using FDL3DI in the literature. 
Additional details on the solver are available in \shortciteA{Sharma2019} and \shortciteA{Visbal2002}.

Table~\ref{tab:datasets} lists the datasets used in the current analysis. 
In all cases, an SD7003 airfoil having a maximum thickness of 8.5\% chord at $x/c= 0.24$ and a camber of 1.4\% chord undergoes a constant-rate, pitch-up motion about its quarter-chord point at a freestream Mach number, $M_{\infty} = 0.1$. 
A smooth hyperbolic tangent function is used to reach the final nondimensional pitch rate, $\Omega_0^+$, as described in \shortciteA{Sharma2019}.
The results reported herein are obtained by averaging the three-dimensional solutions in the spanwise direction.
The datasets consist of two different $Re$ (10,000 and 60,000) at two different $\Omega_0^+$ (0.05 and 0.25), for a total of four cases.
The acronyms used in Table~\ref{tab:datasets} will be used to refer to each case in the rest of the paper.
Each acronym consists of two parts separated by a hyphen.
In the first part, the numbers following `R' represent the $Re$ in thousands; for example, `R10' refers to $Re =  10,000$. 
In the second part, the numbers following `p' refer to the nondimensional pitch rate in hundredths; for example, `p05' refers to $\Omega_0^+ = 0.05$.
\begin{table}
	\centering
	\caption{Datasets used in the present work. In all cases, an SD7003 airfoil is pitched up at a constant rate about the quarter-chord point, at $M_{\infty} = 0.1$.}
	\label{tab:datasets}
	\begin{tabular}{c c c c}
		Case\# & Acronym & $Re$ & $\Omega_0^+$\\ [3pt]
		\toprule
		1 & R60-p05 & 60,000 & 0.05\\ 
		2 & R60-p25 & 60,000 & 0.25\\ 
		3 & R10-p05 & 10,000 & 0.05\\ 
		4 & R10-p25 & 10,000 & 0.25\\ 
		\bottomrule
	\end{tabular}
\end{table}

\section{Results \& Discussion}
\label{sec:results_discussion}

\subsection{Definitions of \texorpdfstring{\(BEF\)}{BEF} and \texorpdfstring{\(LESP\)}{LESP}}
\label{sec:definitions}
We first provide the definitions of the $BEF$ and $LESP$ parameters for reference.
For a 2D flow field, the $BEF$ is the flux of the squared spanwise vorticity at the wall scaled by $Re$, as shown in Eq.~\eqref{eq:bef_def}. 
That is, $\partial(\omega^2/2) / \partial n$, which is written as a product of vorticity ($\omega$) and vorticity flux ($\partial \omega / \partial n$), and scaled by $Re$.
\begin{equation}
    BEF = \frac{1}{Re}\int_{(x/c)_p}^{(x/c)_s} \omega \frac{\partial \omega}{\partial n} \rm{d}s, \label{eq:bef_def}
\end{equation}
where $\omega$ is normalized by $U_{\infty}/c$. 
The normal and tangential coordinates to the airfoil surface, $n$ and $s$, respectively, are normalized by $c$.
The integral is carried out between $x/c$ on the pressure side to $x/c$ on the suction side, with all quantities calculated in the airfoil frame of reference.
The factor $Re$ can be combined with the vorticity flux to write an equivalence relation with the favorable streamwise pressure gradient for small tangential accelerations (that is, $(1 / Re) (\partial \omega / \partial n) = - (1 / \rho) (\partial p / \partial s)$).
Therefore, large contributions to the \(BEF\) arise from regions where high vorticity coexists with large pressure gradients. 
At moderate to high $Re$, this occurs primarily near the airfoil leading edge, and hence the $BEF$ is nearly independent of integration length at such $Re$ as long as the region very close to the leading edge is included.
The integration region was determined to be about 1\% chord for $Re \ge \mathcal{O}\left(10^5\right)$ by \shortciteA{Sudharsan2022}.

$LESP$, given by Eq.~\eqref{eq:LESP_def}, is a measure of the chord-wise or camber-wise suction force $F_{\rm suction}$ near the leading edge, obtained by integrating surface pressure.
In the forward part of the airfoil (upstream of max thickness location, which is at $x/c = 0.24$ for the SD7003 airfoil), $F_{\rm suction}$ points in the  $-x$ direction, indicating a `suction' force acting on the airfoil, which explains the nomenclature.
The integration to obtain $F_{\rm suction}$ is conventionally carried out from the maximum thickness point on the pressure side to that on the suction side.
Since the SD7003 airfoil has a very small camber ($\sim 0.014\,c$), the chord-wise direction has been used to compute $LESP$, which has negligibly small differences compared to using the camber direction.
\begin{equation}
LESP = \sqrt{\abs{C_{\rm suction}}/(2 \pi)}, \;{\rm where} \; C_{\rm suction} =  F_{\rm suction}/(q_{\infty}c), \; {\rm and} \; q_\infty = \rho_\infty U^2_\infty.
\label{eq:LESP_def}
\end{equation}

In reduced-order models based on unsteady thin airfoil theory, the instance of vortex shedding from the leading edge is determined by the $LESP$ reaching a critical value~\shortcite{Ramesh2014}.
The idea is based on the suction at the leading edge being limited (via vortex shedding) to a certain (critical) value that can be supported.
The $LESP$ is a measure of the $A_0$ coefficient in the Fourier series expansion of vorticity from classical thin airfoil theory~\shortcite{Katz2001}; $A_0$ represents the vorticity at the leading edge. 
A leading-edge vortex is, therefore, shed any time the $LESP$ reaches the critical value, thus limiting the maximum $LESP$ in the model to the critical value.
This approach requires the critical $LESP$ value to be determined a priori (via simulations or measurements) instead of real-time parameter tracking; critical $LESP$ depends on airfoil geometry and operating conditions.
$\max(LESP)$ has been used as a proxy for critical $LESP$ to circumvent this impediment~\shortcite{Deparday2018, Deparday2019}, and will be evaluated in the present work for comparison.

The critical value of $LESP$ is the $LESP$ at the time instant when the profile of the skin friction coefficient ($C_f$) over the suction surface of the airfoil near the leading edge first develops an inflection point~\shortcite{Narsipur2020}.
This is referred to as the $C_f$-signature criterion, which signifies the development of instabilities within the LSB. 
These instabilities lead to the bursting of the LSB, which is followed by DSV formation.
Note that the $C_f$-signature criterion per se is hard to use as a stall indicator because it is spatially localized, and the location of the inflection point is not fixed or known a priori.
Spatially integrated quantities such as the $LESP$ and $BEF$ are therefore preferred.


The results from the numerical simulations are described in detail in the following sections.

\subsection{Case R60-p05}
\label{sec:R60-p05}
\begin{figure}[htb!]
    \hspace*{\fill}
    \subcaptionbox{$-C_p$\label{fig:R60-p05_xt_cp}}{\incfig[width=0.49\textwidth]{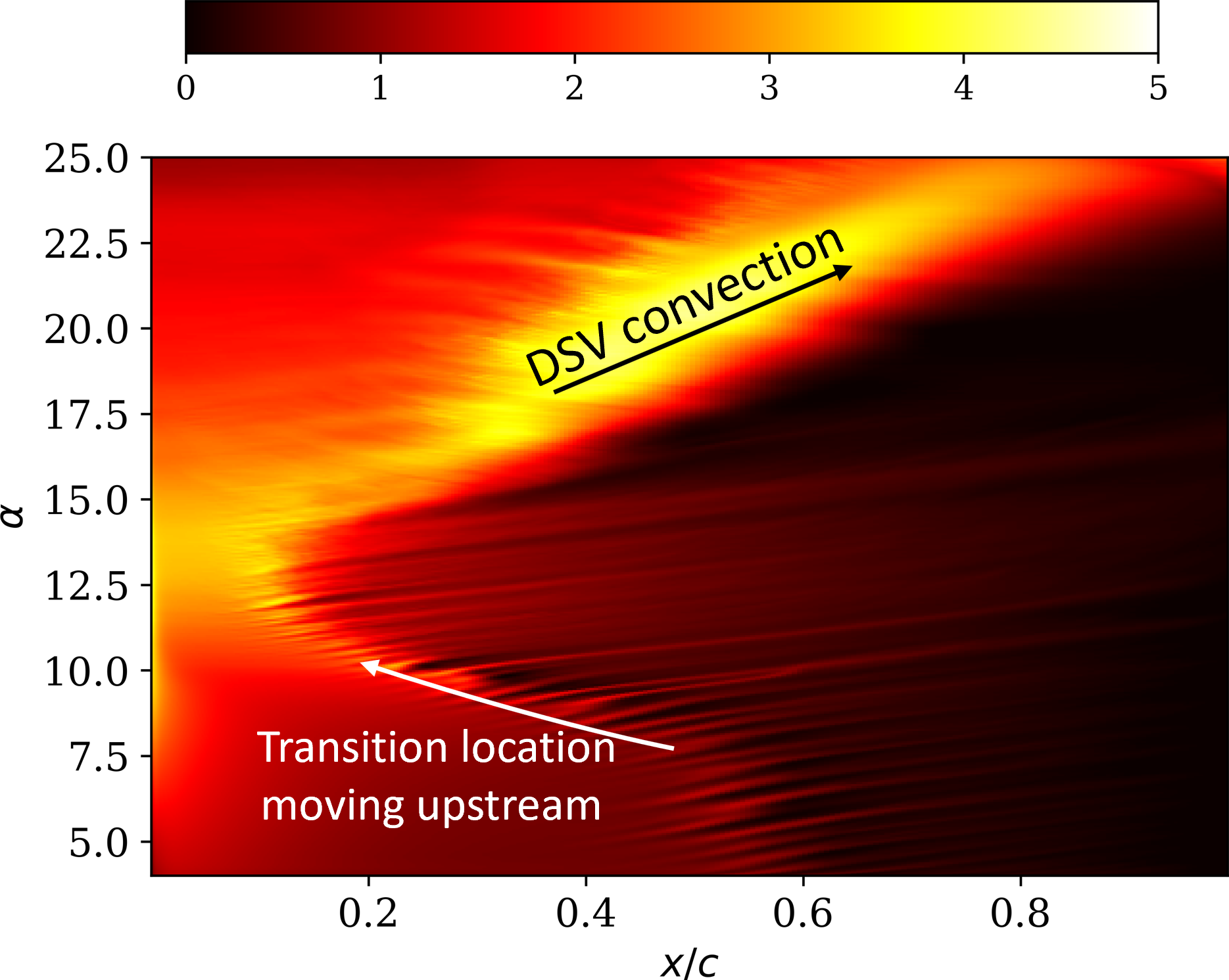}}
    \hfill
    \subcaptionbox{$C_f$\label{fig:R60-p05_xt_cf}}{\incfig[width=0.49\textwidth]{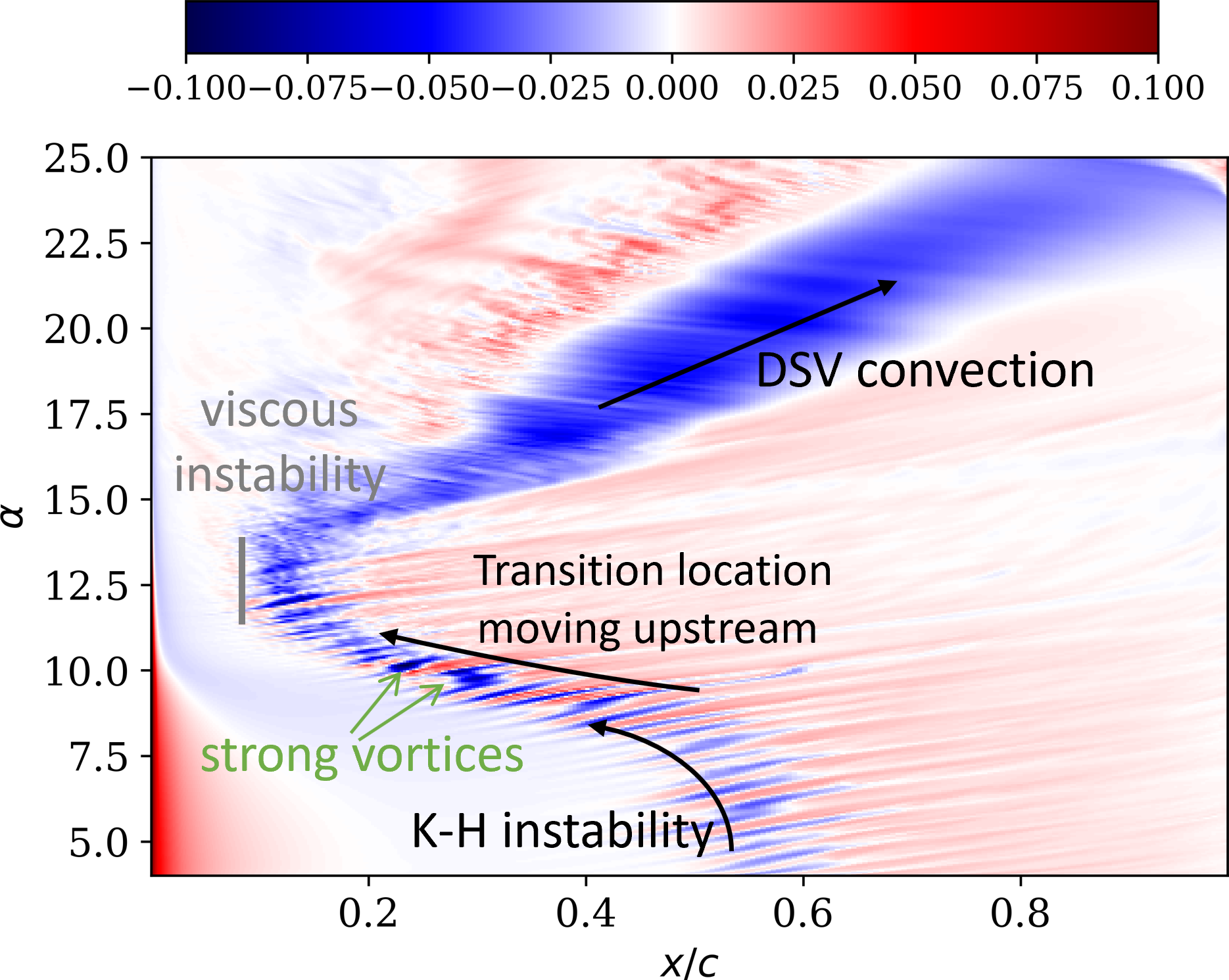}}
    \hspace*{\fill}
    \caption{Space-time contours for Case R60-p05}
    \label{fig:R60-p05_xt}
\end{figure}

We begin by describing the results obtained at $Re$ $60,000$ and pitch rate $0.05$.
Figure~\ref{fig:R60-p05_xt} shows the space-time contours of $-C_p$ and $C_f$ with the normalized chord-wise distance along the airfoil suction surface ($x/c$) as the abscissa and angle of attack ($\alpha$) as the ordinate.
The shear layer that separates from the airfoil leading edge develops inviscid, Kelvin-Helmholtz (K-H)-type instabilities downstream, resulting in the shedding of laminar vortices around 60\% chord.
The shed vortices subsequently transition to turbulence, and the instability and transition locations move upstream as the airfoil pitches up. 
These are pointed out in the $C_f$ contours in Fig.~\ref{fig:R60-p05_xt_cf}.

\begin{figure}
     \centering
     \hspace*{\fill}
     \subfloat[$\alpha = 6.9^{\circ}$]{\incfig[width=0.325\textwidth]{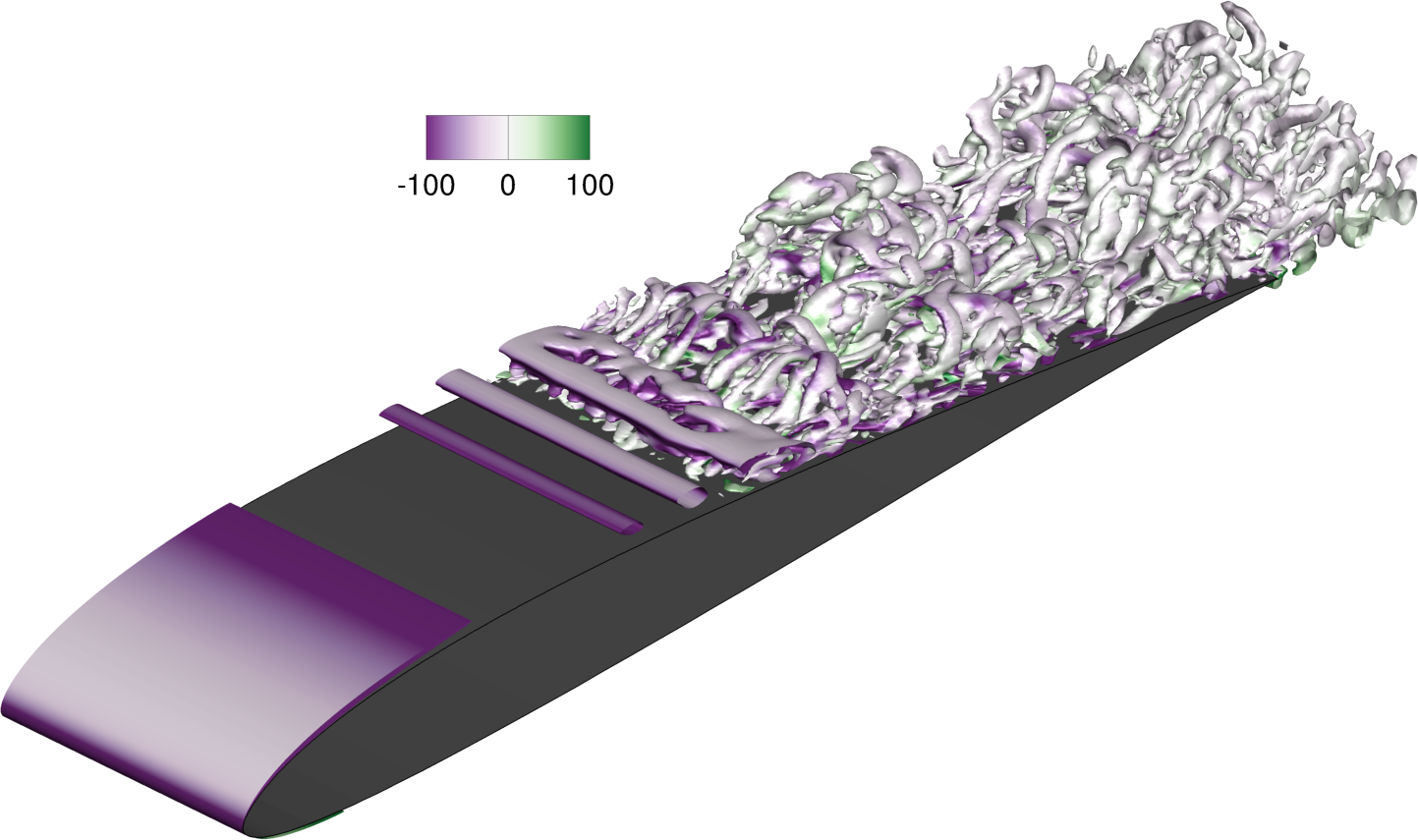}} 
     \hfill
     \subfloat[$\alpha = 9.3^{\circ}$]{\incfig[width=0.325\textwidth]{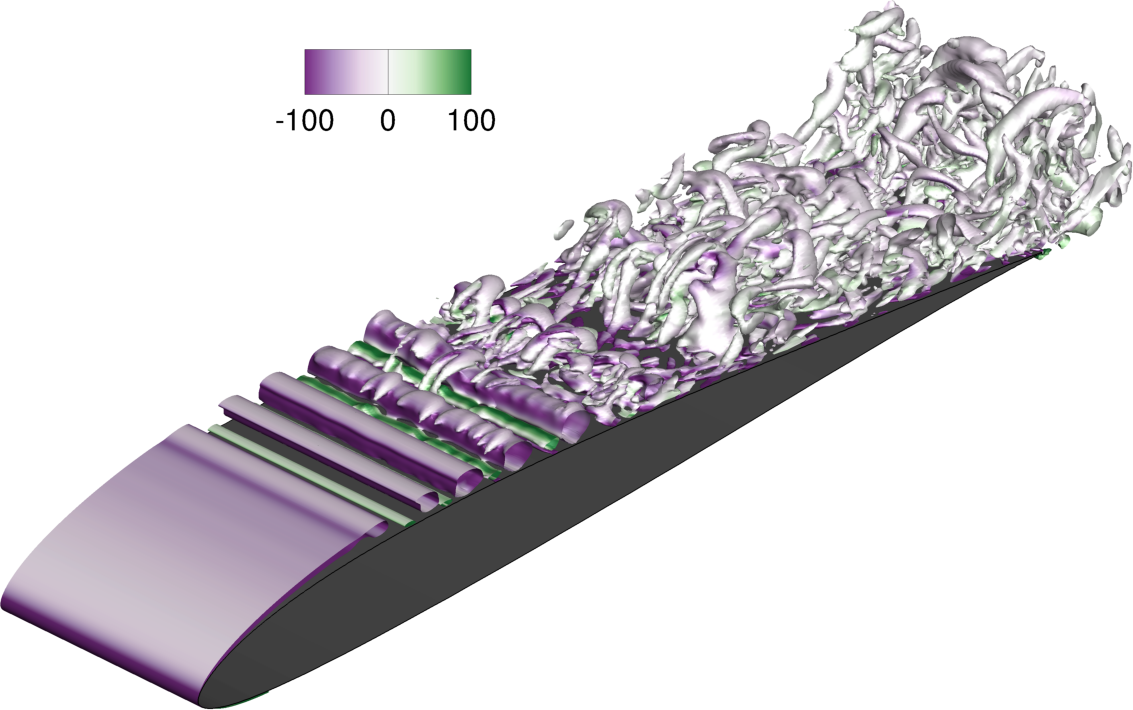}} 
     \hfill
     \subfloat[$\alpha = 10.7^{\circ}$]{\incfig[width=0.325\textwidth]{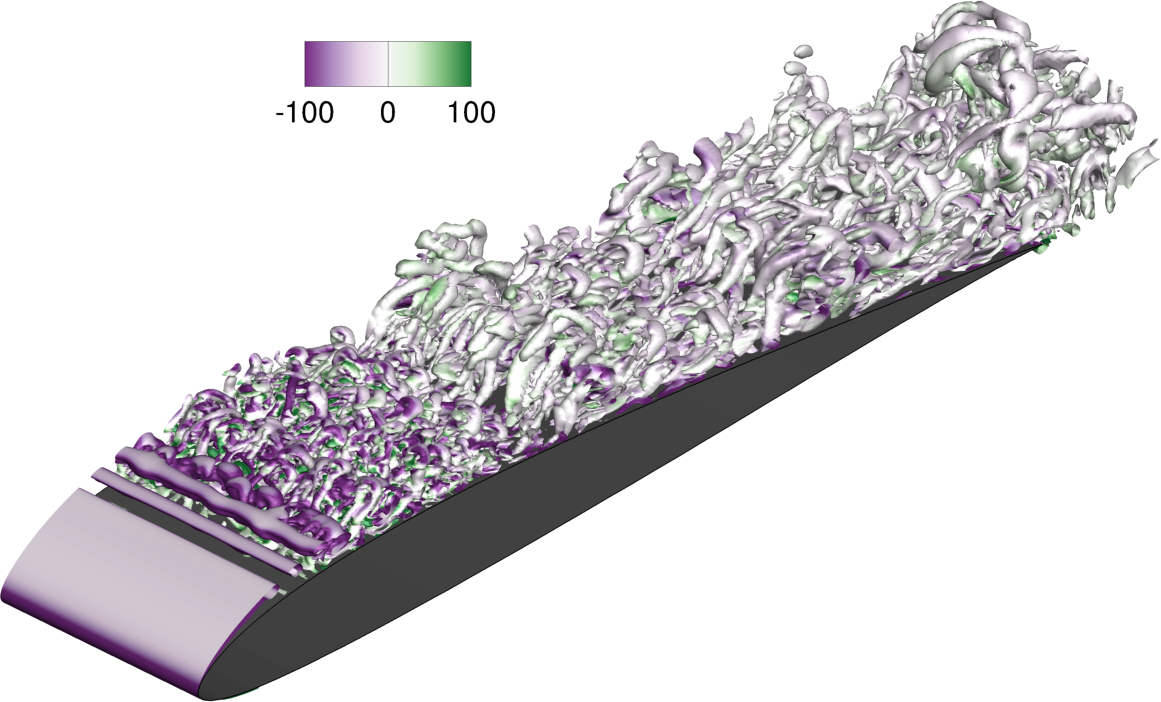}}
     \hspace*{\fill}
     \caption{Isosurfaces of Q-criterion (value 100) colored by spanwise vorticity contours (inlaid legend) showing the shear layer undergoing a K-H instability and subsequent transition to turbulence (a), upstream propagation of the K-H instability (b), and development of viscous instabilities close to the leading edge (c), for Case R60-p05.}
     \label{fig:R60-p05_isosurf}
\end{figure}

Around $\alpha \sim 11^{\circ}$, the initial K-H instabilities give way to viscous instabilities as an LSB is established close to the leading edge.
A couple of strong laminar vortices are shed around this time, as pointed out in Fig.~\ref{fig:R60-p05_xt_cf}.
The flow in the LSB transitions to turbulence and reattaches as the increasing APG amplifies the instabilities.
Figure~\ref{fig:R60-p05_isosurf} shows isosurfaces of the Q-criterion~\shortcite{Hunt1988} colored by spanwise vorticity values at the following instances: (a) initial K-H instability of the shear layer leading to laminar vortex shedding, (b) upstream propagation of the K-H instability, and (c) the development of viscous instabilities near the leading edge.
Stall occurs primarily due to flow breakdown at the leading edge. Contours of spanwise vorticity, $\omega$, as the unsteady motion progresses are included in Movie \dispmcounter.

\begin{figure}
	\centering
	\incfig[width=0.7\textwidth]{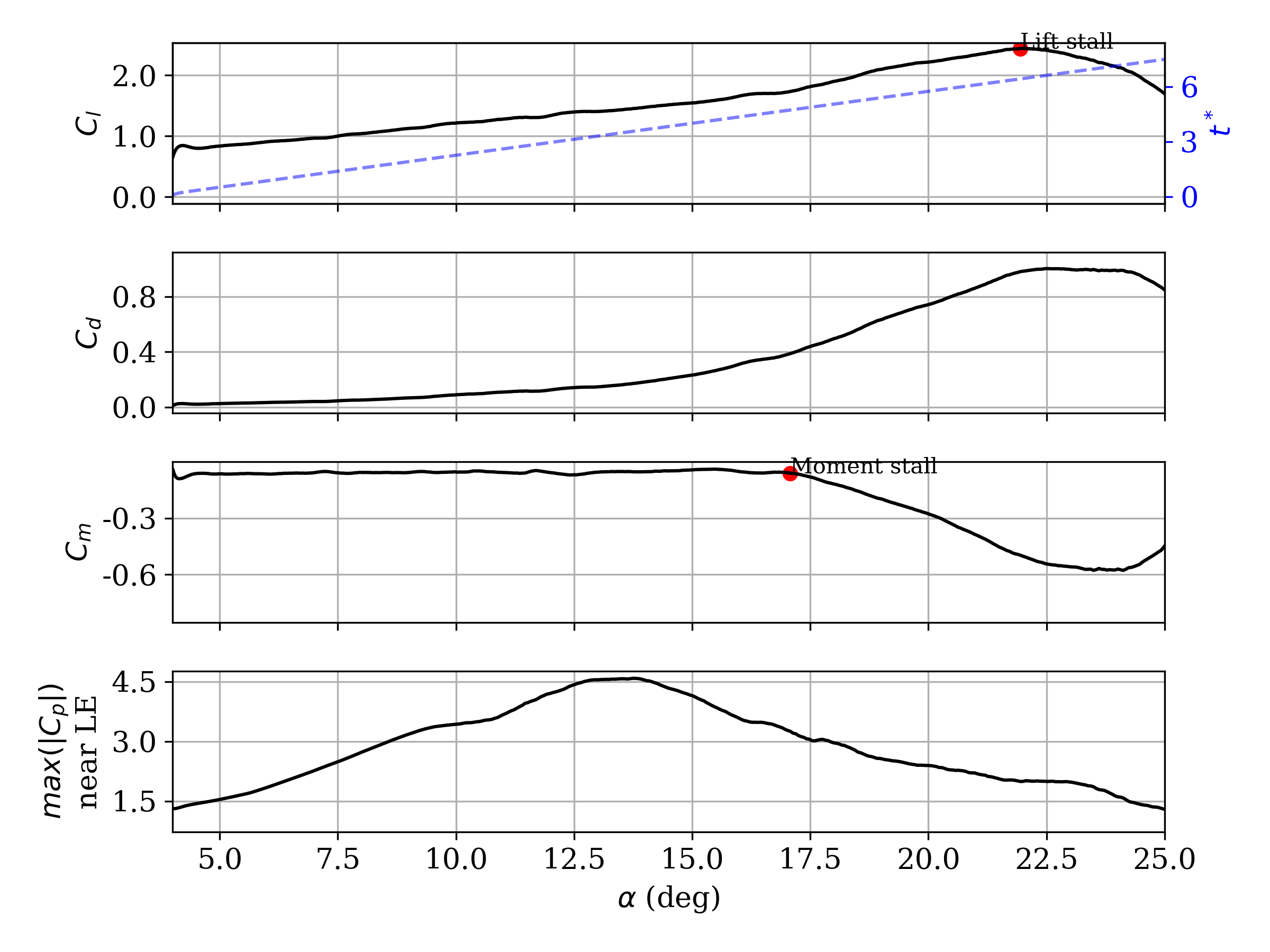}
	\caption{Variation with $\alpha$ of aerodynamic coefficients (top three panels) and $\max(|C_p|)$ near the first 5\% of airfoil chord (bottom panel) for Case R60-p05.}
	\label{fig:R60-p05_aerodyn_coeff}
\end{figure}

A DSV is formed from the shed vorticity near the leading edge. The DSV grows and convects downstream, as seen from the imprint on the airfoil suction surface in the contours shown in Fig.~\ref{fig:R60-p05_xt}.
The downstream convection of the DSV results in a moment stall, and its subsequent shedding results in a lift stall, as observed from the aerodynamic coefficients ($C_l$, $C_d$, and $C_m$) plotted against $\alpha$ in the top three panels of Fig.~\ref{fig:R60-p05_aerodyn_coeff}.
The nondimensional time, $t^*$, is plotted in the top panel of the figure for reference.
The bottom panel of Fig.~\ref{fig:R60-p05_aerodyn_coeff} shows the maximum magnitude of $C_p$ within the first $5\%$ chord from the leading edge.
Note that $C_p$ is a point quantity, in contrast to the aerodynamic coefficients plotted in the top panels, which are integrated quantities.
As the airfoil pitches up, the stagnation point moves downstream on the pressure side, even as the shear layer slowly moves away from the surface on the suction side.
This leads to a net increase in the curvature of the shear layer near the leading edge, leading to a rise in $C_p$.
However, when strong vortices are generated and shed around $\alpha \sim 10^{\circ}$, there is a slight reduction in the slope of the max($\abs{C_p}$) curve.
This reduction occurs because the growth of these vortices induces strong secondary vorticity beneath them, resulting in the shear layer being pushed away from the surface at a faster rate.
This decreases the curvature of the shear layer, leading to a reduction in $C_p$ magnitude.
Once the vortices are shed, the shear layer movement away from the surface again slows down, leading to a further increase in  $C_p$ magnitude.
After the leading edge flow breaks down around $13.5^{\circ}$, the magnitude of $C_p$ begins to reduce.

%
%


\begin{figure}
	\centering
	\incfig[width=0.6\textwidth]{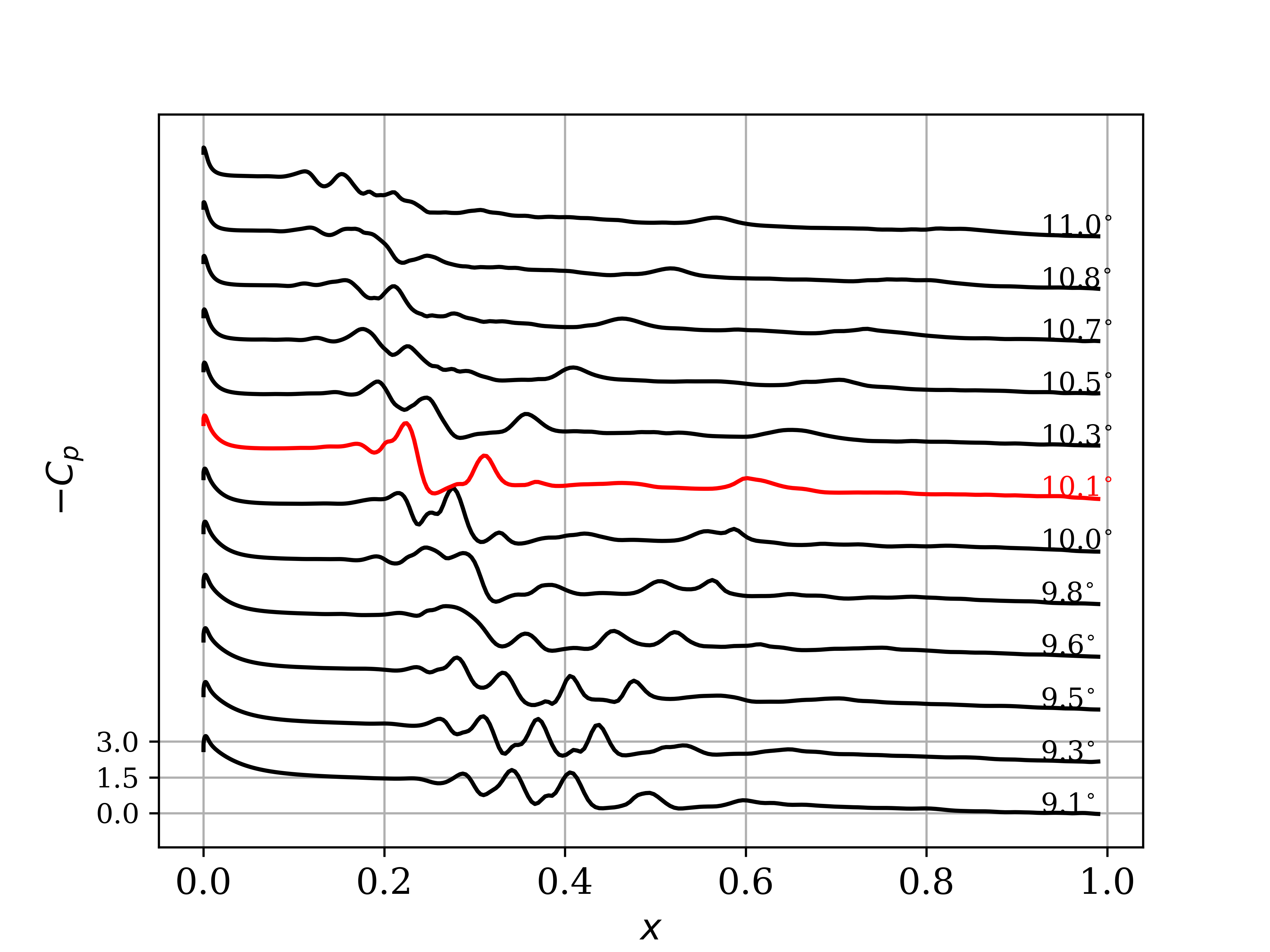}
	\caption{$-C_p$ profiles close to the time of strong laminar vortex shedding. The profile corresponding to relatively strong vortex shedding is highlighted in red.}
	\label{fig:R60-p05_cpprofs}
\end{figure}

Figure~\ref{fig:R60-p05_cpprofs} shows $C_p$ profiles around the time when strong laminar vortices are shed.
Around $\alpha = 10.1^{\circ}$, the strengths of the shed vortices markedly increase, as pointed out in the $C_f$ contours in Fig.~\ref{fig:R60-p05_xt_cf}. 
In addition to the instance of DSV formation, we are also interested in the instances of strong laminar vortex shedding.
At low $Re$, these vortex cores have $C_p$ magnitudes comparable to peak $|C_p|$ at the leading edge, and they induce a large viscous response at the wall, significantly impacting the onset of dynamic stall and its progression.
Therefore, effective criteria for vortex shedding need to demonstrate critical behavior around these events.

Our prior studies at moderate-to-high $Re$ have shown that $BEF$ calculated by integrating over 1\% chord is sufficient to indicate imminent DSV formation.
The small region of integration is justified because the effect of DSV formation is strongly felt close to the airfoil leading edge through the collapse of leading-edge $C_p$.
In contrast, laminar vortex shedding, which occurs at low $Re$ is a spatially localized event not limited to the leading-edge region.
Therefore, we calculate $BEF$ by integrating over different chord lengths to capture localized vortex-shedding events.
The lowest integration length is set to $0.05\,c$ since the spatial scales are larger at these low $Re$.
$LESP$ is typically calculated by integrating up to the point of maximum thickness, according to the definition provided by \shortciteA{Narsipur2020}.
For comparison, we integrate \textit{both} parameters over a range of fractional chord lengths.
The suction force used in the definition of $LESP$ given by Eq.~\eqref{eq:LESP_def} has a length scale $c$ in the denominator.
We retain the same length scale $c$ while varying the integration length used.
Setting this length scale to the integration length would simply scale the $LESP$ curves by a constant.
We also use a low-pass filter to remove the high-frequency fluctuations in time that are captured in the LES simulations for $LESP$, $BEF$, and max($\abs{C_p}$) near the leading edge.
A Gaussian filter with a half-width equal to the inverse of a nondimensional frequency $f^+ = f c / U_{\infty}$ of 20 is used.

\begin{figure}
    \hspace*{\fill}
    \subcaptionbox{$LESP$\label{fig:R60-p05_LESP}}{\incfig[width=0.49\textwidth]{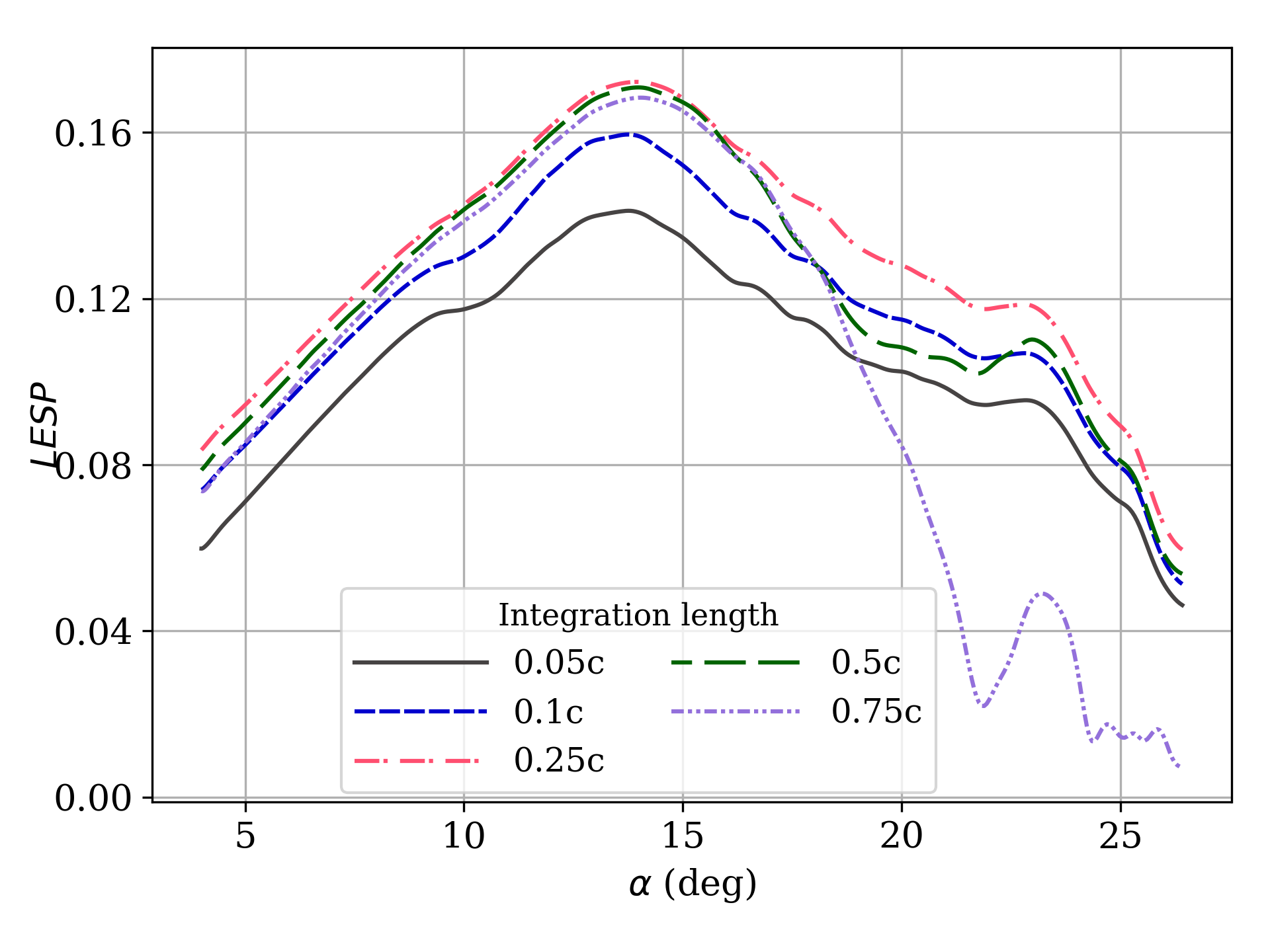}}
    \hfill
    \subcaptionbox{$\abs{BEF}$\label{fig:R60-p05_BEF}}{\incfig[width=0.49\textwidth]{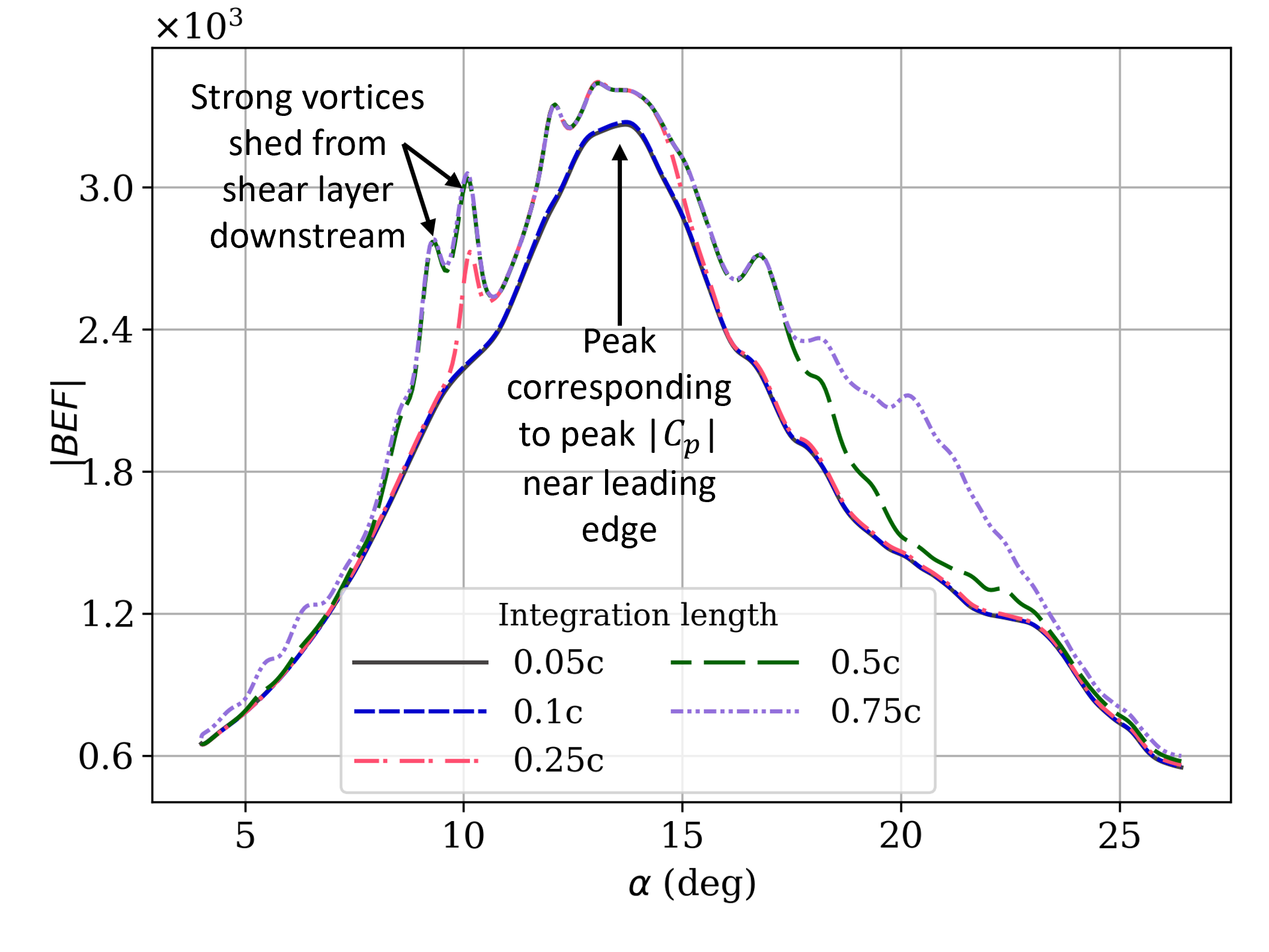}}
    \hspace*{\fill}
    \caption{$LESP$ (a) and $BEF$ (b) integrated over different chord lengths, plotted against $\alpha$, for Case R60-p05.}
    \label{fig:R60-p05_LESP_BEF}
\end{figure}

Figure~\ref{fig:R60-p05_LESP_BEF} plots the variation of $LESP$ and $\abs{BEF}$ obtained by integrating over different chord lengths.
Both parameters reach their respective global maxima between $13^\circ - 14^{\circ}$, which is ahead of DSV formation.
This point reflects the leading edge $\abs{C_p}$ reaching a maximum, following which DSV formation begins.
In addition, two secondary peaks, corresponding to strong laminar vortices that are shed, are observed around $\alpha \sim 10^{\circ}$ for the $\abs{BEF}$ curves with integration length $\ge 0.25\,c$.
Note that there is no change in $BEF$ across integration lengths unless there is an additional downstream source of vorticity at a given time.
That is, the $BEF$ curves all coincide at initial times and vary only as the flow develops.
This is in contrast to the $LESP$, where the values vary with integration length even at the initial times (note the vertical displacement of the $LESP$ curves at low $\alpha$).
Another aspect of $BEF$ behavior is the contribution from vortex shedding being in the same sense as the contribution from the leading edge (see Fig.~9 and the related description in \shortciteA{Sudharsan2022} for a detailed explanation on the sense of $BEF$ contributions over the airfoil).

Since the effect of the laminar vortices is felt locally, they are indicated by the $BEF$ when the pertinent area is included in the integration region.
Calculating $BEF$ over different integration lengths and comparing the difference can also identify the region over the airfoil where vortex shedding occurs.
This is demonstrated in detail in Section~\ref{sec:results_comparisons}.
The $LESP$ curves do not show a peak around the instance of laminar vortex shedding since only the camber-wise component of the suction force features in the $LESP$ definition (see Eq.~\eqref{eq:LESP_def}). 
A vortex shed away from the leading edge has a more pronounced effect on the normal component (as opposed to the camber-wise component) of the suction force.
The curves corresponding to $LESP$ and $\abs{BEF}$ integrated up to less than $0.25\,c$ show a slight decrease in slope around the instance of laminar vortex shedding.
This can be attributed to a temporary reduction in leading-edge $C_p$ magnitude due to the change in the curvature of the shear layer caused by the downstream vortex shedding.

These results demonstrate that the $BEF$ can be used to identify imminent vortex shedding occurring away from the airfoil leading edge.

\subsection{Case R60-p25}
\label{sec:R60-p25}
\begin{figure}
    \hspace*{\fill}
    \subcaptionbox{$-C_p$\label{fig:R60-p25_xt_cp}}{\incfig[width=0.49\textwidth]{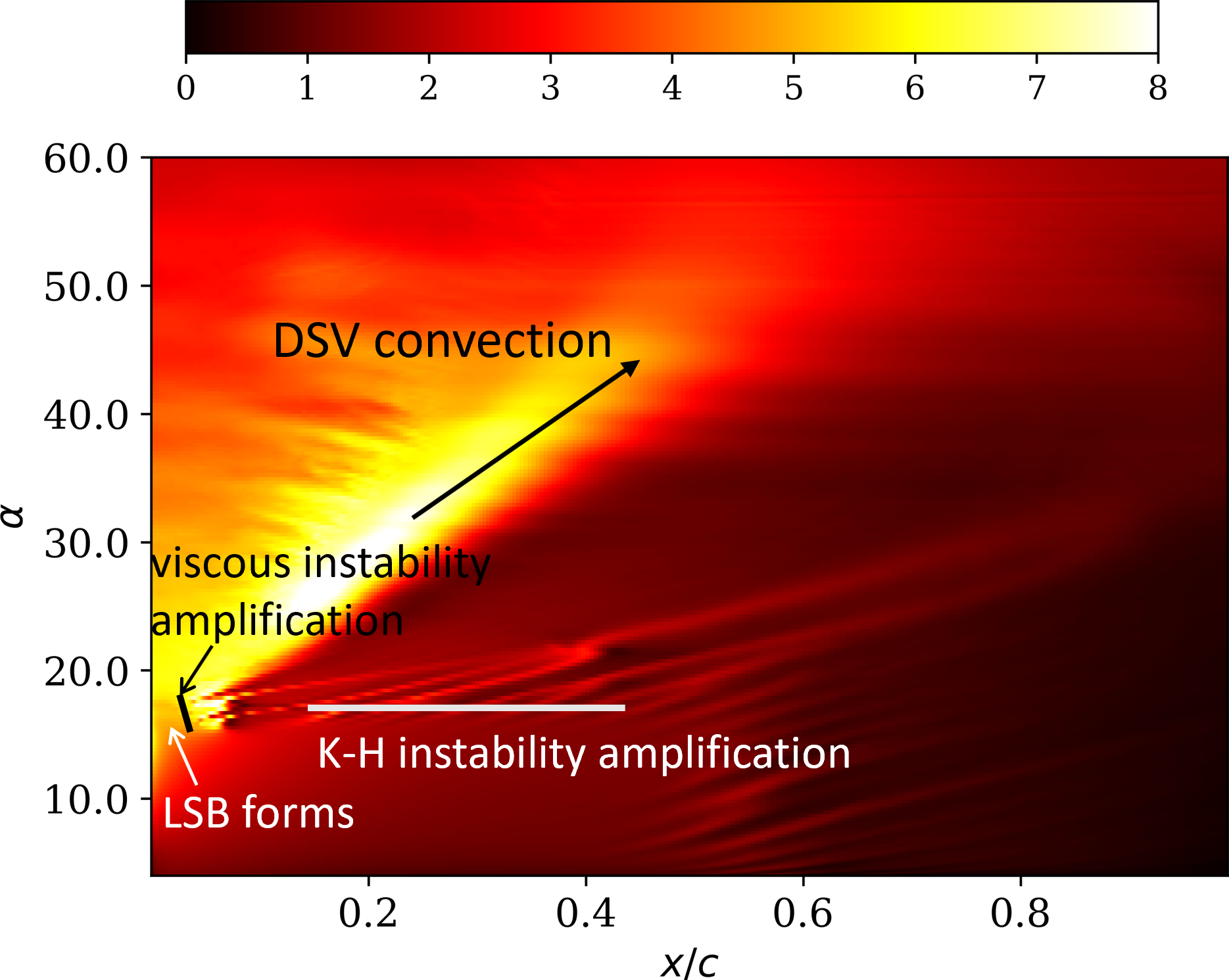}}
    \hfill
    \subcaptionbox{$C_f$\label{fig:R60-p25_xt_cf}}{\incfig[width=0.49\textwidth]{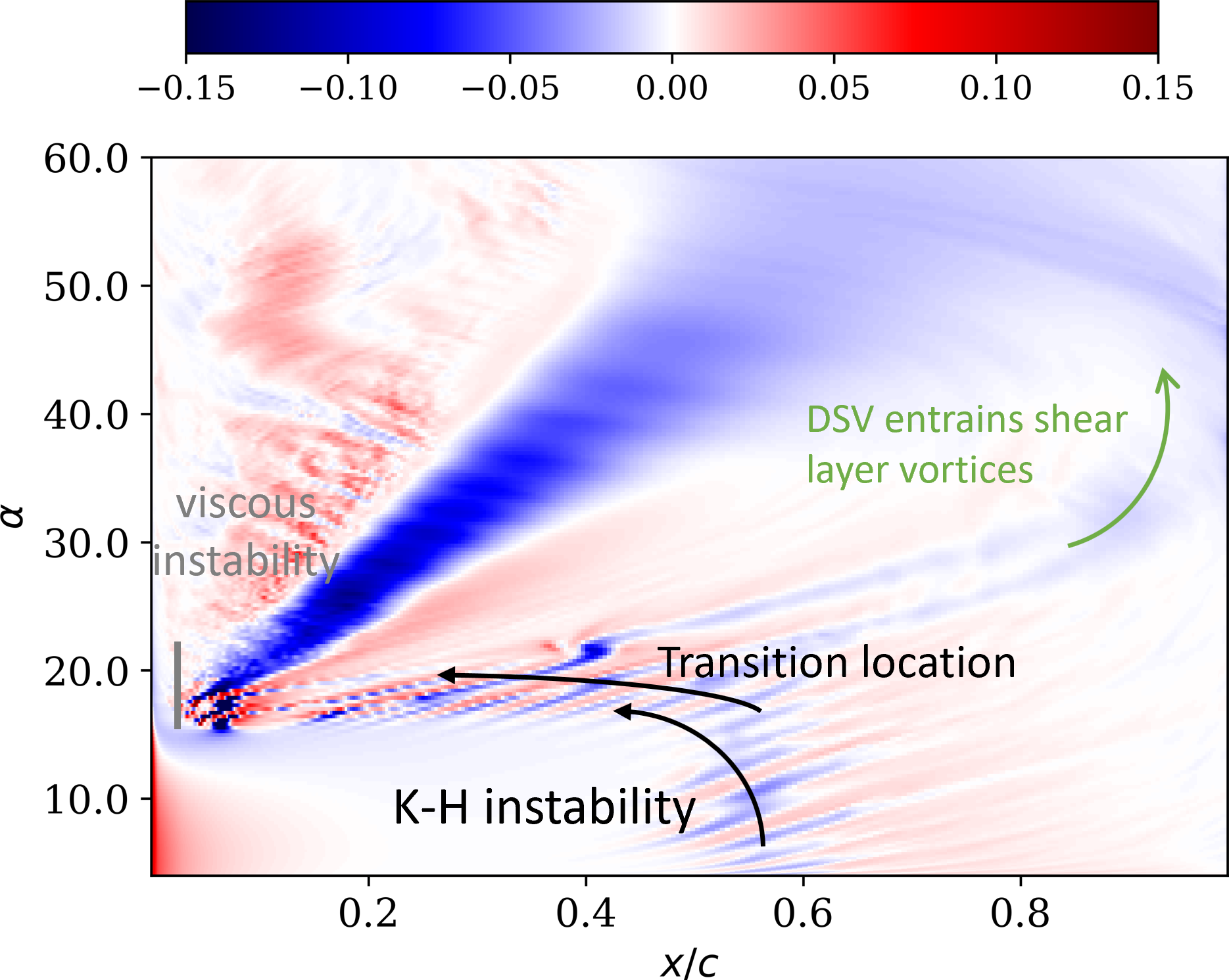}}
    \hspace*{\fill}
    \caption{Space-time contours for Case R60-p25.}
    \label{fig:R60-p25_xt}
\end{figure}

Next, we discuss the results obtained at the same $Re$ of 60,000, albeit with a higher pitch rate of 0.25.
The unsteady lag effects due to the higher pitch rate serve to delay the angle of attack at which stall occurs, in comparison with Case R60-p05.
Moreover, the higher pitch rate also promotes a more conventional leading-edge stall, characterized by the breakdown of the LSB and a drop in leading-edge $\abs{C_p}$.
Space-time contours of $-C_p$ and $C_f$ are shown in Fig.~\ref{fig:R60-p25_xt}.
As in the previous case, the inflectional velocity profiles in the shear layer subjected to an APG develop an inviscid K-H instability, leading to vortex roll-up and transition to turbulence.
This occurs close to $0.6\,c$ initially, as pointed out in the $C_f$ contours (Fig.~\ref{fig:R60-p25_xt_cf}).

\begin{figure}
    \centering
    \subcaptionbox{$\alpha = 15.0^{\circ}$}{\incfig[width=0.485\textwidth]{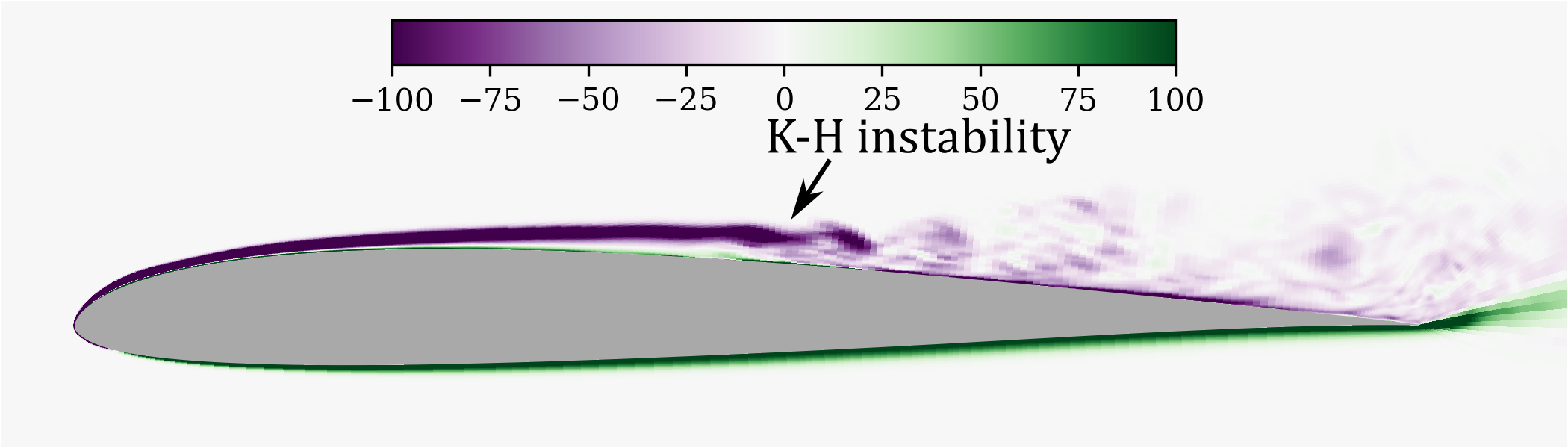}} 
    \hfill
    \subcaptionbox{$\alpha = 15.5^{\circ}$}{\incfig[width=0.485\textwidth]{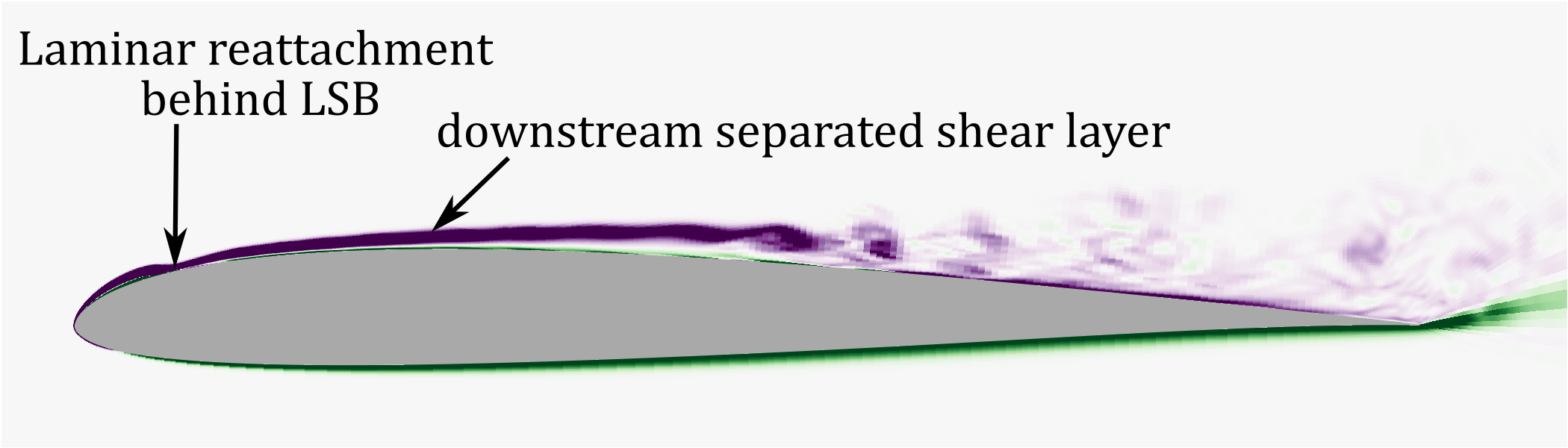}}
    \par\bigskip
    %
    \subcaptionbox{$\alpha = 16.7^{\circ}$}{\incfig[width=0.485\textwidth]{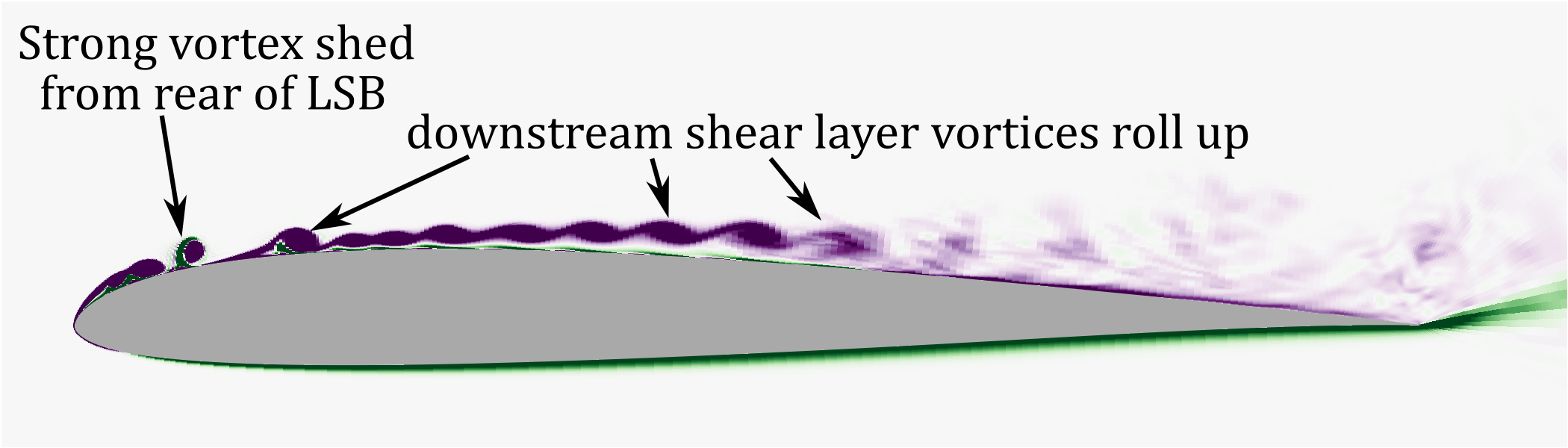}}
    \hfill
    \subcaptionbox{$\alpha = 18.4^{\circ}$}{\incfig[width=0.485\textwidth]{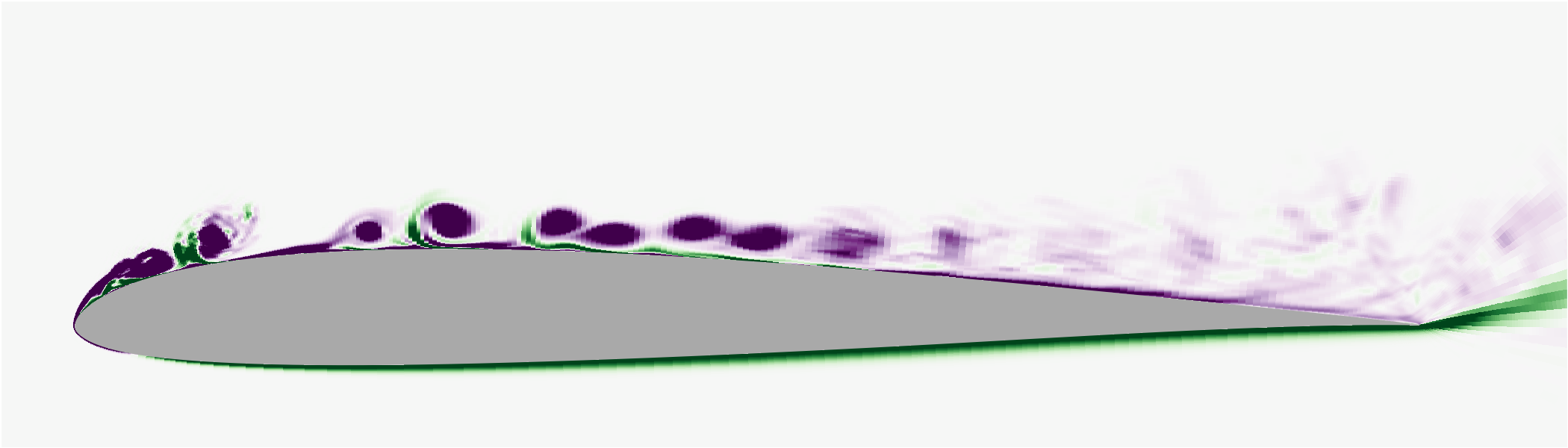}}
    \par\bigskip
    %
    \subcaptionbox{$\alpha = 19.6^{\circ}$}{\incfig[width=0.485\textwidth]{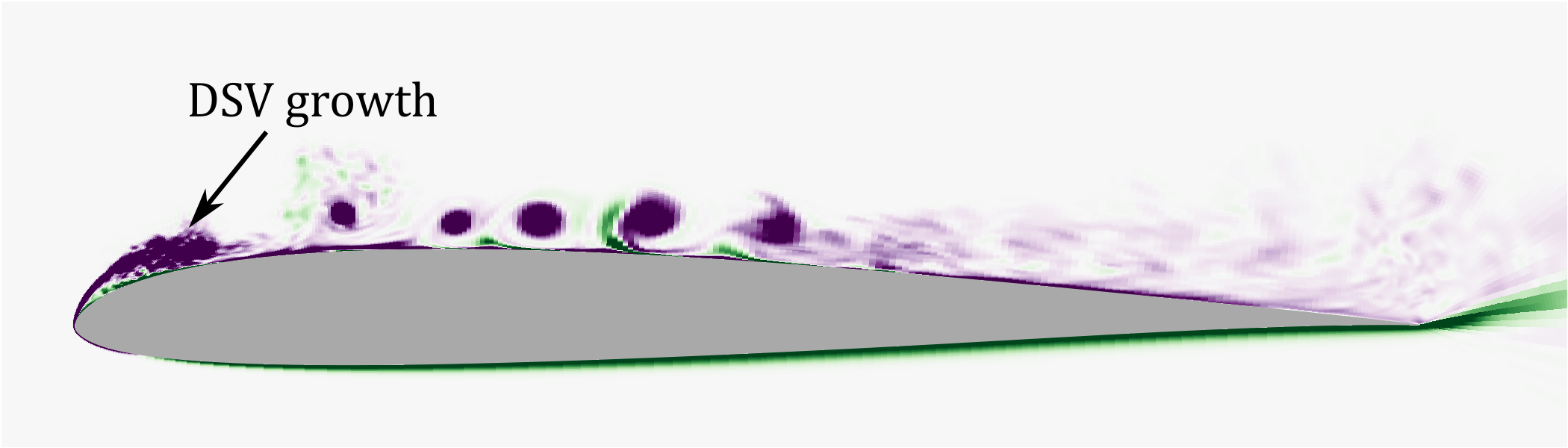}}
    \hfill
    \subcaptionbox{$\alpha = 20.7^{\circ}$}{\incfig[width=0.485\textwidth]{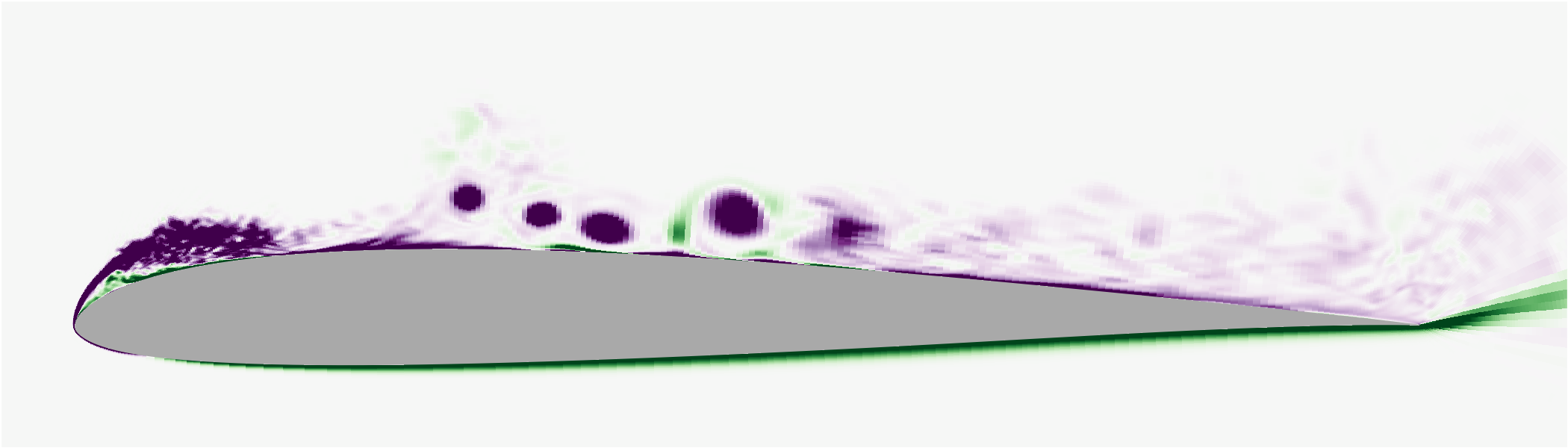}}
    %
    \caption{Case R60-p25: Vorticity contours showing the simultaneous amplification of the K-H instability in the shear layer and the viscous instabilities within the LSB (panels a-c) leading to the roll-up of vortices. Panels d-f show the growth of the leading-edge DSV and shear layer vortices. Purple represents clockwise vorticity, while green represents counter-clockwise vorticity.
    \label{fig:R60-p25_vortcontours}}
\end{figure}

Between approximately $15^{\circ}$ and $20^{\circ}$, a sequence of events transpires in rapid succession.
Spanwise vorticity contours in Fig.~\ref{fig:R60-p25_vortcontours} show the dynamic flow field during this time.
In panel a of the figure, the shedding of vortices due to K-H instabilities in the shear layer and transition to turbulence on the aft section of the airfoil are evident.
Panel b shows a time instance after the establishment of the LSB with laminar reattachment.
In the downstream region of this reattached flow, a second separated shear layer develops, experiencing K-H instabilities that lead to the roll-up of laminar vortices and downstream transition to turbulence.
After its initial establishment, the LSB's existence is short-lived, as it quickly succumbs to viscous instabilities magnified by the increasing APG. 
During this time, strong vortices are shed from the rear of the LSB.
Almost simultaneously, the K-H instabilities also undergo substantial amplification, giving rise to a series of shear layer vortices that roll up downstream (panel c).
This case serves as an illustration of a transition mechanism wherein both viscous instabilities within the LSB and the inviscid K-H instabilities are simultaneously amplified.
This phenomenon is clearly discernible from the $C_p$ and $C_f$ contours in Fig.~\ref{fig:R60-p25_xt}, where a horizontal row of vortices is observed shortly after the formation of the LSB.
Similar behavior has previously been observed in water tunnel experiments for a pitching-plunging airfoil undergoing deep stall at $Re$=20,000 (see Fig. 9 of \shortciteA{ol2009shallow}).

Following the onset of instabilities, the LSB collapses and a coherent leading-edge DSV is formed.
The subsequent growth of the DSV and the shear layer vortices are showcased in panels d-f of Fig.~\ref{fig:R60-p25_vortcontours}.
The DSV begins to convect downstream as it grows, with a higher convection speed compared to the downstream shear layer vortices.
It entrains a pair of previously-shed shear layer vortices further downstream (marked in Fig.~\ref{fig:R60-p25_xt_cf}).
During its downstream propagation, the DSV maintains its proximity to the airfoil surface until $\alpha \sim 45^{\circ}$.
A more comprehensive visual representation of this sequence is included in the accompanying video (see Movie~\dispmcounter).

\begin{figure}
	\incfig[width=0.6\textwidth]{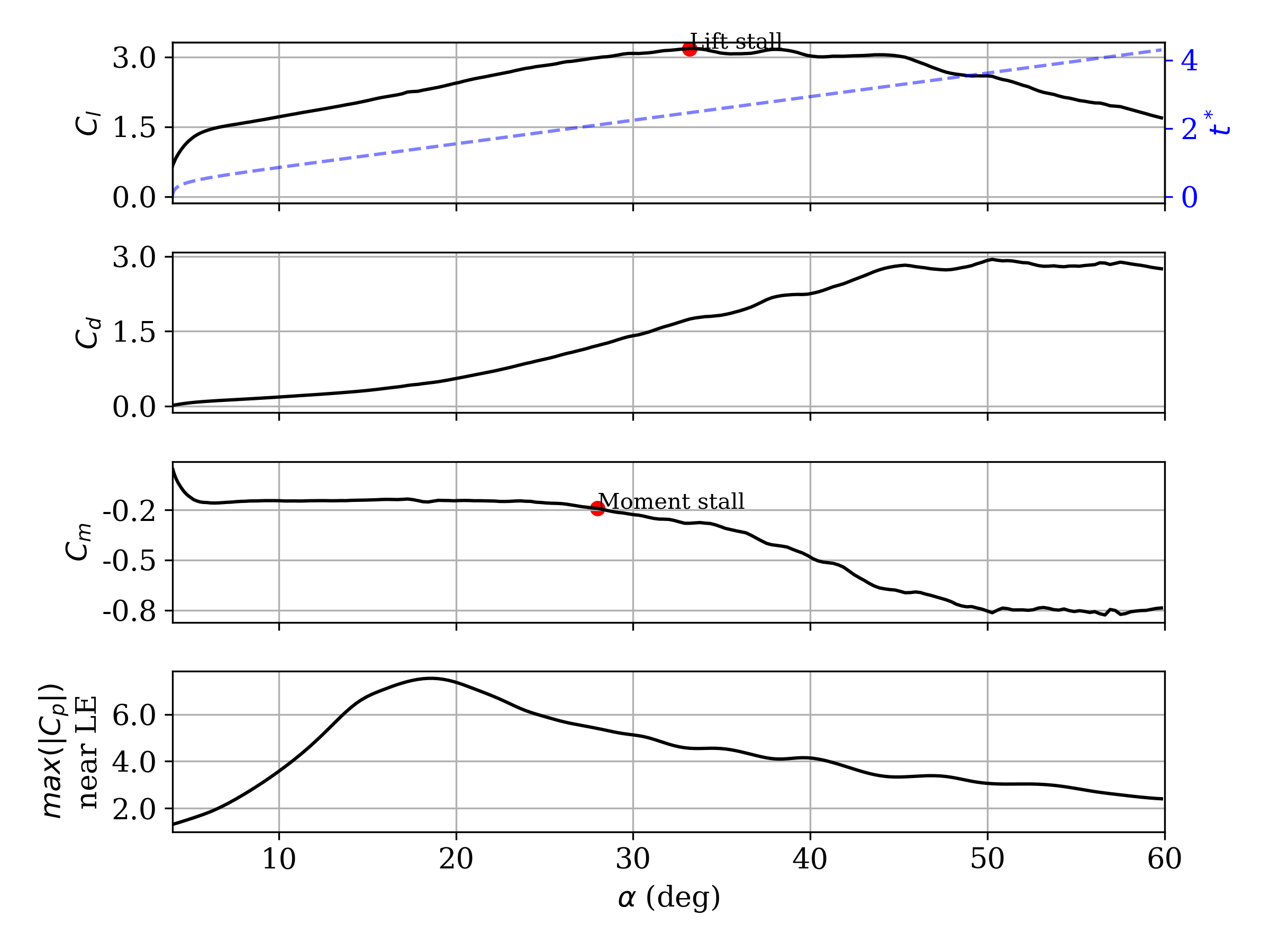}
	\caption{Variation with $\alpha$ of aerodynamic coefficients (top three panels) and $\max(|C_p|)$ near the first 5\% of airfoil chord (bottom panel) for Case R60-p25.\label{fig:R60-p25_aerodyn_coeff}}
\end{figure}

The top three panels of Fig.~\ref{fig:R60-p25_aerodyn_coeff} show the variation of the unsteady aerodynamic coefficients with $\alpha$.
Moment stall occurs around $\alpha \sim 25^{\circ}$ when the DSV convects past the quarter-chord point.
Maximum lift occurs around $\alpha \sim 33^{\circ}$, which is marked as the lift stall point in the figure.
However, the lift remains elevated until $45^{\circ}$, attributed to the proximity of the DSV to the airfoil surface. 
As the DSV progressively convects away from the airfoil surface, the reduction in vortex-induced contribution to the bound circulation leads to a decline in $C_l$.
The bottom panel of Fig.~\ref{fig:R60-p25_aerodyn_coeff} shows the variation of peak leading-edge $\abs{C_p}$.
The collapse of the LSB is accompanied by a reduction max($\abs{C_p}$) near the leading edge around $\alpha \sim 18^{\circ}$, as is typical in a bubble-bursting, leading-edge stall.

%

\begin{figure}
    \hspace*{\fill}
    \subcaptionbox{$LESP$\label{fig:R60-p25_LESP}}{\incfig[width=0.49\textwidth]{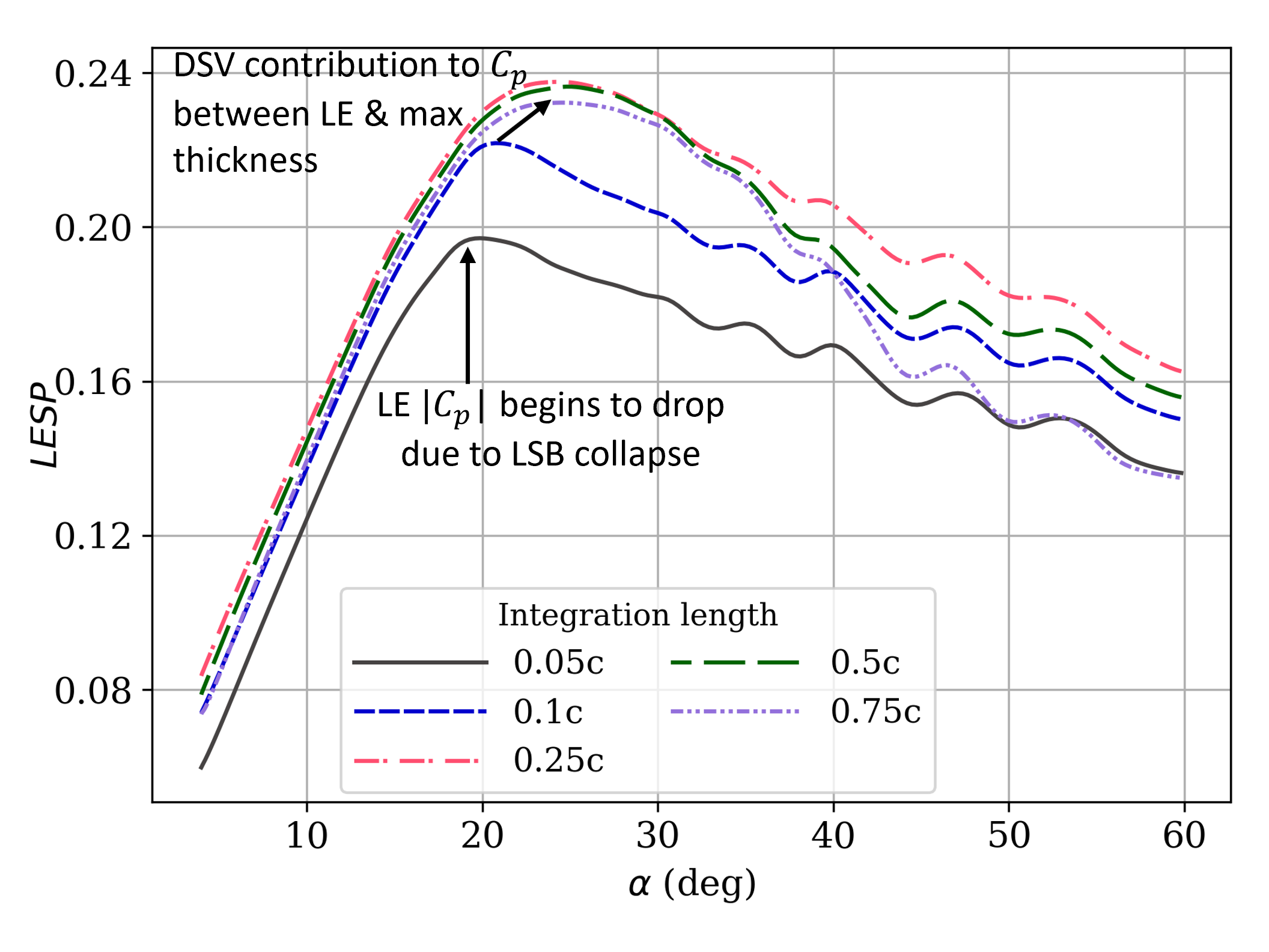}}
    \hfill
    \subcaptionbox{$\abs{BEF}$\label{fig:R60-p25_BEF}}{\incfig[width=0.49\textwidth]{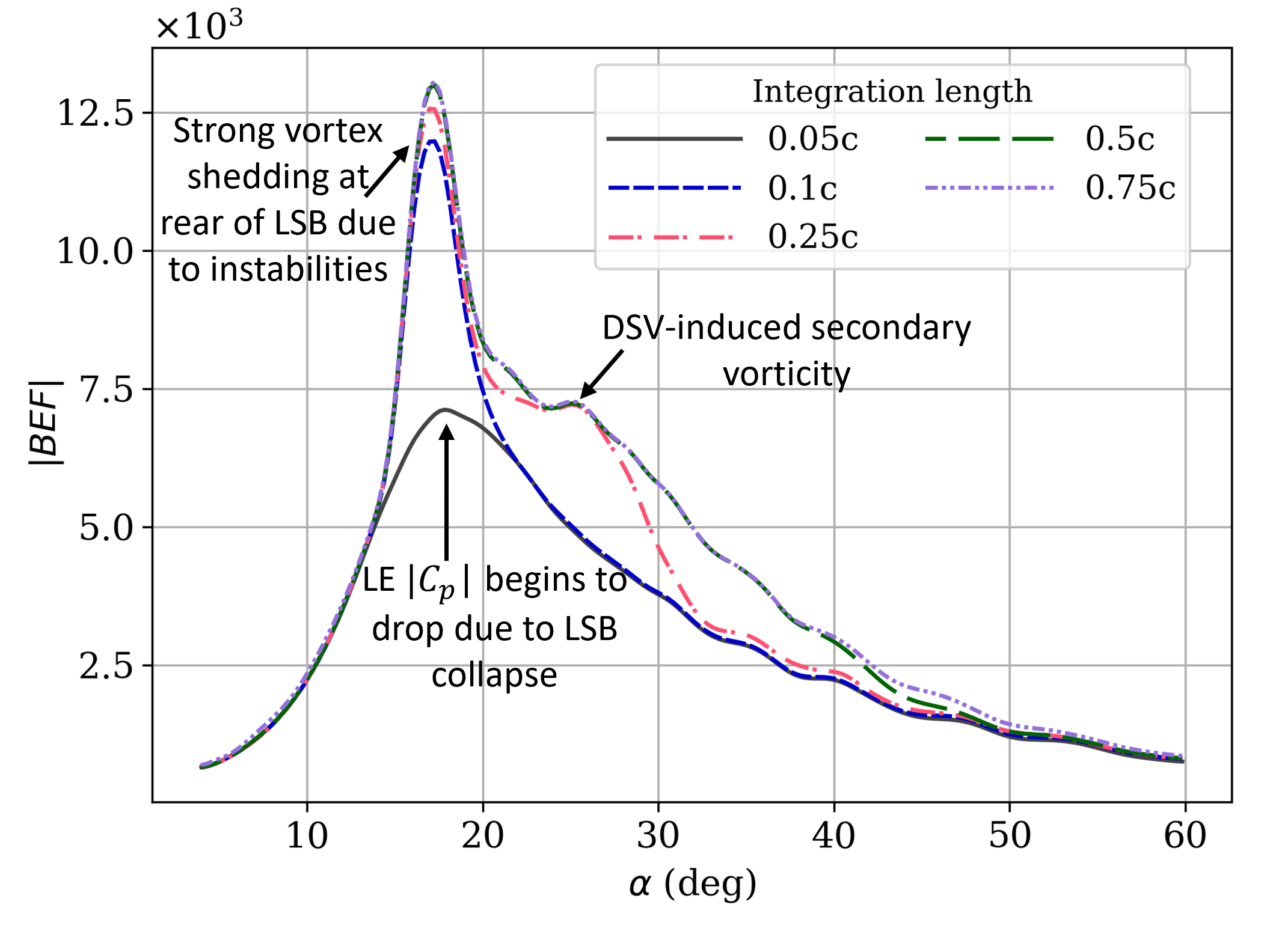}}
    \hspace*{\fill}
    \caption{$LESP$ (a) and $BEF$ (b) integrated over different chord lengths, plotted against $\alpha$, for Case R60-p25.}
    \label{fig:R60-p25_LESP_BEF}
\end{figure}

The variation with $\alpha$ of $LESP$ and $BEF$ is shown in Fig.~\ref{fig:R60-p25_LESP_BEF}.
Both parameters exhibit distinct peaks for all integration lengths, attributed to the decrease in leading-edge $\abs{C_p}$ as the LSB breaks down.
The larger values of $\abs{BEF}$ for integration lengths greater than $0.05\,c$ are due to the large vorticity and vorticity flux associated with the vortices shed downstream of the LSB as it becomes unstable.
Due to the greater spatial distance of the LSB from the geometric leading edge and its extended length, DSV formation occurs around $0.1\,c$ rather than very close to the leading edge. 
This is consistent with the expected behavior at low $Re$~\shortcite{Gaster1967}.

%

Figure~\ref{fig:R60-p25_LESP_BEF} shows that the instance of the peak in $LESP$ moves aft (to higher $\alpha$) with increasing integration length.
This is due to the increasing contribution from the suction induced by the DSV as it grows and influences a larger portion of the airfoil between the leading edge and the maximum thickness point. 
Hence, even though leading-edge $C_p$ magnitude begins to drop in advance of $20^{\circ}$, the $LESP$ curve obtained by integrating over $0.25\,c$ continues to rise until about $22^\circ$. 
In contrast, the peak $\abs{BEF}$ location occurs before  $20^{\circ}$ for all integration lengths.
The leading edge $C_p$ peak representing stall onset is captured by the $BEF$ curve corresponding to integration up to $0.05\,c$.
For integration lengths larger than $0.05\,c$, the peak $BEF$ value is much higher due to the strong vortices shed from the rear of the LSB around the same time.
The vortex shedding event is captured by the $BEF$ for this case as well when appropriate regions of the airfoil are included.
There is another peak around $25^{\circ}$ for integration lengths $\geq 0.25\,c$, which corresponds to an increase in induced secondary vorticity as the DSV gets stronger as it grows.
While $LESP$ captures the stall onset point, the localized vortex shedding event is missed.

\subsection{Case R10-p05}
\label{sec:R10-p05}

For the lower $Re$ cases, the DSV remains laminar through nearly the entire maneuver. These cases are also characterized by a DSV system consisting of several large-scale laminar vortices.

\begin{figure}
    \centering
    \subcaptionbox{$\alpha = 11.0^{\circ}$}{\incfig[width=0.485\textwidth]{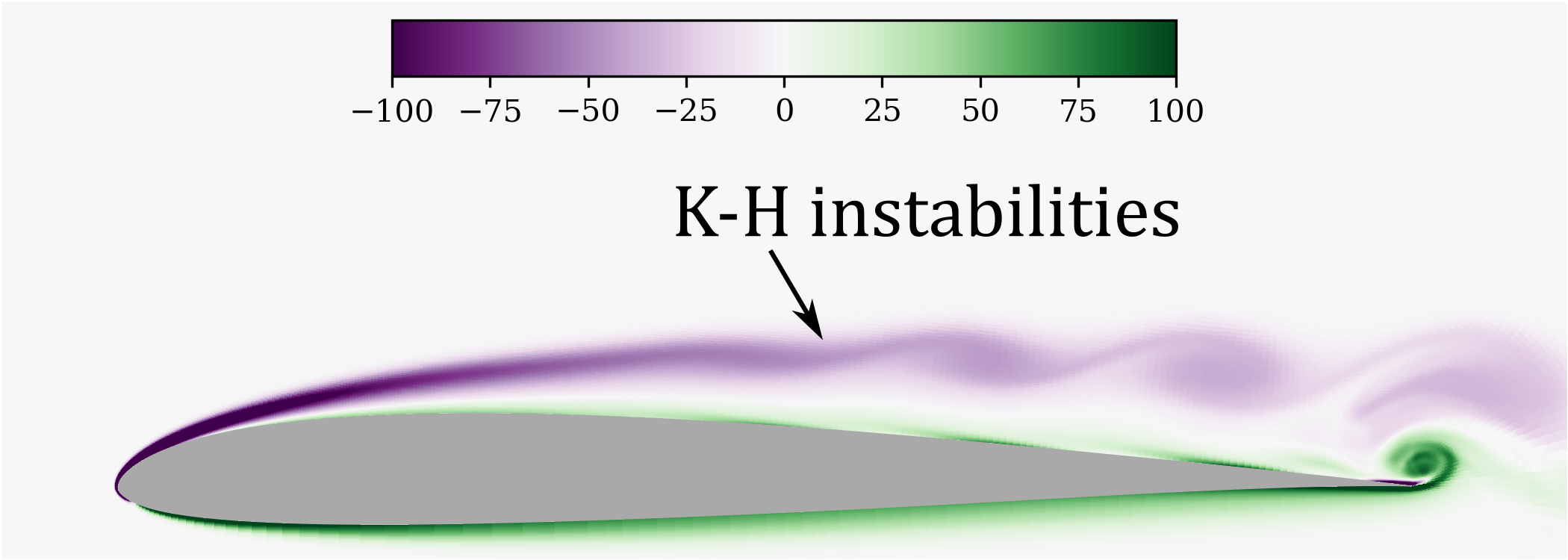}} 
    \hfill
    \subcaptionbox{$\alpha = 13.3^{\circ}$}{\incfig[width=0.485\textwidth]{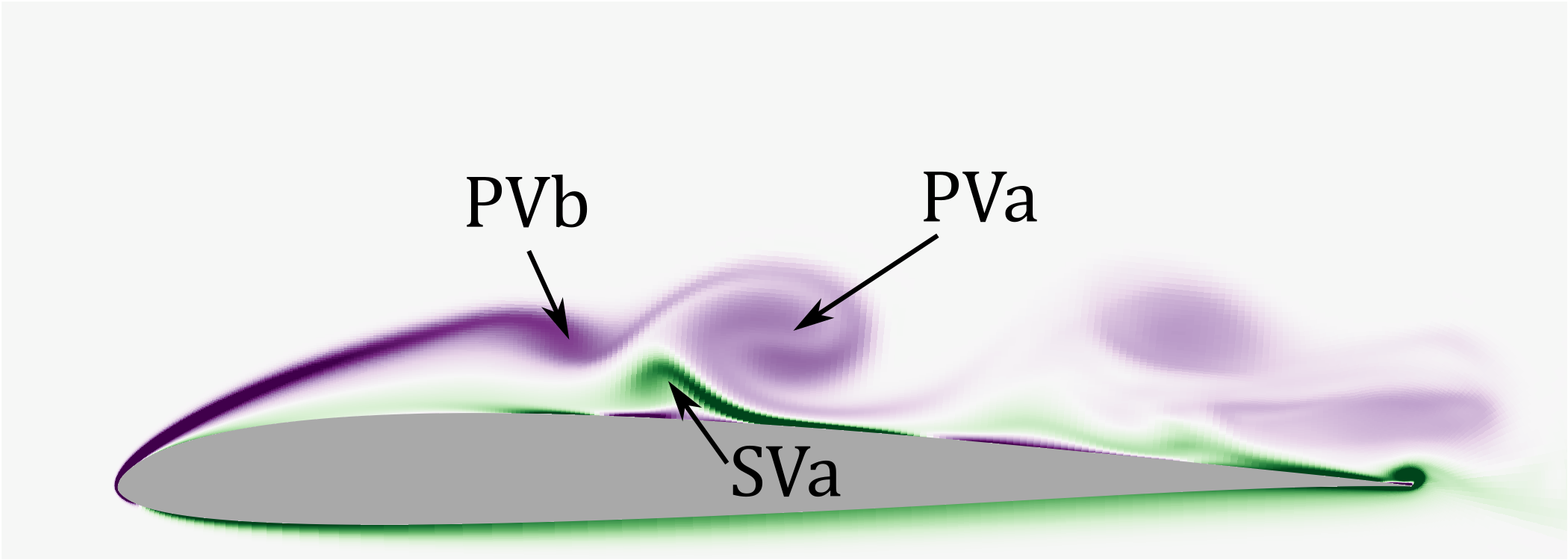}}
    \par\bigskip
    %
    \subcaptionbox{$\alpha = 14.8^{\circ}$}{\incfig[width=0.485\textwidth]{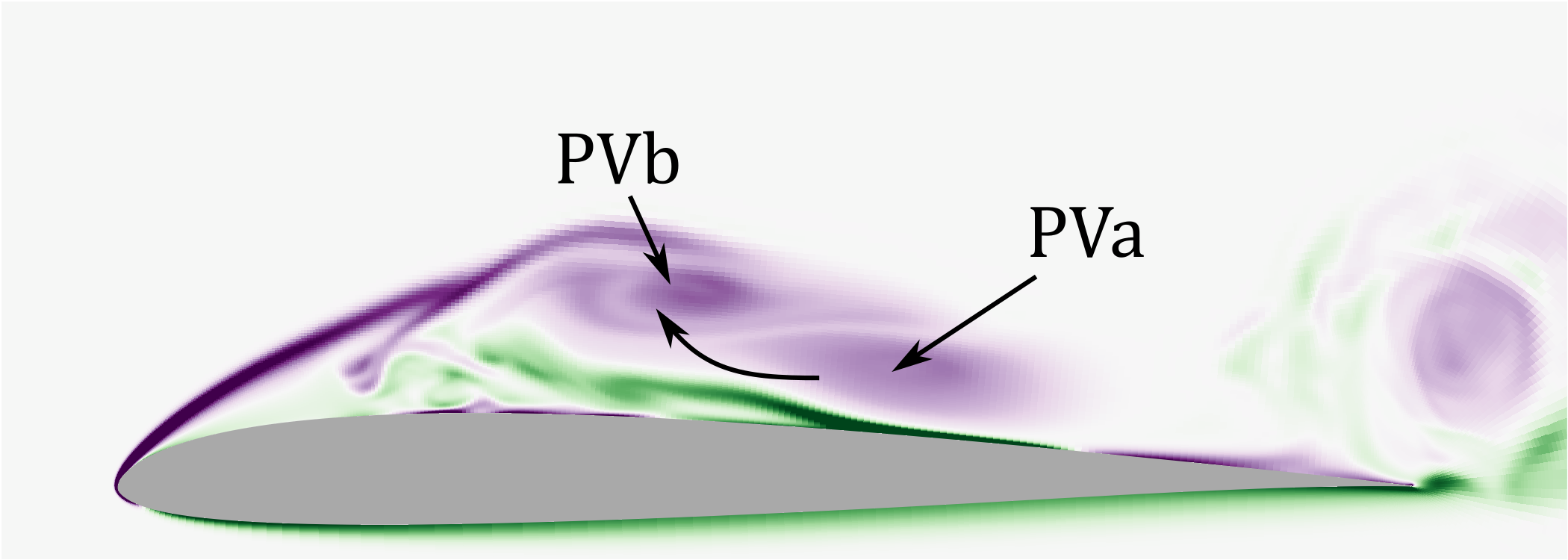}}
    \hfill
    \subcaptionbox{$\alpha = 15.7^{\circ}$}{\incfig[width=0.485\textwidth]{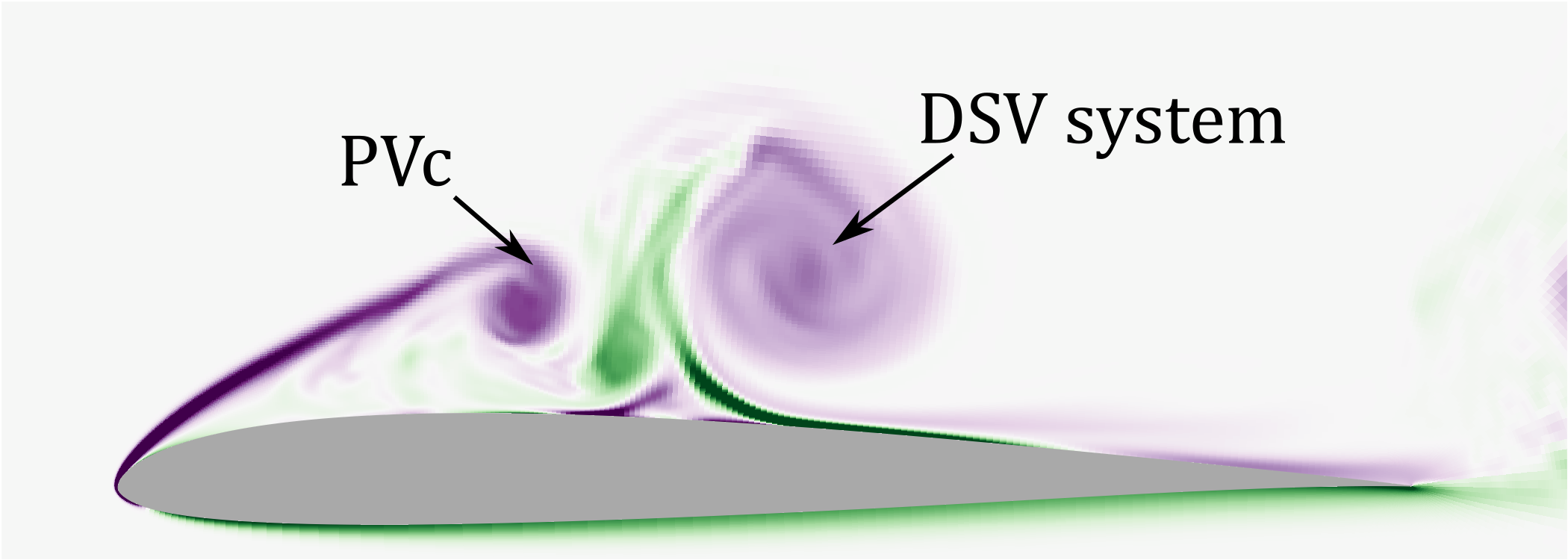}} 
    \par\bigskip
    %
    \subcaptionbox{$\alpha = 18.3^{\circ}$}{\incfig[width=0.485\textwidth]{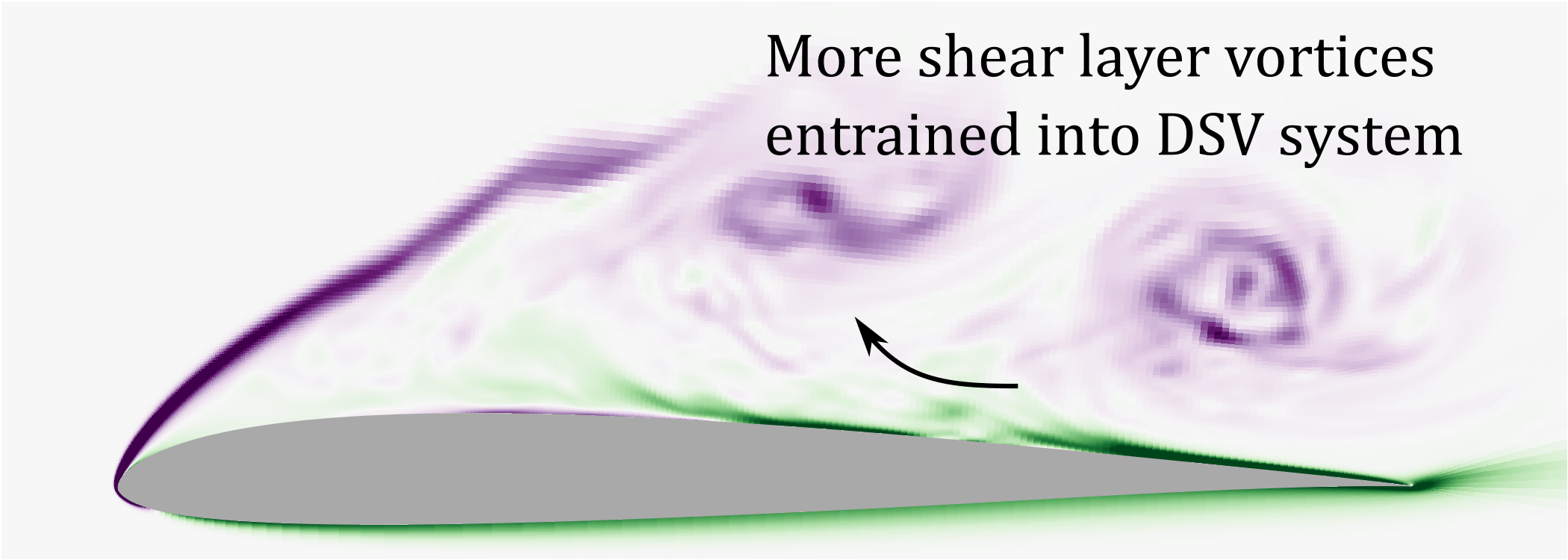}} 
    \hfill
    \subcaptionbox{$\alpha = 20.3^{\circ}$}{\incfig[width=0.485\textwidth]{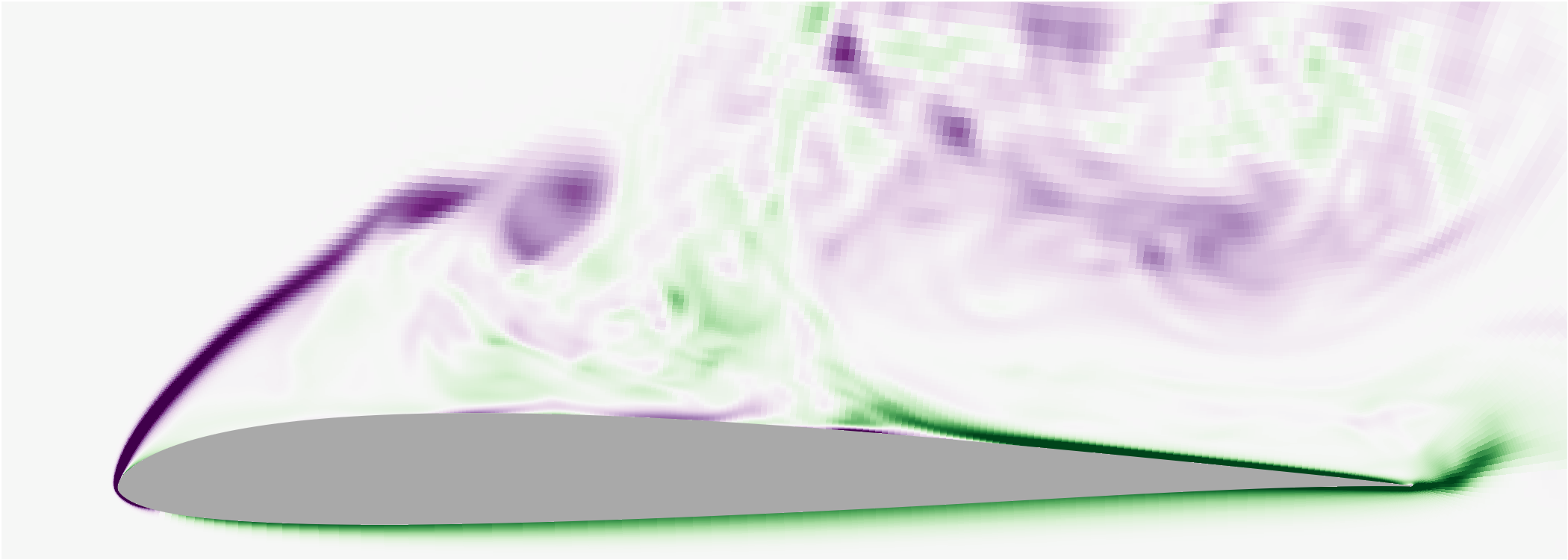}} 
    %
    \caption{Vorticity contours showing the formation of a DSV system through shear layer vortex and wall interactions for Case R10-p05.
    \label{fig:R10-p05_vortcontours}}
\end{figure}

Figure~\ref{fig:R10-p05_vortcontours} shows the vorticity contours at specific instances.
A region of reversed flow at the trailing edge propagates upstream as the airfoil starts to pitch up.
Around $\alpha \sim 10^{\circ}$, the separated shear layer from the leading edge becomes susceptible to instabilities and sheds spanwise vortices.
Panel a ($\alpha \sim 11^{\circ}$) shows the initial stages of the unstable, separated shear layer.
The instability location moves upstream as the airfoil continues to pitch up.
The large-scale shear layer vortices induce secondary vorticity beneath them.
The induced secondary vorticity develops into coherent, counter-clockwise vortices, which act to cut off the primary shear layer vortex that is formed around mid chord.

Panel b ($\alpha \sim 13.3^{\circ}$) shows the first instance of secondary vorticity (SVa) cutting off the primary shear layer vortex (PVa), and the roll-up of the shear layer upstream into another vortex (PVb). 
The shedding of shear layer vortices and induced secondary vortices result in multiple instances of vortex entrainment.
Therefore, the DSV system in the present case comprises multiple shear layer vortices entrained together as $\alpha$ increases.
Note that the DSV system is centered downstream of the quarter-chord point of the airfoil during its incipience.
Panels c ($\alpha \sim 14.8^{\circ}$) and d ($\alpha \sim 15.7^{\circ}$) show the merging of PVa and PVb into a single coherent vortex and its downstream convection, respectively.
During the later part of the unsteady motion (panels e-f), the DSV, continuing to entrain more leading-edge vortices, transitions to turbulence as it propagates downstream. 
The DSV weakens due to the entrained counter-clockwise vorticity (shown in green contours) and viscous dissipation.
The dominant, large-scale vortices located away from the surface of the airfoil generate and influence the motion of smaller-scale secondary vortices and, over time, tertiary vortices close to the wall, that affect surface quantities and make it challenging to interpret the space-time plots of $C_p$ and $C_f$.
Movie \dispmcounter provides a clearer picture of the sequence of events as the airfoil pitches up.

\begin{figure}
	\centering
	\incfig[width=0.6\textwidth]{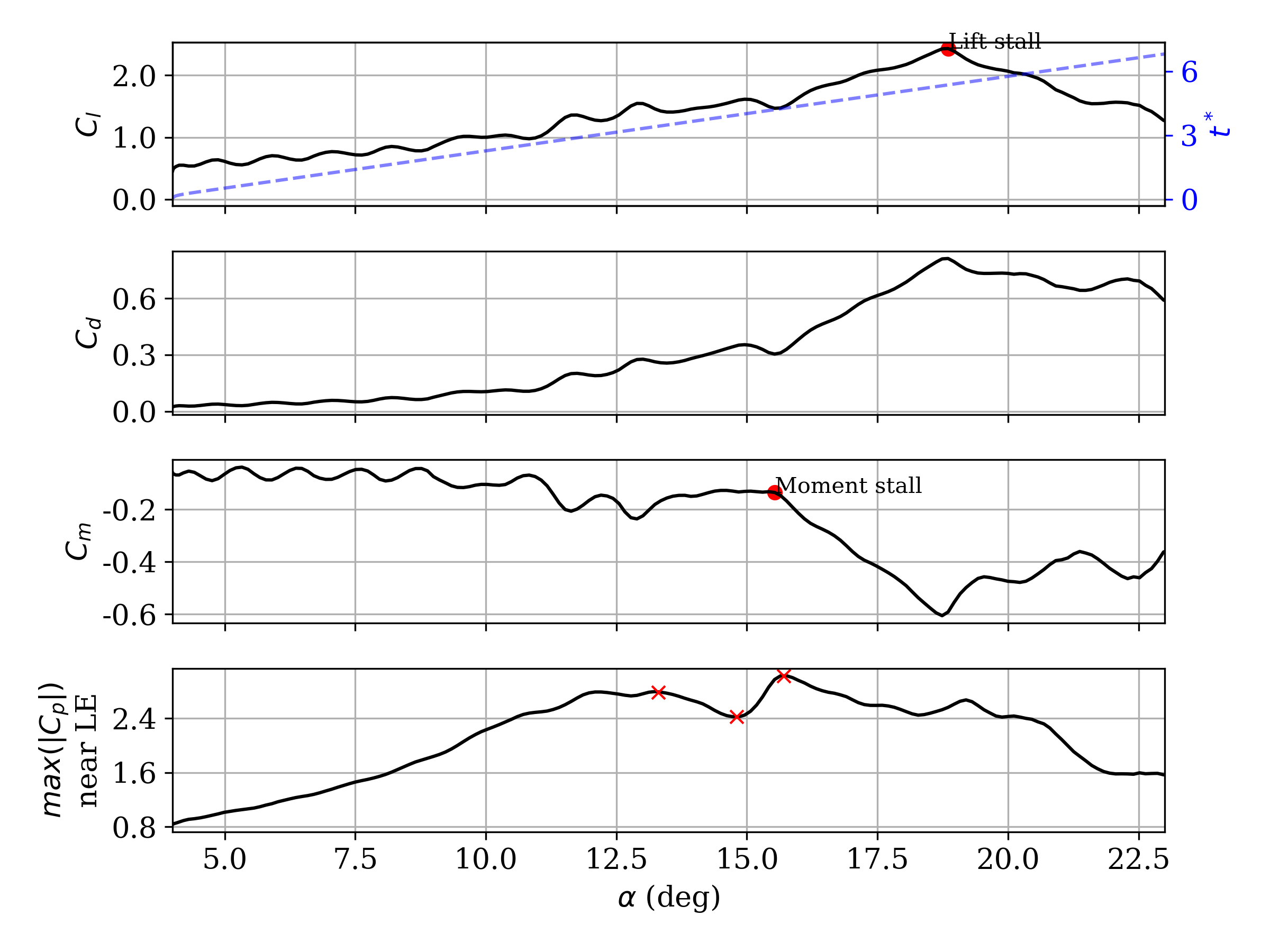}
	\caption{Variation with $\alpha$ of aerodynamic coefficients (top three panels) and $\max(|C_p|)$ near the first 5\% of airfoil chord (bottom panel) for Case R10-p05. The red x's correspond to panels b-d in Fig.~\ref{fig:R10-p05_vortcontours}.}
	\label{fig:R10-p05_aerodyn_coeff}
\end{figure}

The top three panels of Fig.~\ref{fig:R10-p05_aerodyn_coeff} show the variation with $\alpha$ of the unsteady aerodynamic coefficients. 
Large wiggles are observed in $C_l$, $C_d$, and $C_m$ before stall due to the strong laminar vortices shed from the shear layer.
$C_l$ continues to increase until $\alpha \sim 18.8^{\circ}$ due to vortex-induced suction/lift from multiple shear layer vortices.
However, large fluctuations in $\max(\abs{C_p})$ near the leading edge (shown in the bottom panel of Fig.~\ref{fig:R10-p05_aerodyn_coeff}) are observed much earlier as large variations in shed vorticity occur downstream.
The peaks and the valley in $\max(\abs{C_p})$, identified by red x's in Fig.~\ref{fig:R10-p05_aerodyn_coeff}, are due to the vortices highlighted in panels b-d in Fig.~\ref{fig:R10-p05_vortcontours}.
This undulatory behavior of $\max(\abs{C_p})$ is attributed to the changes in the curvature of the shear layer around the leading edge due to the vortices shed downstream.
As $\alpha$ increases beyond $12.5^\circ$ and vortex shedding occurs around mid-chord, the induced secondary vorticity pushes the shear layer away from the airfoil suction surface, reducing its curvature and causing $\max(\abs{C_p})$ to dip.
After $15^\circ$, the (PVa + PVb) DSV system detaches from the shear layer, arresting the movement of the shear layer away from the surface, thereby temporarily but sharply increasing $\max(\abs{C_p})$ near the leading edge.

\begin{figure}
    \hspace*{\fill}
    \subcaptionbox{$LESP$\label{fig:R10-p05_LESP}}{\incfig[width=0.49\textwidth]{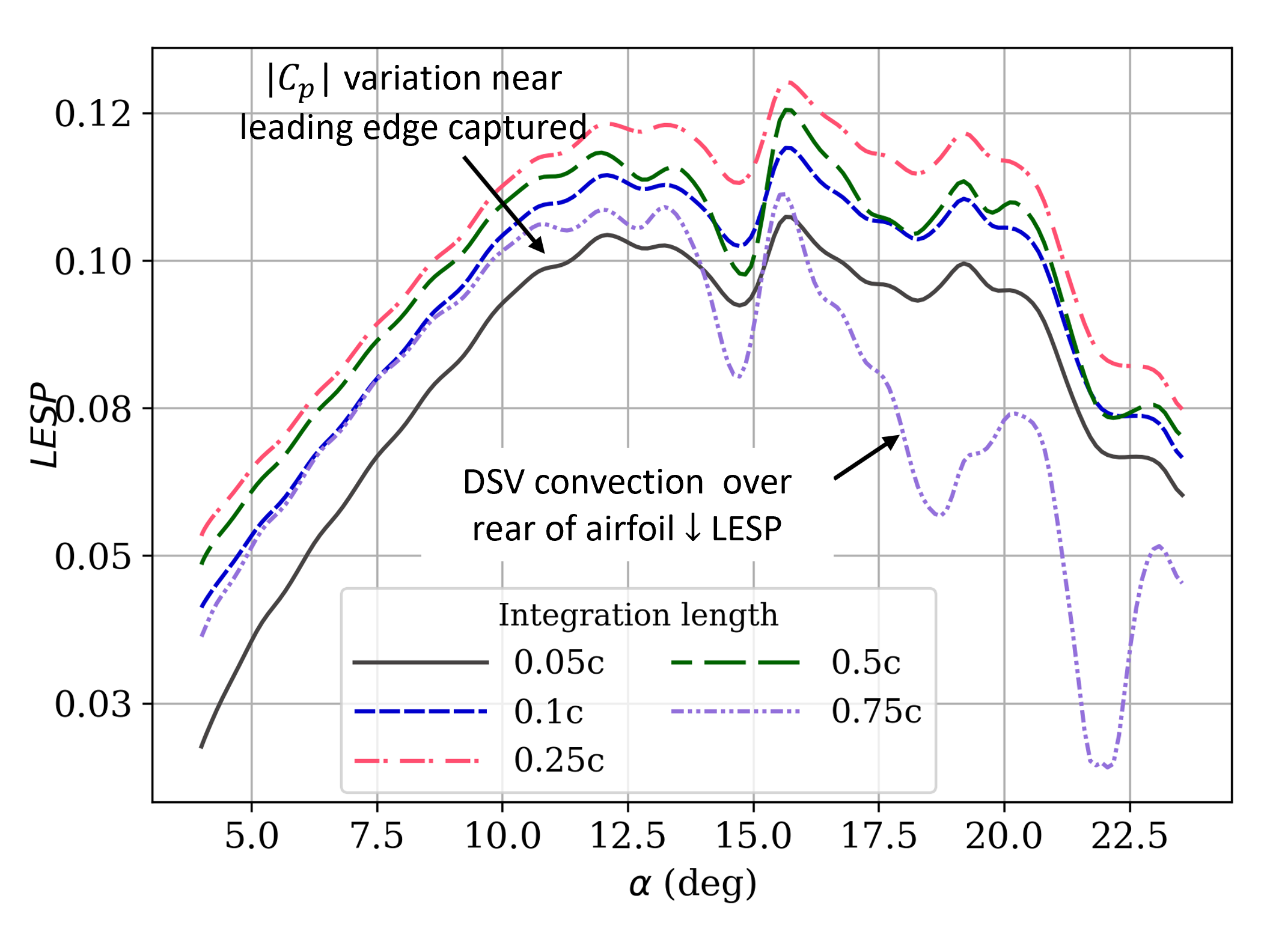}}
    \hfill
    \subcaptionbox{$\abs{BEF}$\label{fig:R10-p05_BEF}}{\incfig[width=0.49\textwidth]{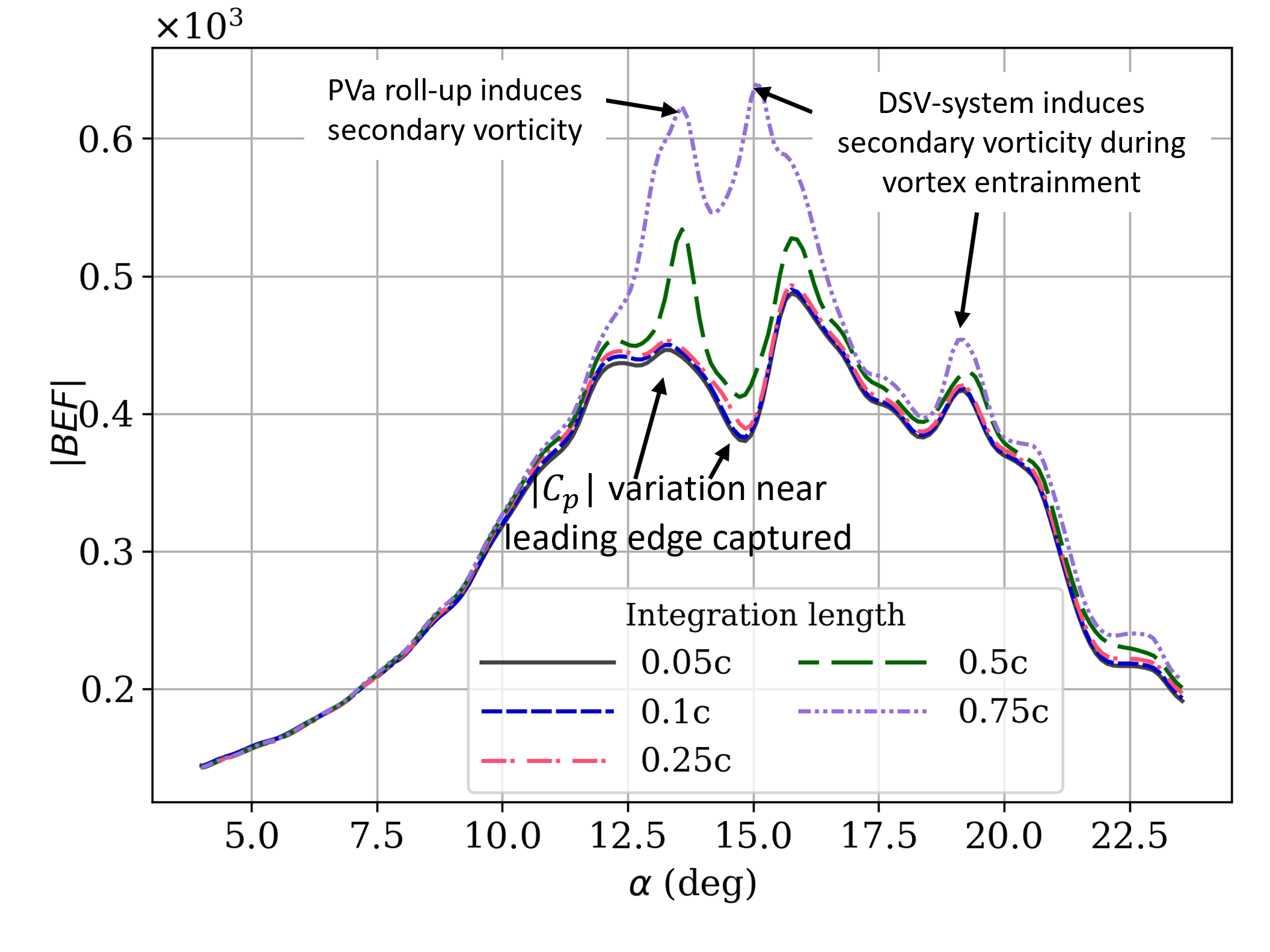}}
    \hspace*{\fill}
    \caption{$LESP$ (a) and $\abs{BEF}$ (b) integrated over different chord lengths, plotted against $\alpha$, for Case R10-p05.}
    \label{fig:R10-p05_LESP_BEF}
\end{figure}

The variation with $\alpha$ of $LESP$ and $BEF$ integrated over various chord lengths is shown in Fig.~\ref{fig:R10-p05_LESP_BEF}.
The shedding of the downstream shear layer vortices that affects leading-edge $C_p$ also impacts the variation in $LESP$ and $BEF$.
For the $BEF$, integrated lengths up to $0.25\,c$ follow the same trend as $\max(\abs{C_p})$ near the leading edge. 
The larger integrated lengths show larger $\abs{BEF}$ values at the peak locations, corresponding to the vorticity shed downstream.
The increase in $BEF$ around $13^{\circ}$ correponds to increased induced secondary vorticity from the roll-up of PVa.
The increase to the second peak around $15.0^{\circ}$ correponds to the entrainment of PVa by PVb during which PVa moves closer to the wall, causing increased secondary vorticity.
The small increase to another peak around $18.3^{\circ}$ corresponds to another instance of vortex merging with the DSV system, which again induces increased secondary vorticity.
The $LESP$ curves for all integration lengths show nearly the same trend as $\max(\abs{C_p})$ near the leading edge.
Note that there is no additional local (in space and time) contribution to $LESP$ from the vortices formed around mid-chord; the curves show the same trend and are almost simply displaced along the ordinate.
The largest integration length shows a deviation in trend at high $\alpha$ since the chord-wise force contribution from the shed vortices is directed in the opposite direction over the rear of the airfoil.
It is the local contributions around vortex formation that distinguish the $BEF$ from the $LESP$.
This is elucidated in section~\ref{sec:results_comparisons} by delineating specific contributions from different airfoil sections. 



\subsection{Case R10-p25}
\label{sec:R10-p25}
The higher pitch rate case at lower $Re$ is also characterized by the shedding of multiple shear layer vortices with the flow over the airfoil remaining laminar throughout.
The difference from the lower pitch rate case (R10-p05) is the formation of a stronger DSV system further upstream with more pronounced lag effects due to unsteadiness.
Movie~\dispmcounter~clearly illustrates the sequence of events during the unsteady maneuver using spanwise vorticity contours.

\begin{figure}
    %
    \subcaptionbox{$\alpha = 20.2^{\circ}$\label{fig:R10-p25_vortcontours_a}}{\incfig[width=0.485\textwidth]{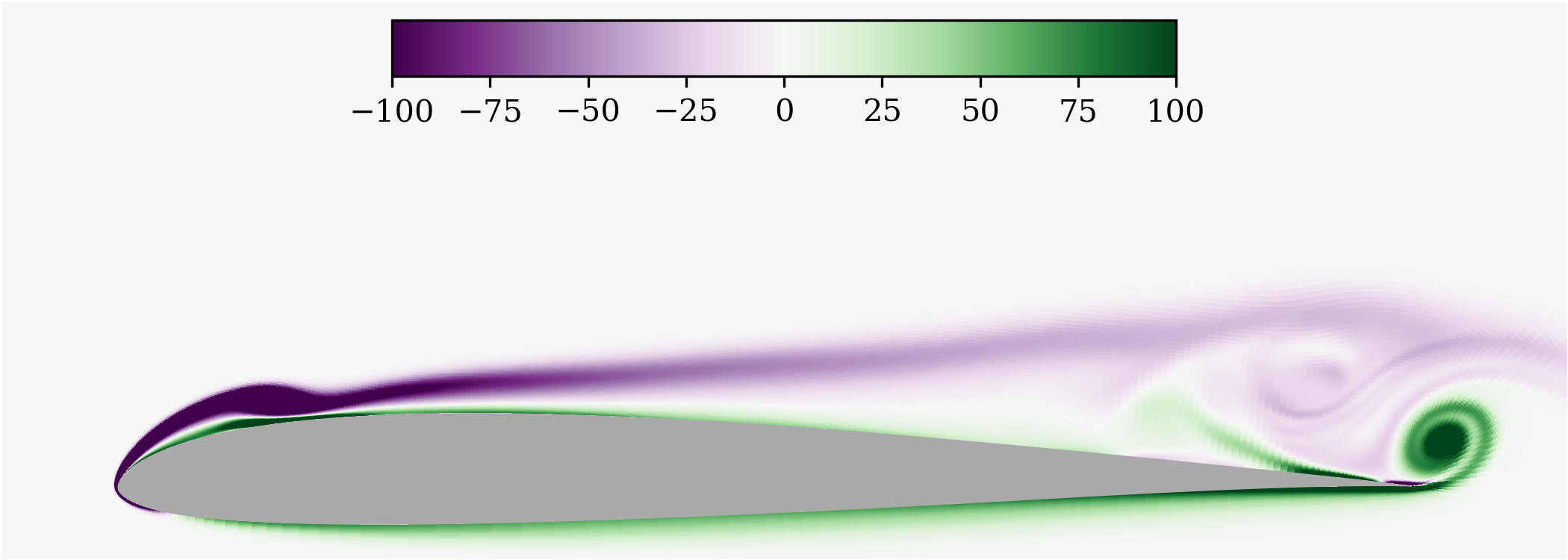}}
    \hfill
    \subcaptionbox{$\alpha = 22.8^{\circ}$\label{fig:R10-p25_vortcontours_b}}{\incfig[width=0.485\textwidth]{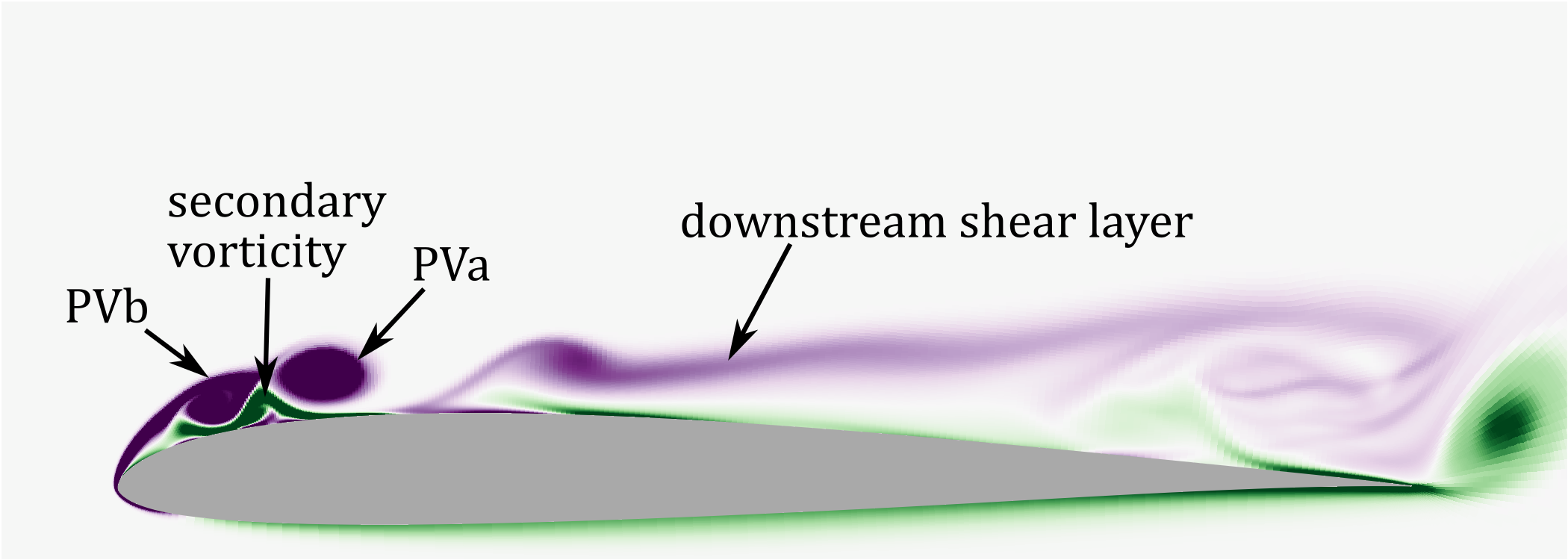}} 
    \par\bigskip
    %
    \subcaptionbox{$\alpha = 25.7^{\circ}$\label{fig:R10-p25_vortcontours_c}}{\incfig[width=0.485\textwidth]{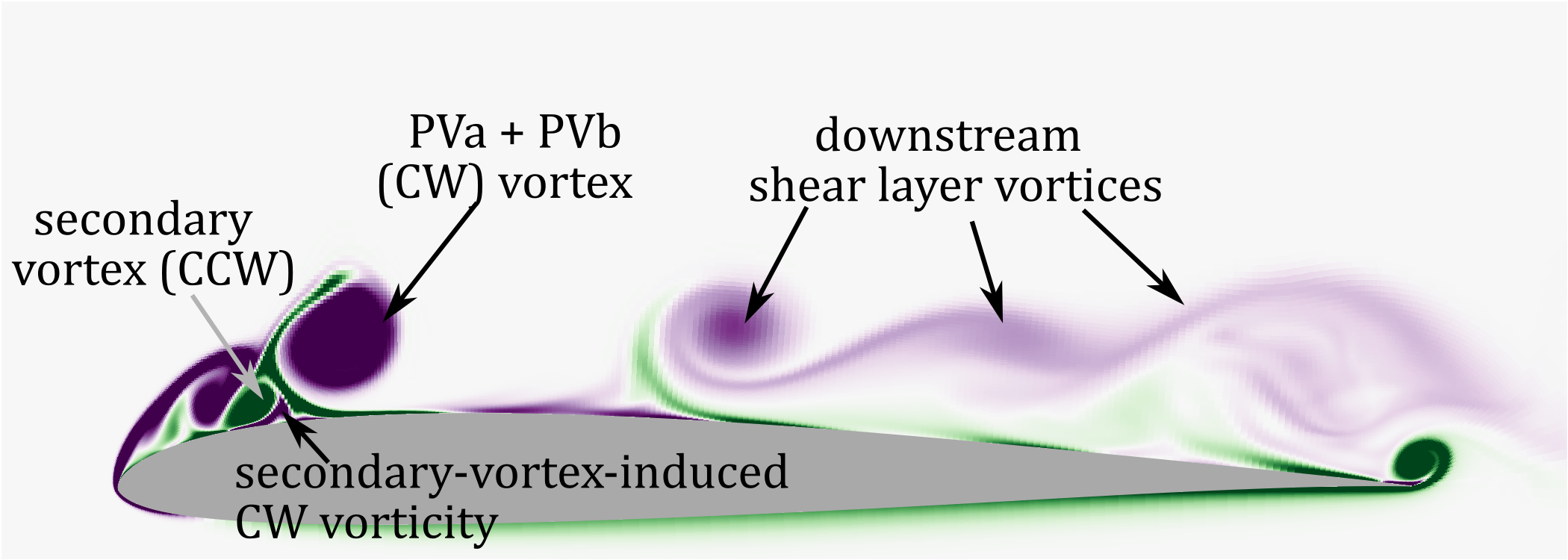}} 
    \hfill
    \subcaptionbox{$\alpha = 29.7^{\circ}$\label{fig:R10-p25_vortcontours_d}}{\incfig[width=0.485\textwidth]{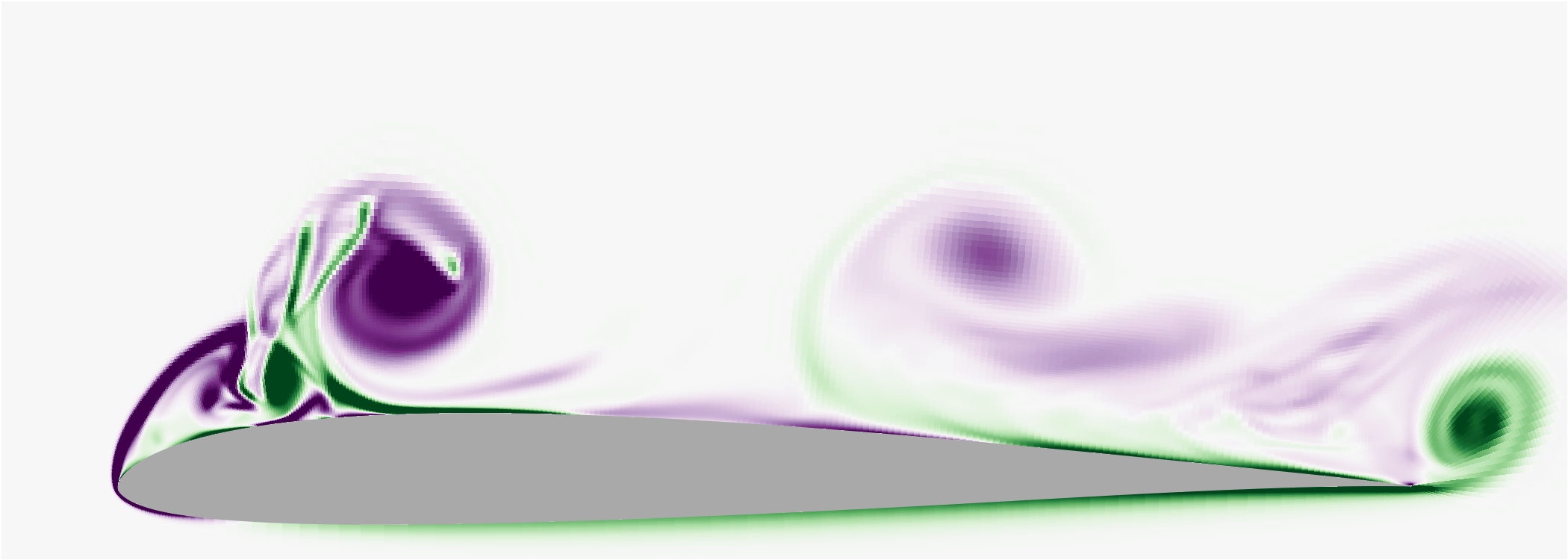}} 
    \par\bigskip
    %
    \subcaptionbox{$\alpha = 33.2^{\circ}$\label{fig:R10-p25_vortcontours_e}}{\incfig[width=0.485\textwidth]{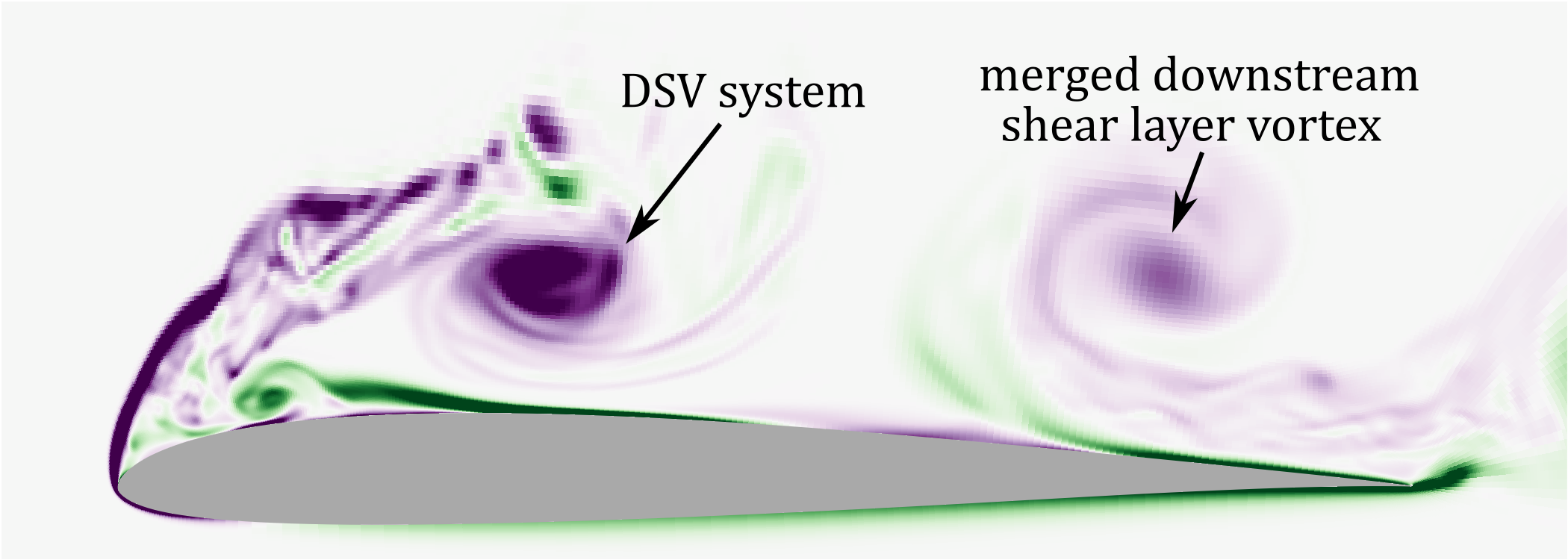}} 
    \hfill
    \subcaptionbox{$\alpha = 49.7^{\circ}$\label{fig:R10-p25_vortcontours_f}}{\incfig[width=0.485\textwidth]{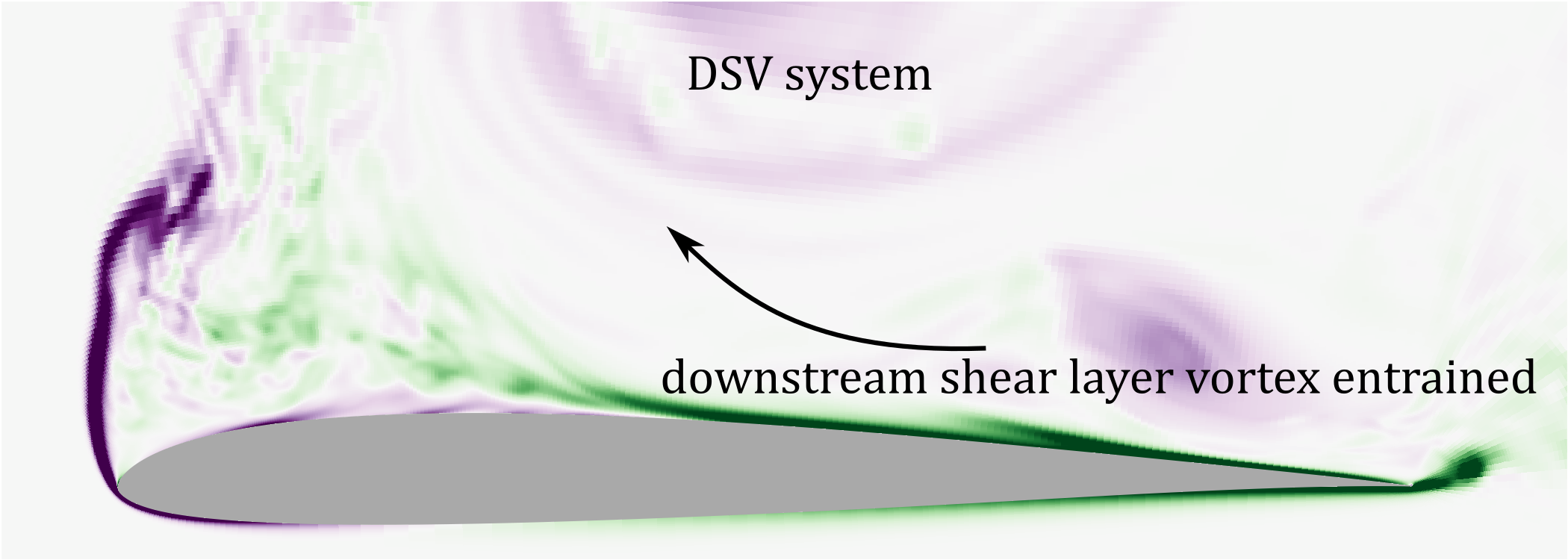}} 
    %
    \caption{Vorticity contours at different instances during the pitch-up maneuver for Case R10-p25. Similar behavior as in Case R10-p05, but is delayed (in $t^*$ and $\alpha$) due to the higher pitch rate, and vortex shedding occurs closer to the leading edge.
    \label{fig:R10-p25_vortcontours}}
\end{figure}

The sequence of events remains similar to Case R10-p05 but is postponed to higher $\alpha$ due to the pronounced unsteady lag. 
Figure \ref{fig:R10-p25_vortcontours} shows span-averaged vorticity contours at a few instances during the pitch-up maneuver.
The upstream propagation of trailing edge reversed flow and the development of instabilities in the shear layer leading to vortex formation are observed in this case as well.
However, due to the larger APG encountered by the flow, stronger leading-edge vortices are shed farther upstream (around $0.12\,c$), beginning around $\alpha \sim 20^{\circ}$ (Fig.~\ref{fig:R10-p25_vortcontours_a}).
Induced secondary (counter-clockwise) vorticity acts to cut off the shear layer, which forms clockwise vortices near the airfoil leading edge (Fig.~\ref{fig:R10-p25_vortcontours} panels b-d).
When the clockwise vortex (marked as `PVa') rolls up, it cuts off the downstream shear layer from the leading edge, leading to its roll-up into vortices downstream (as pointed out in Fig.~\ref{fig:R10-p25_vortcontours_b}).
The secondary vorticity itself lifts up due to the induction from the leading edge vortices, rolls up, and is pinched off by the clockwise vorticity induced by it (Fig.~\ref{fig:R10-p25_vortcontours_c}).
This process repeats a few times, and multiple vortices are shed from the leading edge.
Similar to Case R10-p05, the DSV system continues to entrain these shed vortices and grows in size. 
DSV-induced suction on the airfoil surface is strongest around $\alpha \sim 33^{\circ}$ (Fig.~\ref{fig:R10-p25_vortcontours_e}).
The effect of the DSV system on the surface drops as it moves farther away and becomes more diffuse due to viscosity and annihilation from the entrained secondary vorticity (Fig.~\ref{fig:R10-p25_vortcontours_f}).
The DSV system also entrains the downstream shear layer vortex around this time.

\begin{figure}[h!]
	\incfig[width=0.6\textwidth]{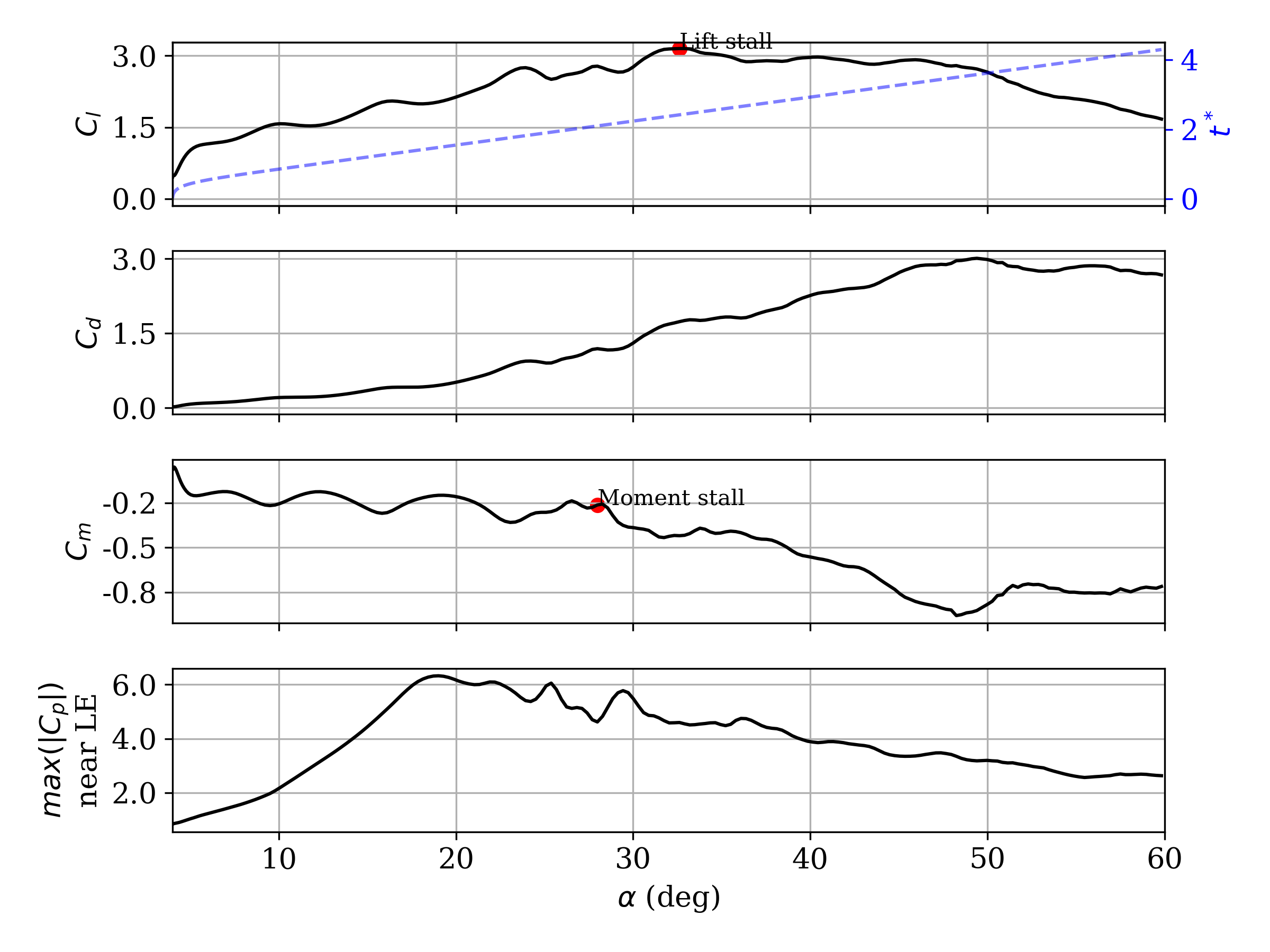}
	\caption{Variation with $\alpha$ of aerodynamic coefficients (top three panels) and $\max(|C_p|)$ near the first 5\% of airfoil chord (bottom panel) for Case R10-p25.}
	\label{fig:R10-p25_aerodyn_coeff}
\end{figure}

The top three panels of Fig.~\ref{fig:R10-p25_aerodyn_coeff} show the variation of the unsteady aerodynamic coefficients. 
Maximum lift occurs around $\alpha \sim 33^{\circ}$, when the DSV system induces the strongest suction over the airfoil surface.
The lift continues to remain high until around $45^{\circ}$, owing to the DSV system remaining relatively close to the airfoil surface and continuing to interact with the newly shed leading-edge vortices and induced secondary vorticity.
The magnitude of leading edge $C_p$ shows undulations (see bottom panel of Fig.~\ref{fig:R10-p25_aerodyn_coeff}) due to the effect of vortices shed downstream, similar to the previous case.
 
\begin{figure}[h!]
    \hspace*{\fill}
    \subcaptionbox{$LESP$\label{fig:R10-p25_LESP}}{\incfig[width=0.49\textwidth]{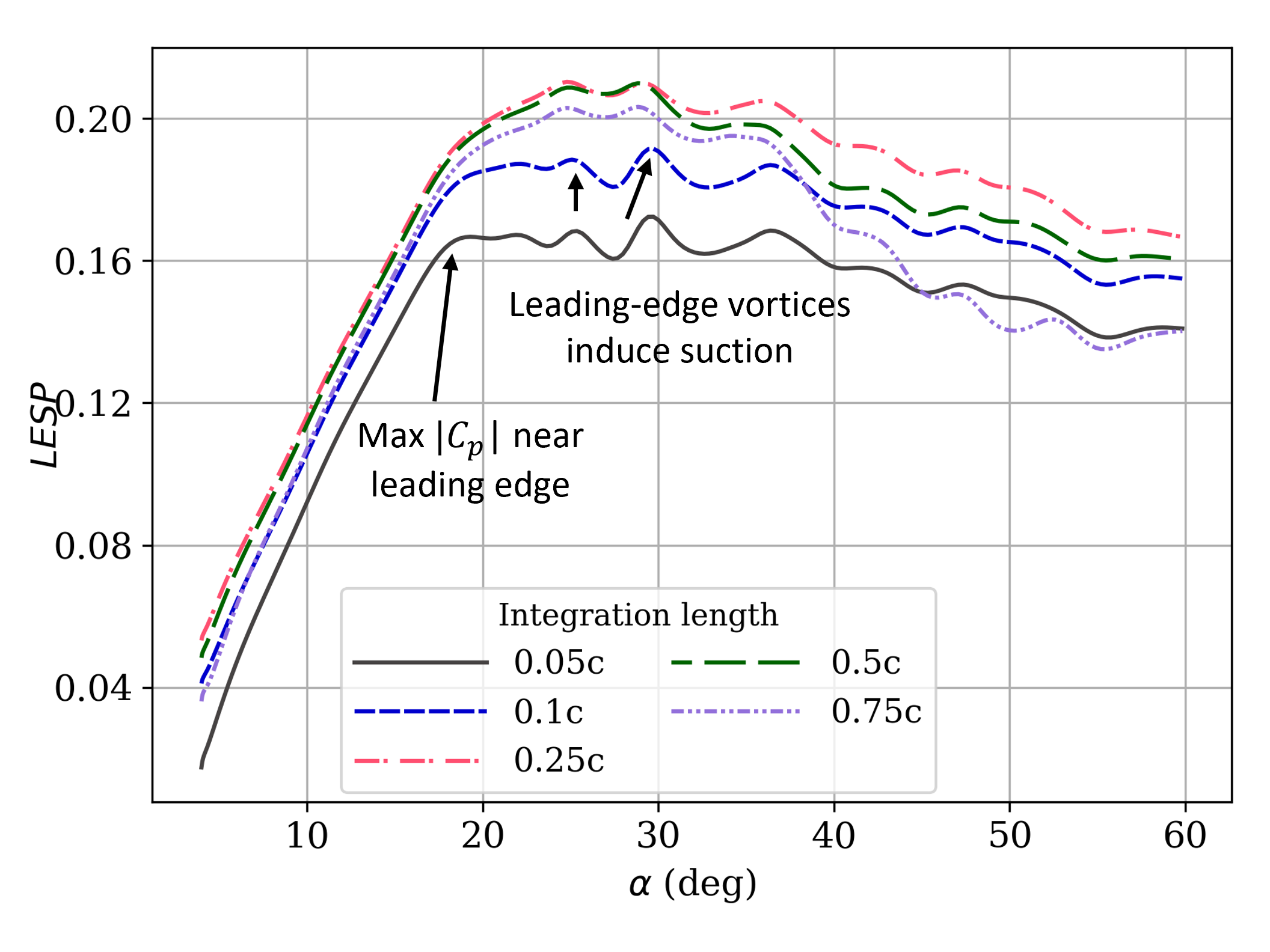}}
    \hfill
    \subcaptionbox{$\abs{BEF}$\label{fig:R10-p25_BEF}}{\incfig[width=0.49\textwidth]{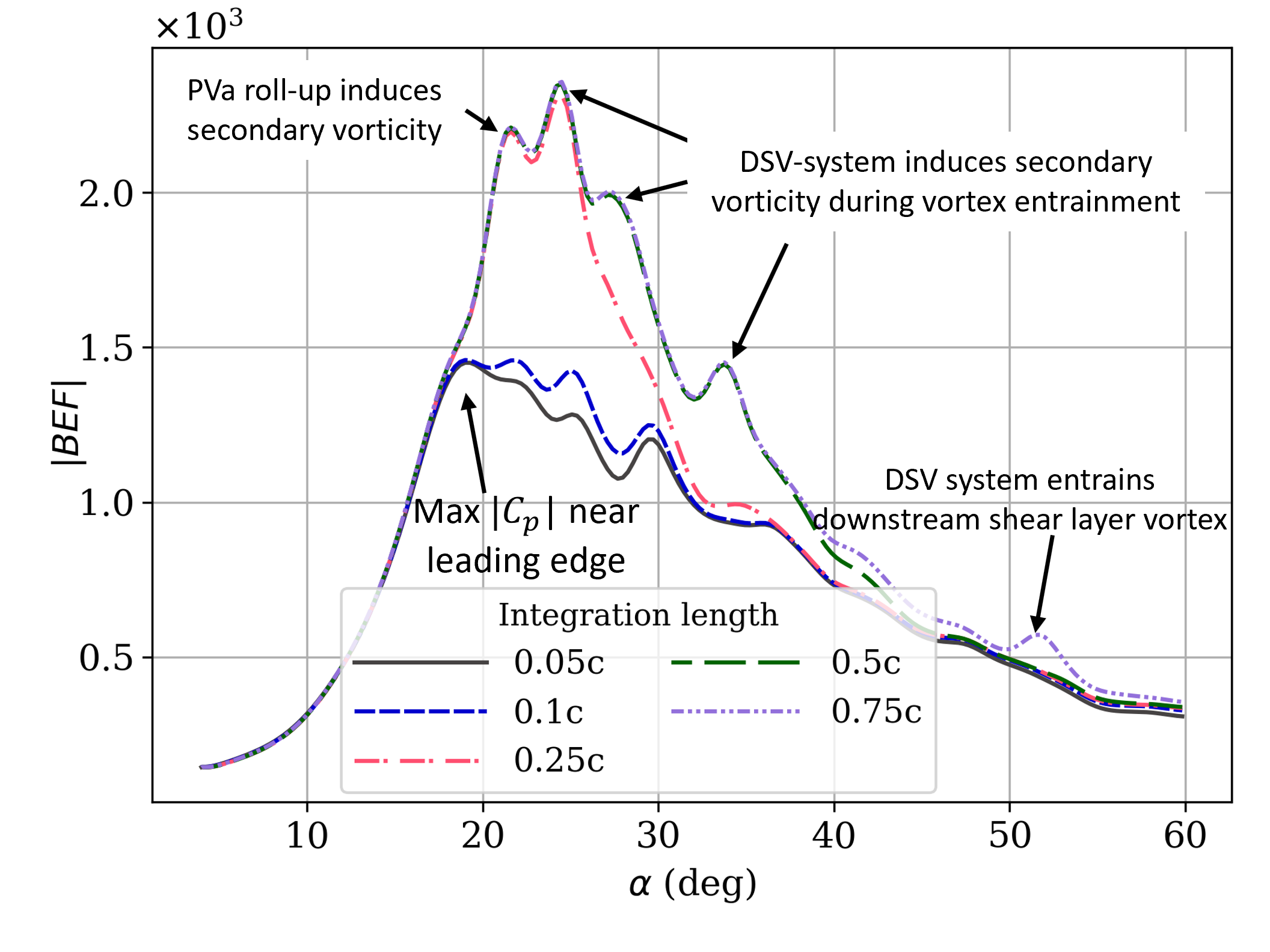}}
    \hspace*{\fill}
    \caption{$LESP$ (a) and $BEF$ (b) integrated over different chord lengths, plotted against $\alpha$, for Case R10-p25.}
    \label{fig:R10-p25_LESP_BEF}
\end{figure}

Figure~\ref{fig:R10-p25_LESP_BEF} shows the variation of $LESP$ and $\abs{BEF}$ for the present case for different integration lengths.
Both sets of curves are characterized by a series of peaks for all integration lengths.
The peak corresponding to the maximum leading edge $|C_p|$ is clearly observed from $BEF$ at the lowest integration length.
In contrast to the previous case, the newly shed shear layer vortices remain very close to the leading edge (for example, see PVb in Fig.~\ref{fig:R10-p25_vortcontours_b}).
Therefore, the contribution to $LESP$ from the suction induced by these vortices is significant, leading to multiple peaks.
A series of significant spikes in $\abs{BEF}$ is observed for larger integration lengths.
The first instance of such a peak around $22^{\circ}$ corresponds to the roll-up of vortex PVa (shown in Fig.~\ref{fig:R10-p25_vortcontours_a}).
The subsequent instances correspond to the DSV system entraining a new leading edge vortex, during which it moves closer to the airfoil surface, inducing strong secondary vorticity.
The trend of the $BEF$ localizing the points corresponding to peak $\abs{C_p}$ and DSV-induced secondary vorticity peaks holds for the present case as well.


\subsection{Comparison of results}
\label{sec:results_comparisons}
\begin{figure}
	\centering
	\includegraphics[width=0.6\textwidth]{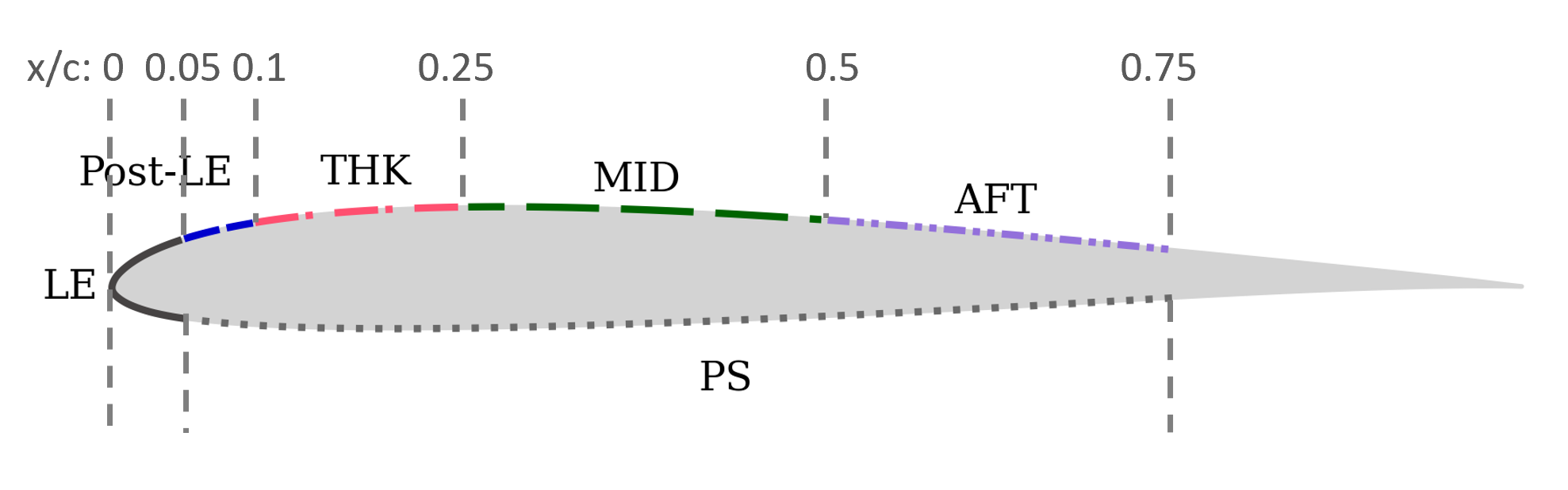}
	\caption{Different sections of the airfoil referred to in subsequent figures.}
	\label{fig:intlimits}
\end{figure}

We next distinguish between the max($LESP$) and max($\abs{BEF}$) criteria.
We divide the airfoil into multiple sections as shown in Fig.~\ref{fig:intlimits} and identify contributions from different regions of the airfoil to the two parameters.
Since the definition of $LESP$ involves a square root, we instead compare the suction force coefficient, $C_{\rm suction}$ (see Eq.~\ref{eq:LESP_def}).
The contributions to both $C_{\rm suction}$ and $BEF$ are primarily negative, so we plot the contributions to the negative of these quantities.

\begin{figure}
    \hspace*{\fill}
    \subcaptionbox{$C_{\rm suction}$\label{fig:R60-p05_CSseg}}{\incfig[width=0.49\textwidth]{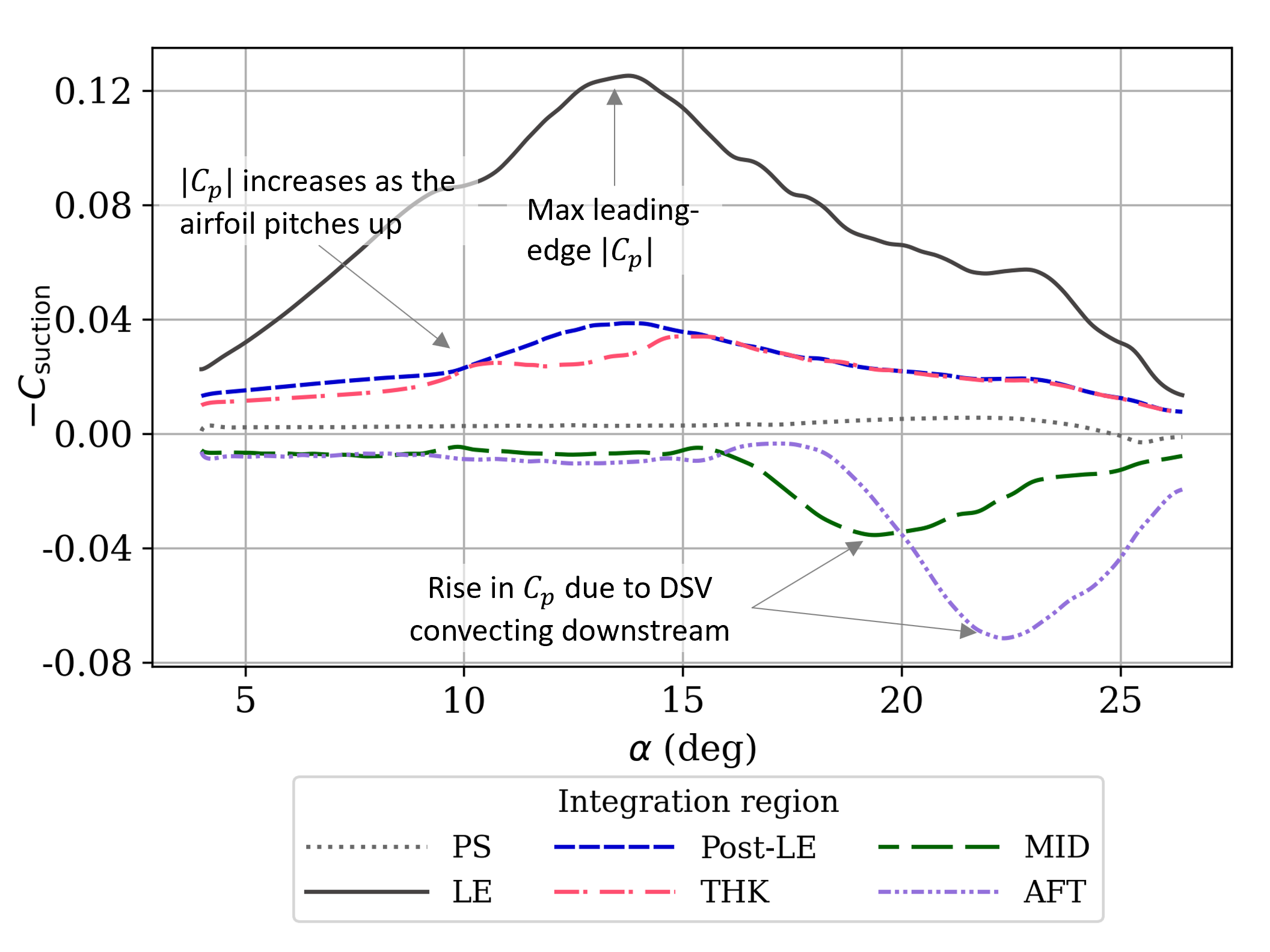}}
    \hfill
    \subcaptionbox{$BEF$\label{fig:R60-p05_BEFseg}}{\incfig[width=0.49\textwidth]{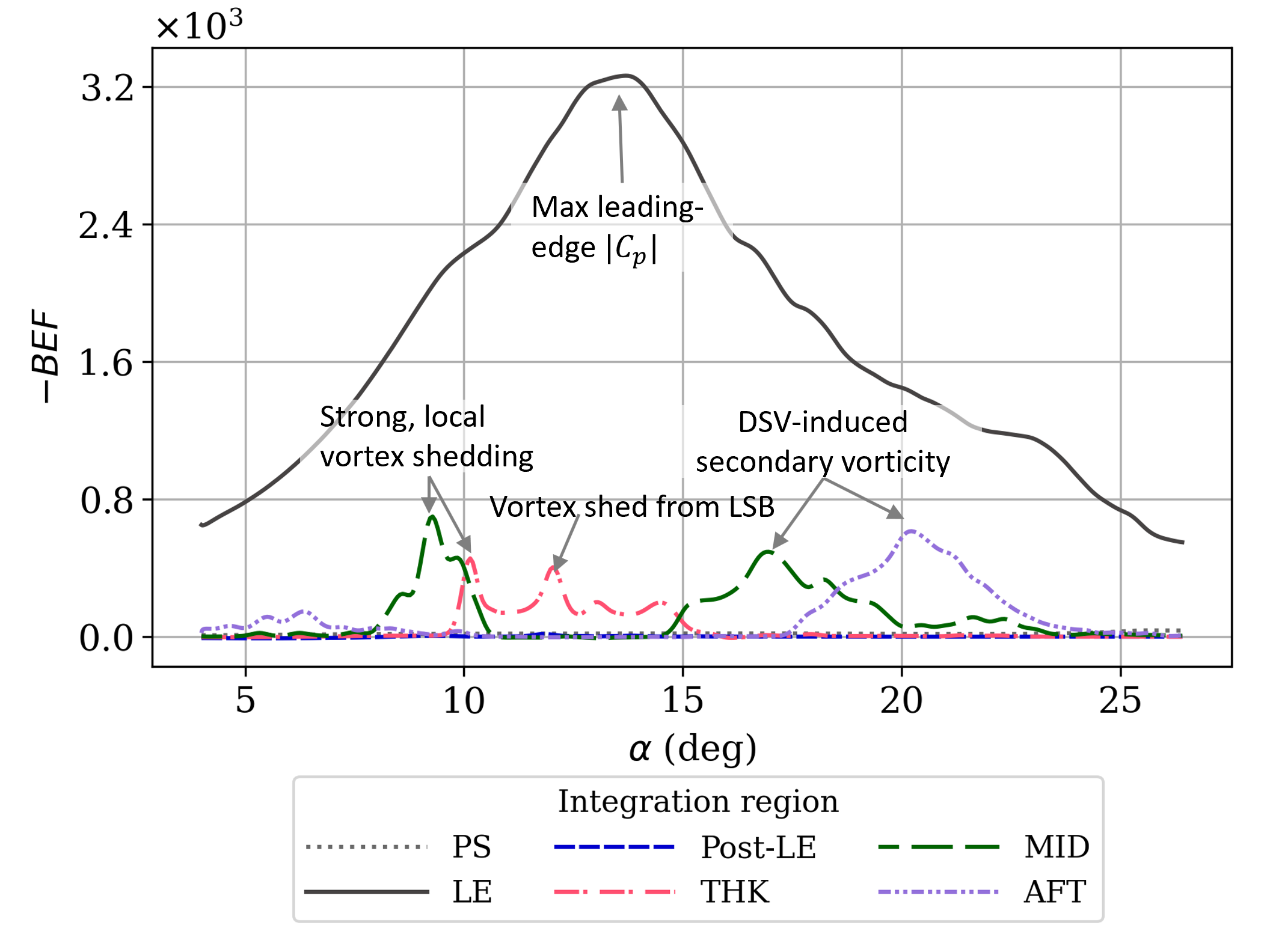}}
    \hspace*{\fill}
    \caption{Contributions to $C_{\rm suction}$ (a) and $BEF$ (b) from different airfoil segments, for Case R60-p05.}
    \label{fig:R60-p05_afseg}
\end{figure}

Figure~\ref{fig:R60-p05_afseg} shows the contributions to $C_{\rm suction}$ and $BEF$ from the airfoil sections marked in Fig.~\ref{fig:intlimits} for Case R60-p05.
The contribution from the pressure side (PS) to either parameter is negligible.
The largest contributions are from the leading edge region (marked `LE') for both $C_{\rm suction}$ and $BEF$.
The Post-LE and THK regions contribute to $C_{\rm suction}$ throughout the unsteady maneuver.
The effect of increasing (or decreasing) leading edge $\abs{C_p}$ is felt at these regions since there is a significant chord-wise component of the surface normal.
In contrast, the contribution to $BEF$ from variation in leading-edge $\abs{C_p}$ is limited to the LE region.
The Post-LE region does not contribute to $BEF$ because, even though $\abs{C_p}$ varies with $\alpha$ in this region, the pressure gradient (similar behavior to vorticity flux) drops to zero.
The local peaks in the $BEF$ curves that occur before the peak in the LE curve correspond to local vortex shedding, while those following it correspond to the DSV convecting downstream and growing stronger.
As noted in section~\ref{sec:R60-p05}, strong laminar vortices are shed from the THK and MID regions around $\alpha \sim 10^{\circ}$.
These are reflected as peaks in the corresponding $BEF$ curves in Fig.~\ref{fig:R60-p05_BEFseg}.
A strong vortex shed from the rear of the LSB is also captured in the THK region.
At higher $\alpha$, the MID and AFT regions show an opposite contribution to $C_{\rm suction}$ relative to other regions.
This arises from the convection of the DSV center downstream, which induces high $\abs{C_p}$ in the MID and AFT regions where the chord-wise component of the surface normal points in the $+x$ direction.
The peaks in $BEF$ from the MID and AFT regions at higher angles of attack($\alpha \sim 16.5^{\circ}$ and $20^{\circ}$) occur due to strong secondary vorticity induced by the DSV as it grows stronger while convecting downstream. 
These peaks occur later in time for $LESP$ compared to $BEF$ since the chordwise component of the surface normal is more significant towards the rear of the airfoil.

\begin{figure}
    \hspace*{\fill}
    \subcaptionbox{$C_{\rm suction}$\label{fig:R10-p25_CSseg}}{\incfig[width=0.49\textwidth]{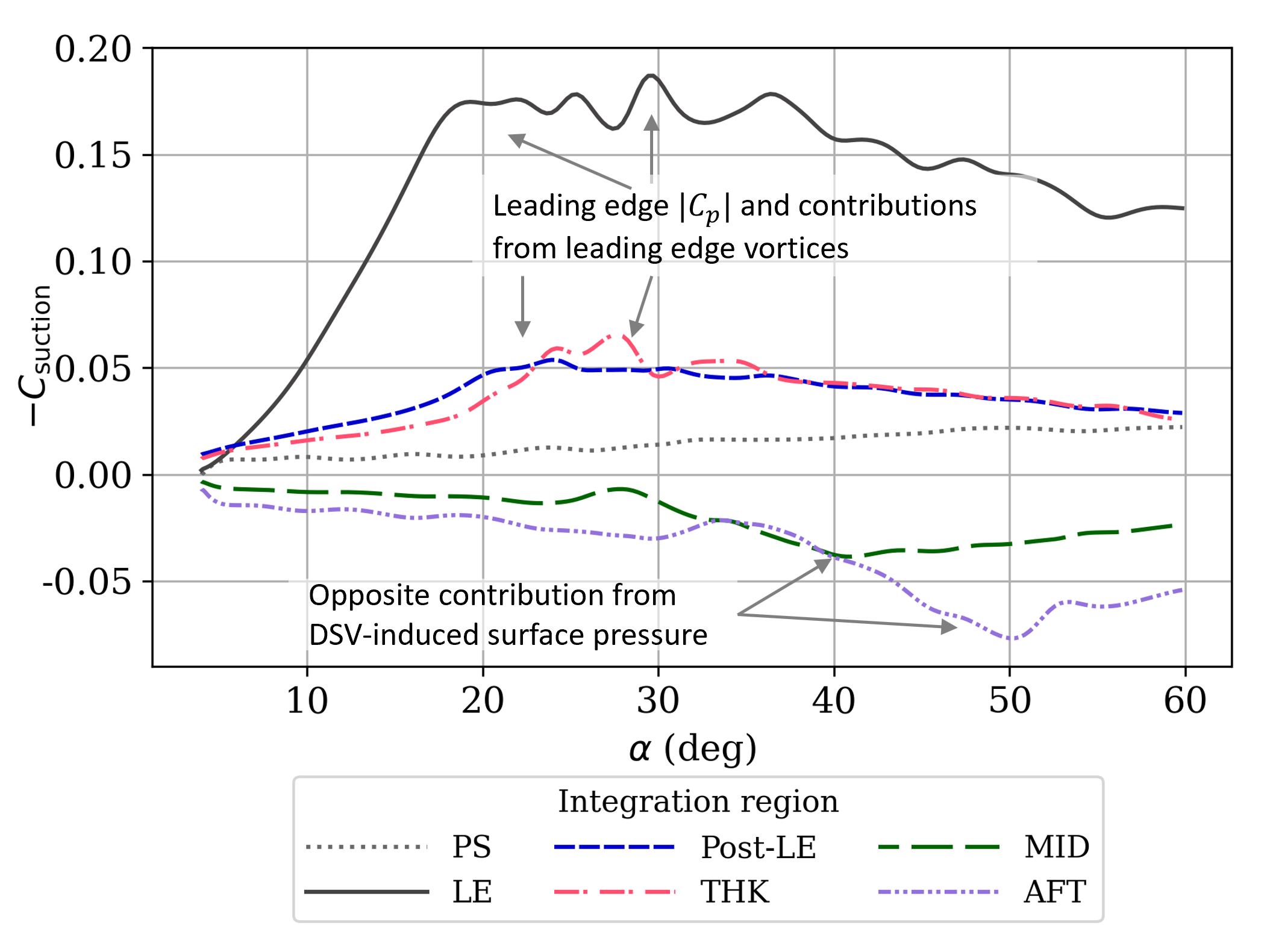}}
    \hfill
    \subcaptionbox{$BEF$\label{fig:R10-p25_BEFseg}}{\incfig[width=0.49\textwidth]{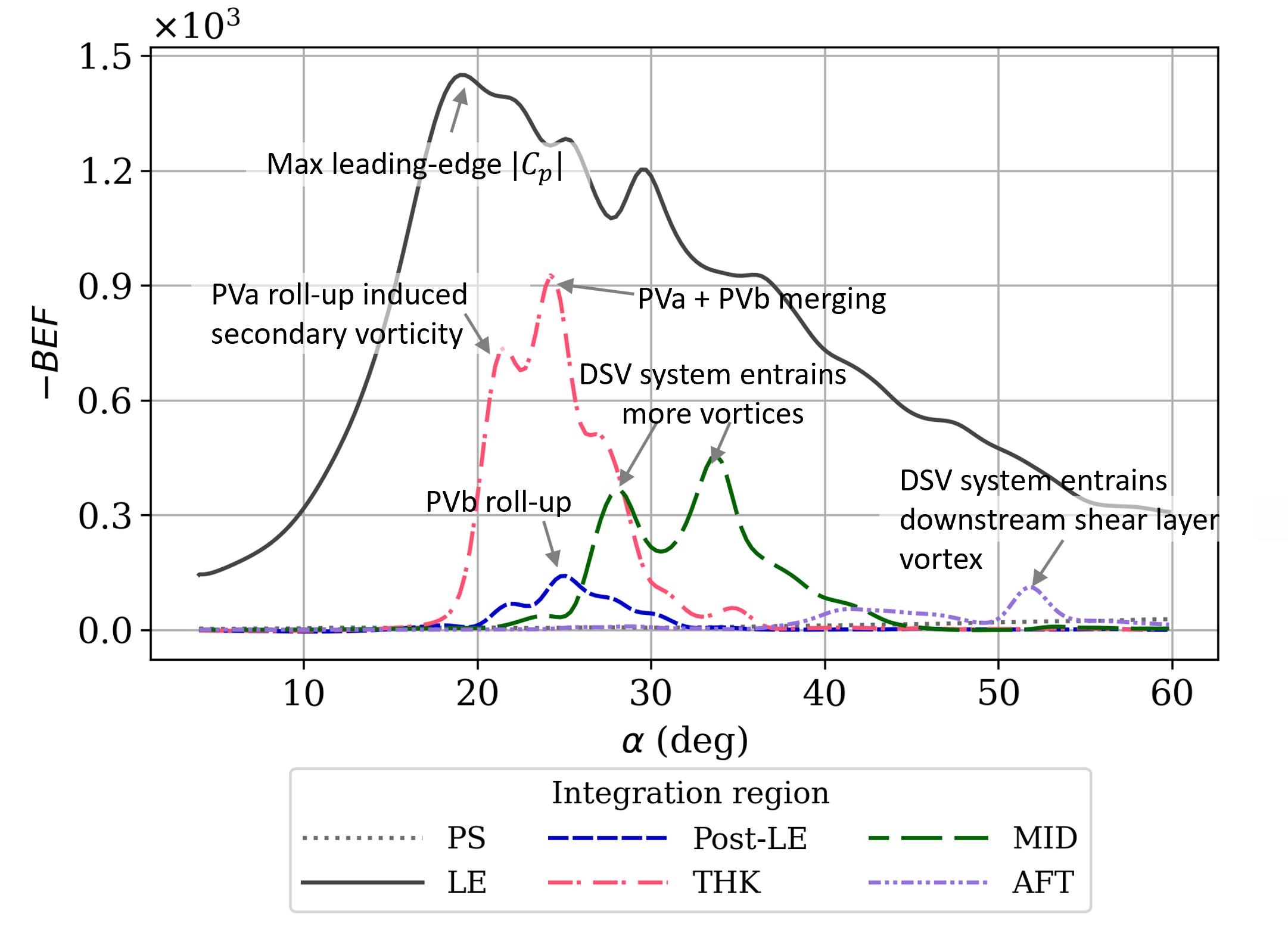}}
    \hspace*{\fill}
    \caption{Contributions to $C_{\rm suction}$ (a) and $BEF$ (b) from different airfoil segments, for Case R10-p25.}
    \label{fig:R10-p25_afseg}
\end{figure}

Case R10-p25 is also briefly presented in a similar way.
Figure~\ref{fig:R10-p25_afseg} shows the contributions to $C_{suction}$ and $BEF$ from different airfoil sections shown in Fig.~\ref{fig:intlimits}.
The contribution to $C_{suction}$ from LE includes multiple peaks, due to the leading edge vortices remaining farther upstream for this case.
In contrast, the LE contribution to $BEF$ shows a clear peak, following the same trend as max($\abs{C_p}$) near the leading edge (see bottom panel of Fig.~\ref{fig:R10-p25_aerodyn_coeff}).
Specific instances of vortex-roll-up/vortex entrainment, which induce increased secondary vorticity into the DSV system are observed from the other regions.
These include the roll-up of PVa, PVb, merging of PVa and PVb, and the entrainment of further leading edge vortices and the downstream shear layer vortex into the DSV system.
The contributions to $C_{suction}$ from the Post-LE and THK regions are influenced by the variation of leading edge $C_p$ and leading edge vortices.
Opposite contributions to $C_{suction}$ due to DSV convection are captured in the MID and AFT regions.
A similar breakdown of contributions to $C_{suction}$ and $BEF$ for the remaining cases, namely, R60-p25 and R10-p05 is presented in Appendix~\ref{app:BEF_LESP_contributions}.

\begin{figure}
	\centering
 	\incfig[width=1.0\textwidth]{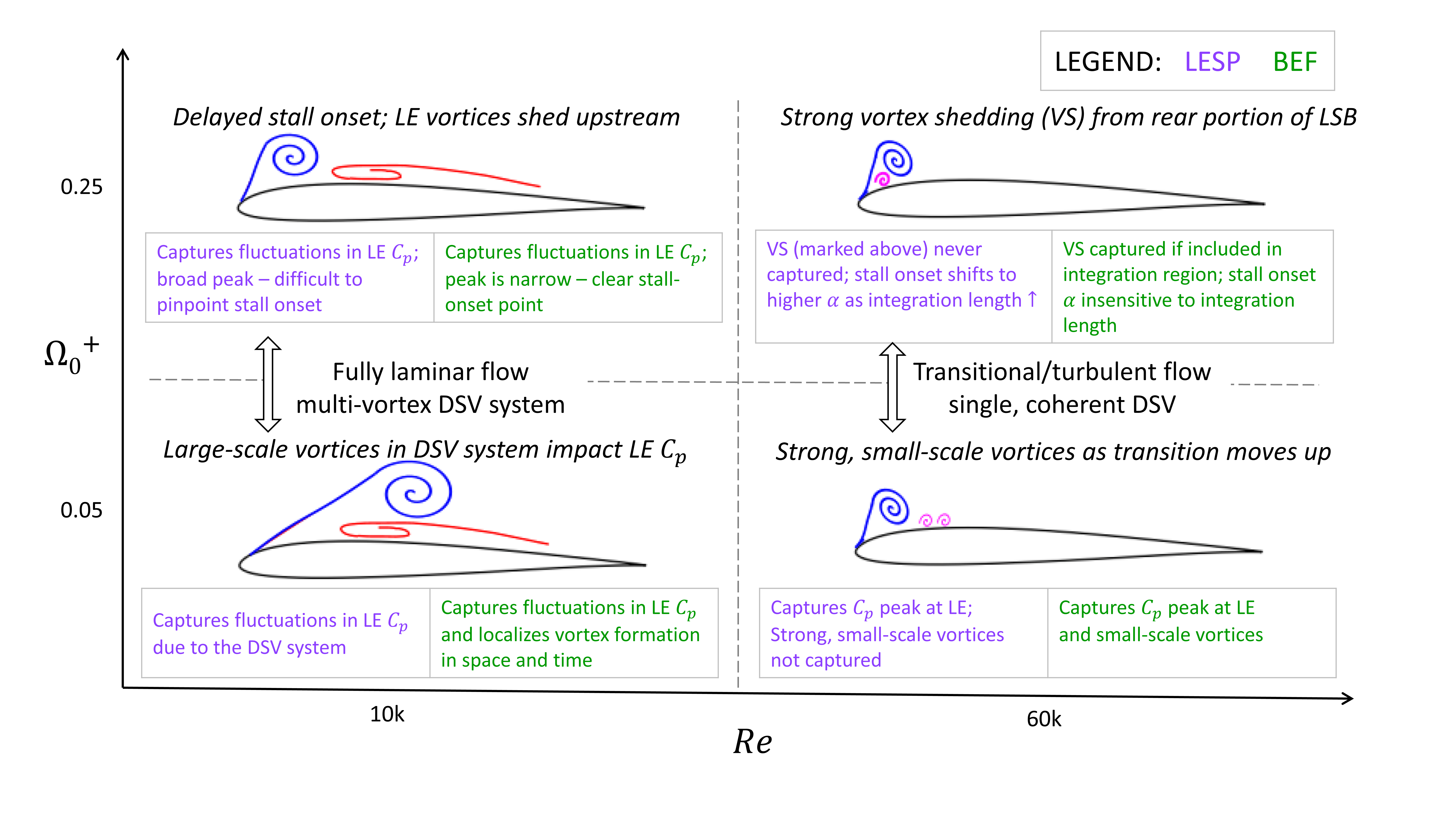}
	\caption{Illustration of observations from all four cases considered in the present study}
	\label{fig:schematic_allcases}
\end{figure}

Figure~\ref{fig:schematic_allcases} illustrates the overall picture that emerges from the results of the four simulations, including the performance of $LESP$ and $BEF$ in identifying critical flow events.
At $Re = 10,000$ (cases R10-p05 and R10-p25), the flow remains laminar almost through the entire maneuver.
There is a large region of reversed flow, and the $C_f$ is negative over nearly the entire suction surface of the airfoil. 
Due to the larger viscous response, the DSV system comprises multiple large-scale laminar vortices. 
These vortices are shed further upstream (close to the airfoil leading edge) for the high pitch-rate case.

Cases with $Re = 60,000$ (R60-p05 and R60-p25) demonstrate a more typical LSB-bursting, leading-edge stall.
During the establishment of the LSB, strong, small-scale vortices are shed.
For the higher pitch rate case, the DSV forms farther upstream and is stronger, as reported in the literature~\shortcite{Acharya1992}. 
The DSV is fed by the vorticity shed from the leading edge, but due to the higher $Re$, it does not organize into large coherent vortices before merging with the DSV.

Of the two stall criteria under investigation, max($LESP$) mainly captures variations in leading edge $C_p$.
The max($\abs{BEF}$) criterion captures the variation in leading edge $C_p$ and also enables the localization of vortex shedding events occurring anywhere on the airfoil surface in space and time.
This includes strong, small-scale vortex shedding that is not reflected in the leading edge $C_p$ as well as large-scale vortex shedding events located away from the leading edge.
These events are directly captured in the spatial region where they occur.
Specific instances of vortex roll-up and entrainment, which trigger an increase in induced secondary vorticity are captured by the $BEF$.
The versatility of the $BEF$ parameter makes it a suitable candidate for flow control at low $Re$.

\section{Conclusion}
\label{sec:conclusion}
We investigate airfoil dynamic stall at low Reynolds numbers where laminar, transitional, and turbulent regimes can coexist, giving rise to rich fluid dynamics.
The dynamic stall process is strongly influenced by multiple vortices that shed from different regions of the airfoil not limited to the leading edge.
Current state-of-the-art stall onset criteria based on the $LESP$ and $BEF$ parameters, are calculated by integrating these quantities around the leading edge and hence may not directly capture vortex shedding occurring away from it.

We evaluate the max($LESP$) and max($\abs{BEF}$) criteria over an extended integration region for low-$Re$ ($\sim \mathcal{O}\left( 10^4 \right) $) dynamic stall for an LES dataset consisting of the SD7003 airfoil undergoing a constant-rate, pitch-up maneuver at two $Re$ values at two pitch rates.
The highly-resolved LES results provide insights into the unsteady flow phenomena (instabilities, shear layer dynamics, vortex formation, pairing, shedding, dissipation, etc.), which are commented on. 
The ability of the $LESP$ and $BEF$ parameters to capture vortex-shedding events is analyzed using these results.

Our analyses indicate that the max($\abs{BEF}$) criterion captures strong vortex-shedding events when the integration region includes the locations of these events.
These events are reflected as local maxima in the $BEF$ curves.
The $LESP$, being based on the camber-wise component of pressure force, captures only the effect of vortex shedding events that significantly affect $C_p$ near the leading edge.
The $BEF$ parameter differs from  $LESP$ in that it \textit{directly} captures the flow events such as (a) strong, small-scale vortex shedding events that do not significantly impact leading edge $C_p$, and (b) large-scale shed vortices, localizing them in space in time.
It captures instances of vortex roll-up and entrainment, which induce increased secondary vorticity.
Therefore, the $BEF$ parameter can be used to localize stall onset and vortex-shedding events in space and time for effective flow control at low $Re$.

\section*{Acknowledgments}
\label{sec:acknowledgments}
This material is based upon work supported by the National Science Foundation (Grants CBET-1935255 and 1554196) and the US Air Force Office of Scientific Research (Award \# FA9550-23-1-0016). We also acknowledge the computational resources provided by Argonne Leadership Computing Facility and Iowa State University.

\section*{Declaration of Interests}
The authors report no conflict of interest.

\bibliographystyle{apacite}
\bibliography{main}

\appendix
\appendixpage
\section{Static simulations}
\label{app:staticsim}
\begin{figure}[h!]
    \hspace*{\fill}
    \subcaptionbox{surface pressure coeff., $C_p$\label{fig:static_cp_re60k}}{\incfig[width=0.49\textwidth]{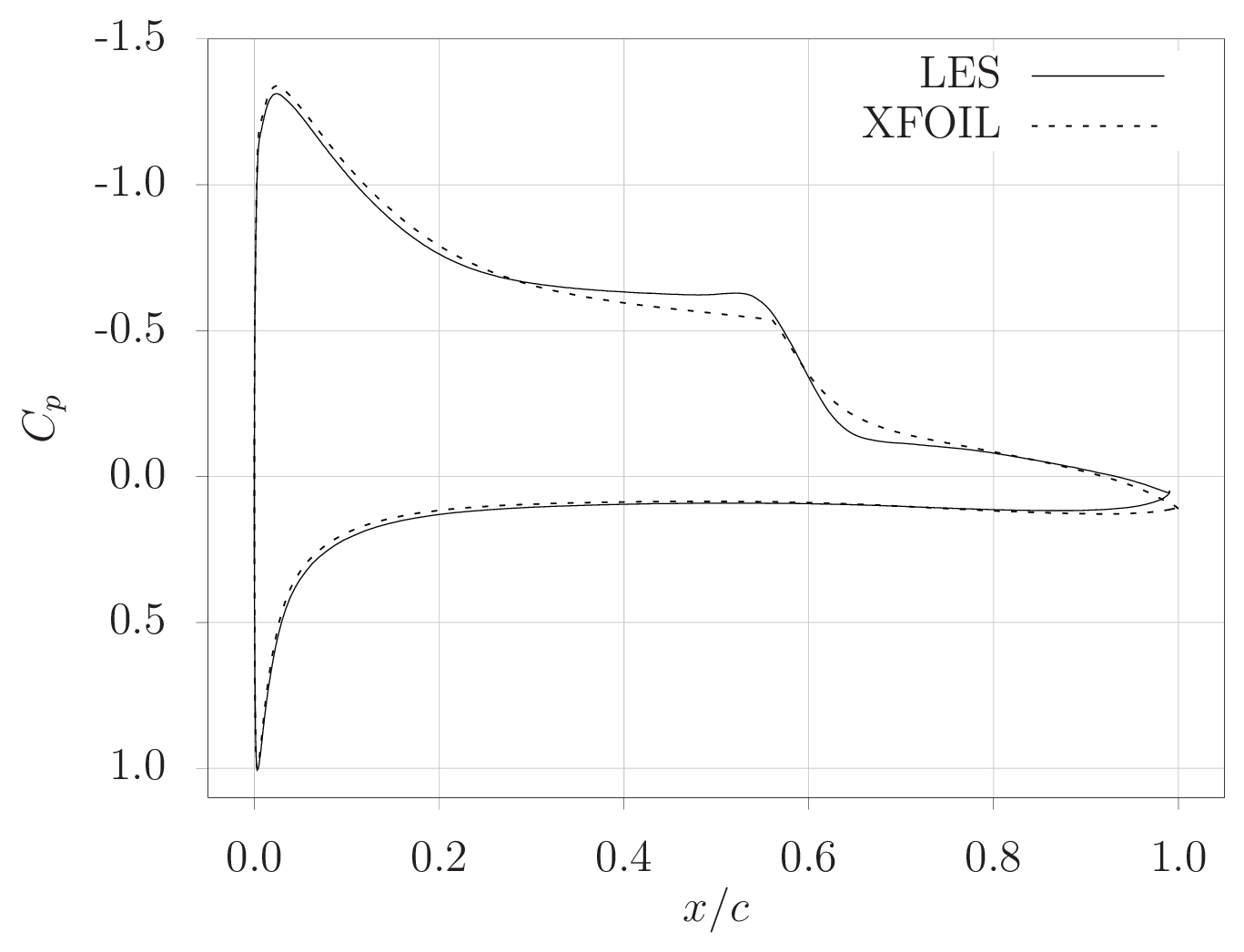}}
    \hfill
    \subcaptionbox{skin friction coeff., $C_f$\label{fig:static_cf_re60k}}{\incfig[width=0.49\textwidth]{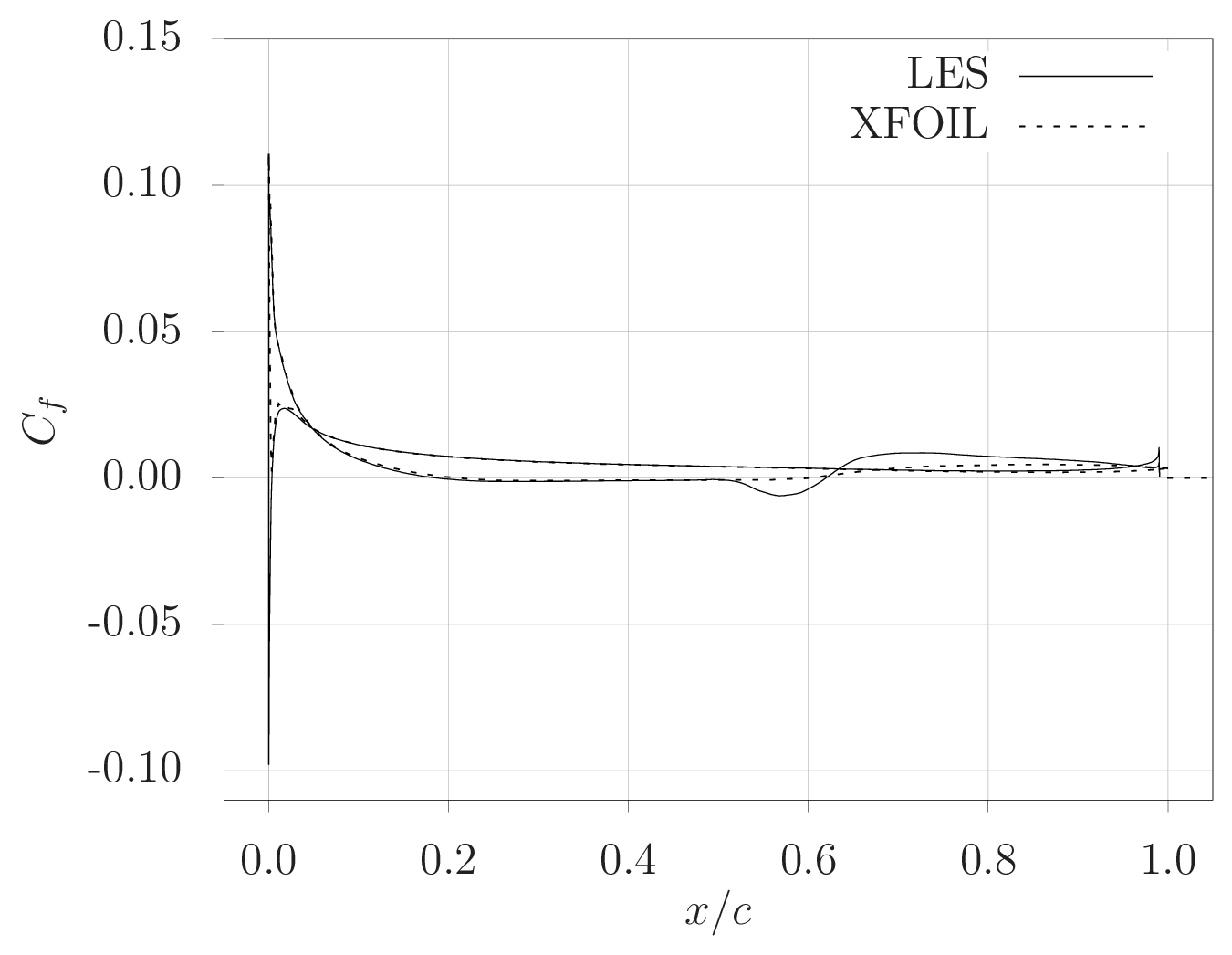}}
    \hspace*{\fill}
	\caption{Comparison of $C_p$ and $C_f$ distributions between static LES and XFOIL for $Re$ 60,000 at $\alpha = 4^{\circ}$.}
	\label{fig:static_results_re60k}
\end{figure}

Static simulations were performed using FDL3DI at $\alpha = 4^{\circ}$ for both $Re$; only the results for $Re=60,000$ are shown for brevity.
The simulations were run for about $40$ characteristic convective times (${c}/{U_{\infty}}$) to ensure that the forces and moments reached statistical stationarity.
Comparisons of surface pressure and skin friction coefficient distributions, $C_p$ and $C_f$, repectively, are made with XFOIL \shortcite{drela1989xfoil}. 
XFOIL results are obtained with the Ncrit parameter set to $9$, corresponding to a low freestream turbulence intensity.
Figure~\ref{fig:static_results_re60k} shows good agreement between LES and XFOIL results; the transition on the suction surface occurs around $60\%$ chord in both.

\section{Contributions to \texorpdfstring{$C_{\rm suction}$}{Csuction} and \texorpdfstring{$BEF$}{BEF} from different airfoil regions}
\label{app:BEF_LESP_contributions}
\begin{figure}[h!]
    \hspace*{\fill}
    \subcaptionbox{$C_{\rm suction}$, R60-p25\label{fig:R60-p25_CSseg}}{\incfig[width=0.48\textwidth,trim={0cm 2.3cm 0cm 0cm},clip]{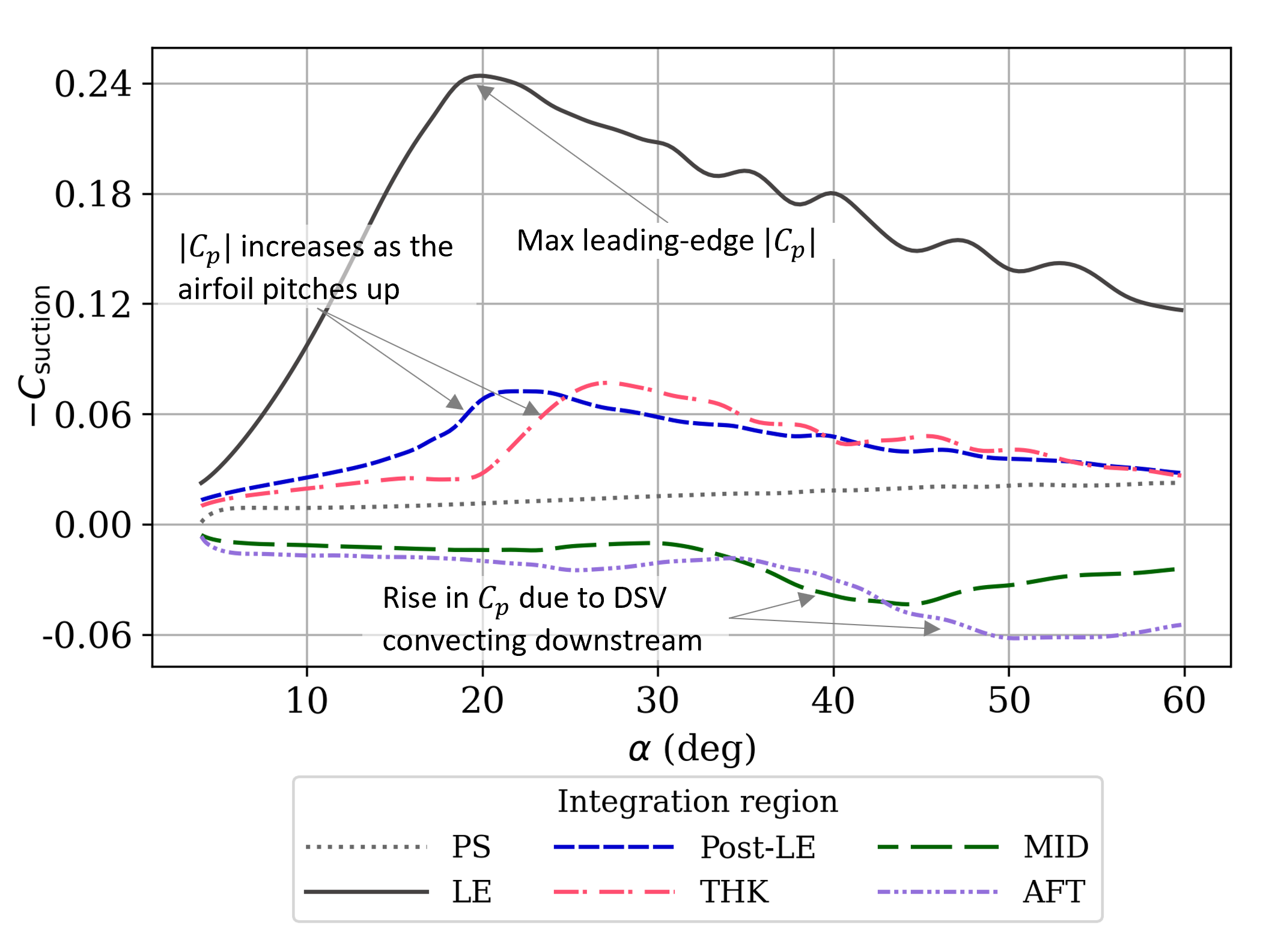}}
    \hfill
    \subcaptionbox{$BEF$, R60-p25\label{fig:R60-p25_BEFseg}}{\incfig[width=0.48\textwidth,trim={0cm 2.3cm 0cm 0cm},clip]{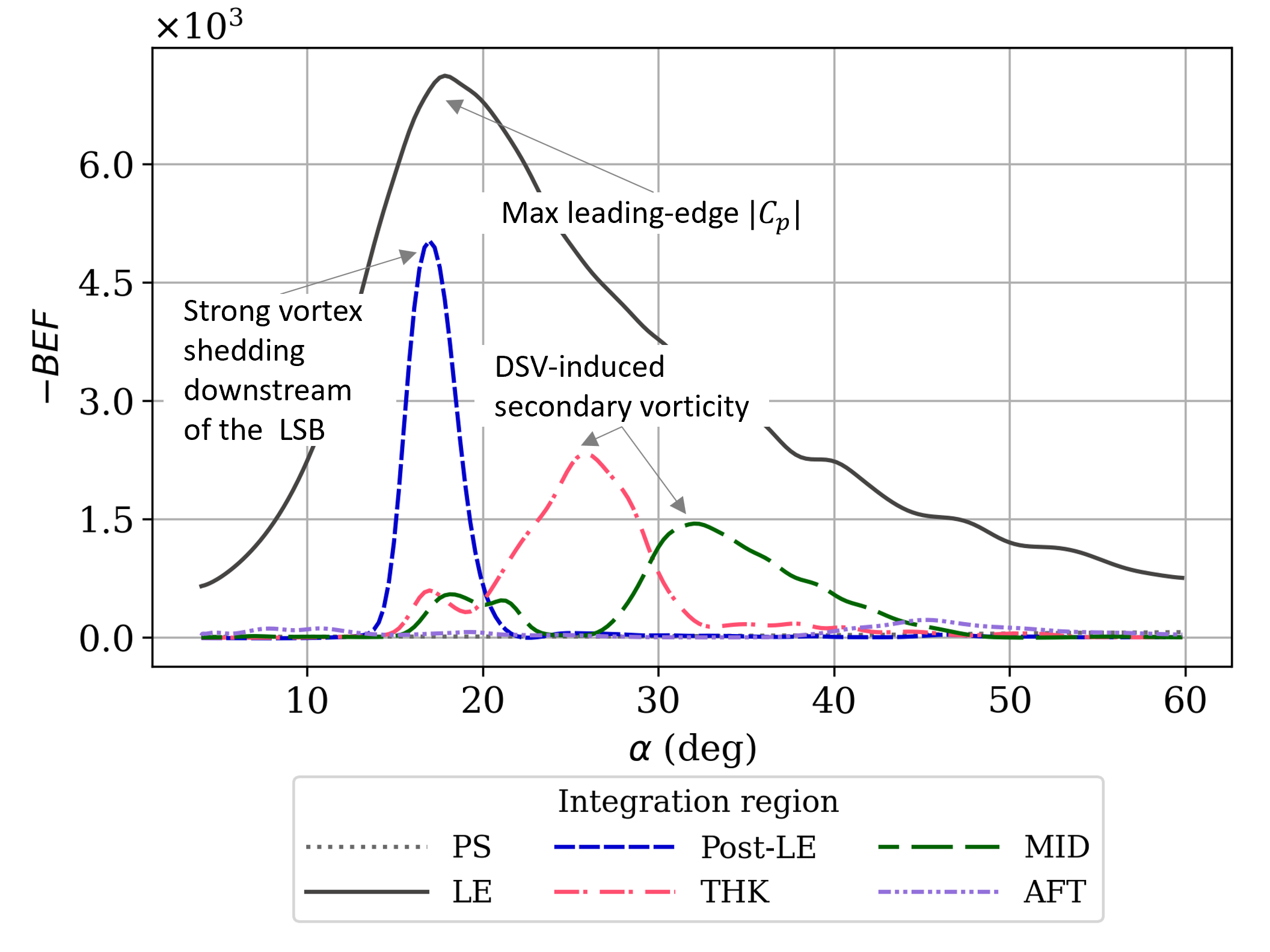}}
    \hspace*{\fill} \\
    \hspace*{\fill}
    \subcaptionbox{$C_{\rm suction}$, R10-p05\label{fig:R10-p05_CSseg}}{\incfig[width=0.48\textwidth,trim={0cm 2.3cm 0cm 0cm},clip]{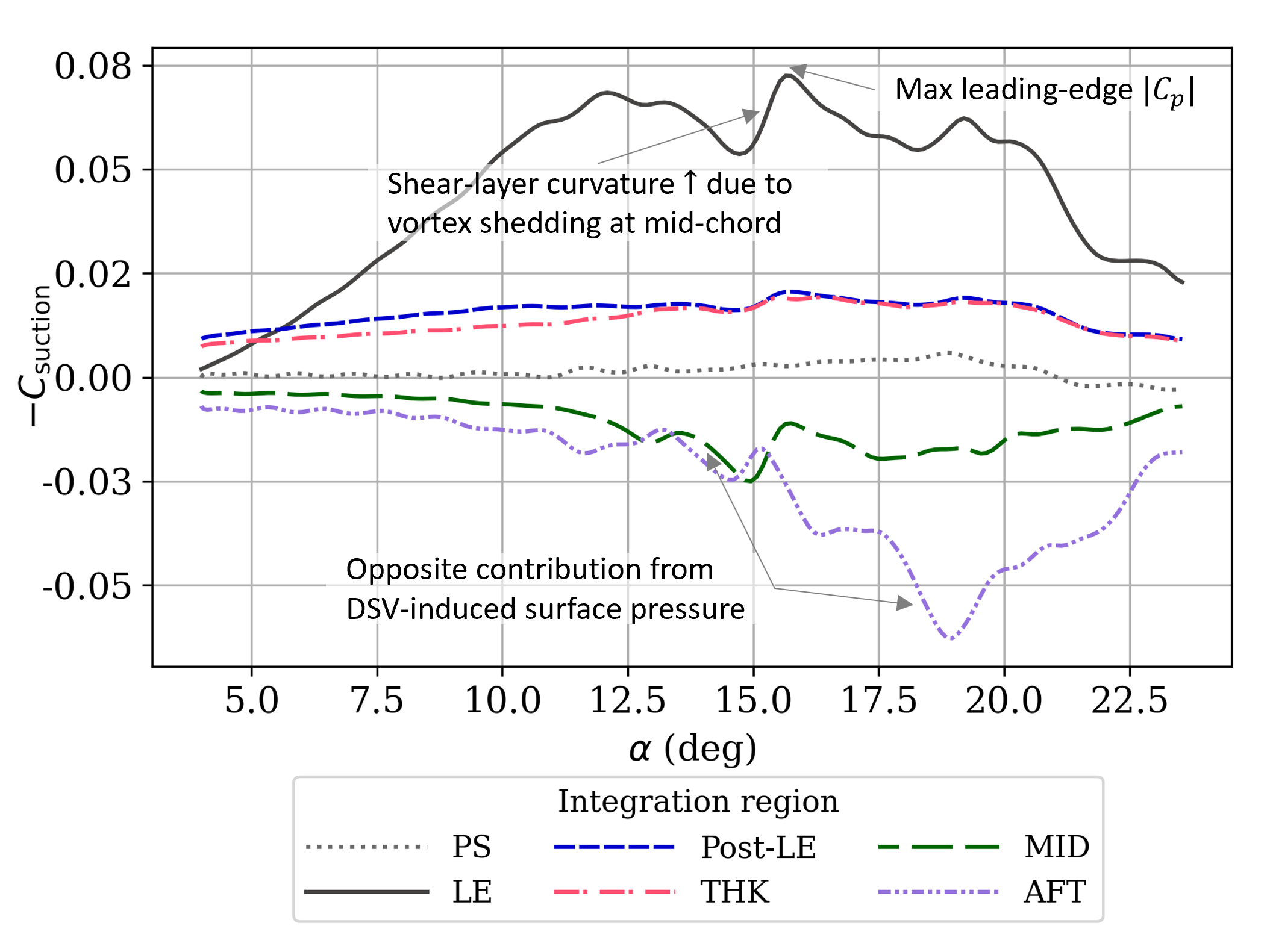}}
    \hfill
    \subcaptionbox{$BEF$, R10-p05\label{fig:R10-p05_BEFseg}}{\incfig[width=0.48\textwidth,trim={0cm 2.3cm 0cm 0cm},clip]{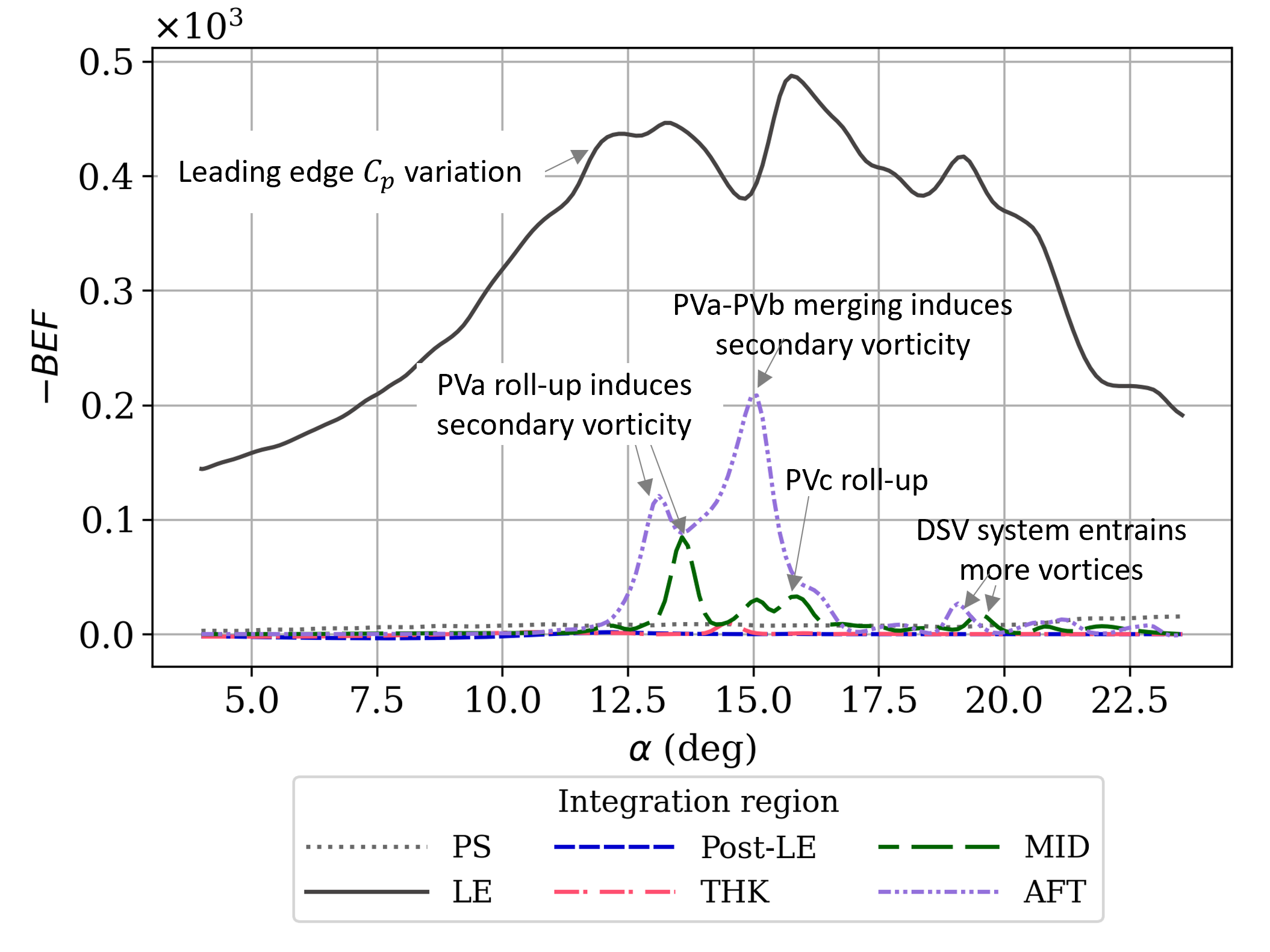}}
    \hspace*{\fill} 
    %
    \caption{Contributions to $C_{\rm suction}$ and $BEF$ from different airfoil segments, for cases R60-p25 (a,b) and R10-p05 (c,d).}
    \label{fig:afsegs}
\end{figure}

The contributions to $C_{\rm suction}$ and $BEF$ from different sections of the airfoil (see Fig.~\ref{fig:intlimits}) for Cases R60-p25 and R10-p05 are shown in Fig.~\ref{fig:afsegs}. Critical flow events identified by the two parameters are annotated in the figure.

\end{document}